\documentclass[twocolumn]{aastex63} 

\usepackage{graphicx} %Required for including pictures
\usepackage{float} % Allows putting an [H] in \begin{figure} to specify the exact location of the figure
\usepackage{wrapfig} % Allows in-line images such as the example fish picture
\usepackage{lipsum} % Used for inserting dummy 'Lorem ipsum' text into the template
\usepackage{hyperref}
\usepackage{times}
\usepackage{amsmath}
\usepackage{graphicx}
\usepackage{subfigure}
\usepackage{placeins}
\usepackage{hyperref}
\usepackage{gensymb}
\usepackage{upgreek}
\usepackage{natbib}

\usepackage{graphicx}
\usepackage{subfigure}
\usepackage{multirow}
\usepackage{comment}
\usepackage{natbib}
\usepackage{hyperref}
\usepackage{mathtools}
\usepackage{mathrsfs}
\usepackage{fontenc}
\usepackage{color}
\usepackage{url}
\usepackage{hyperref}
\usepackage{gensymb}
\usepackage{pifont}
\graphicspath{{./}}

\newcommand       \K            {\,{\rm K}}
\newcommand       \mum          {{\rm \mu m}}

\usepackage{afterpage}

\shorttitle{Ionized Gas Outflows in GATOS Seyferts}
\shortauthors{Zhang et al.}

\begin{document}

\title{The Galaxy Activity, Torus, and Outflow Survey (GATOS). (IV): Exploring Ionized Gas Outflows in Central Kiloparsec Regions of GATOS Seyferts}

\author[0000-0003-4937-9077]{Lulu Zhang}
\affiliation{The University of Texas at San Antonio, One UTSA Circle, San Antonio, TX 78249, USA; lulu.zhang@utsa.edu; l.l.zhangastro@gmail.com}

\author[0000-0001-7827-5758]{Chris Packham}
\affiliation{The University of Texas at San Antonio, One UTSA Circle, San Antonio, TX 78249, USA; lulu.zhang@utsa.edu; l.l.zhangastro@gmail.com}
\affiliation{National Astronomical Observatory of Japan, National Institutes of Natural Sciences (NINS), 2-21-1 Osawa, Mitaka, Tokyo 181-8588, Japan}

\author[0000-0002-4457-5733]{Erin K. S. Hicks}
\affiliation{Department of Physics and Astronomy, University of Alaska Anchorage, Anchorage, AK 99508-4664, USA}
\affiliation{The University of Texas at San Antonio, One UTSA Circle, San Antonio, TX 78249, USA; lulu.zhang@utsa.edu; l.l.zhangastro@gmail.com}
\affiliation{Department of Physics, University of Alaska, Fairbanks, Alaska 99775-5920, USA}

\author{Ric I. Davies}
\affiliation{Max-Planck-Institut für extraterrestrische Physik, Postfach 1312, D-85741, Garching, Germany}

\author{Taro T. Shimizu}
\affiliation{Max-Planck-Institut für extraterrestrische Physik, Postfach 1312, D-85741, Garching, Germany}

\author{Almudena Alonso-Herrero}
\affiliation{Centro de Astrobiolog\'{\i}a (CAB), CSIC-INTA, Camino Bajo del Castillo s/n, E-28692 Villanueva de la Ca\~nada, Madrid, Spain}

\author{Laura Hermosa Mu{\~n}oz}
\affiliation{Centro de Astrobiolog\'{\i}a (CAB), CSIC-INTA, Camino Bajo del Castillo s/n, E-28692 Villanueva de la Ca\~nada, Madrid, Spain}

\author{Ismael Garc{\'i}a-Bernete}
\affiliation{Centro de Astrobiolog\'{\i}a (CAB), CSIC-INTA, Camino Bajo del Castillo s/n, E-28692 Villanueva de la Ca\~nada, Madrid, Spain}
\affiliation{Department of Physics, University of Oxford, Keble Road, Oxford OX1 3RH, UK}

\author{Miguel Pereira-Santaella}
\affiliation{Instituto de F{\'i}sica Fundamental, CSIC, Calle Serrano 123, 28006 Madrid, Spain}

\author{Anelise Audibert}
\affiliation{Instituto de Astrof{\'i}sica de Canarias, Calle V{\'i}a L{\'a}ctea, s/n, E-38205, La Laguna, Tenerife, Spain}
\affiliation{Departamento de Astrof{\'i}sica, Universidad de La Laguna, E-38206, La Laguna, Tenerife, Spain}

\author{Enrique L{\'o}pez-Rodríguez}
\affiliation{Department of Physics \& Astronomy, University of South Carolina, Columbia, SC 29208, USA}
\affiliation{Kavli Institute for Particle Astrophysics \& Cosmology (KIPAC), Stanford University, Stanford, CA 94305, USA}

\author{Enrica Bellocchi}
\affiliation{Departmento de F{\'i}sica de la Tierra y Astro{\'i}sica, Fac. de CC F{\'i}sicas, Universidad Complutense de Madrid, E-28040 Madrid, Spain}
\affiliation{Instituto de F{\'i}sica de Partículas y del Cosmos IPARCOS, Fac. CC F{\'i}sicas, Universidad Complutense de Madrid, E-28040 Madrid, Spain}

\author{Andrew J. Bunker}
\affiliation{Department of Physics, University of Oxford, Keble Road, Oxford OX1 3RH, UK}

\author{Francoise Combes}
\affiliation{LERMA, Observatoire de Paris, Coll{\`e}ge de France, PSL University, CNRS, Sorbonne University, Paris}

\author{Tanio D{\'i}az-Santos}
\affiliation{Institute of Astrophysics, Foundation for Research and Technology-Hellas (FORTH), Heraklion 70013, Greece}
\affiliation{School of Sciences, European University Cyprus, Diogenes street, Engomi, 1516 Nicosia, Cyprus}

\author{Poshak Gandhi}
\affiliation{School of Physics \& Astronomy, University of Southampton, Hampshire SO17 1BJ, Southampton, UK}

\author{Santiago Garc{\'i}a-Burillo}
\affiliation{Observatorio Astron{\'o}mico Nacional (OAN-IGN)-Observatorio de Madrid, Alfonso XII, 3, 28014, Madrid, Spain}

\author{Bego{\~n}a Garc{\'i}a-Lorenzo}
\affiliation{Instituto de Astrof{\'i}sica de Canarias, Calle V{\'i}a L{\'a}ctea, s/n, E-38205, La Laguna, Tenerife, Spain}
\affiliation{Departamento de Astrof{\'i}sica, Universidad de La Laguna, E-38206, La Laguna, Tenerife, Spain}

\author{Omaira Gonz{\'a}lez-Mart{\'i}n}
\affiliation{Instituto de Radioastronom{\'i}a and Astrof{\'i}sica (IRyA-UNAM), 3-72 (Xangari), 8701, Morelia, Mexico}
%\affiliation{Instituto de Radioastronomía y Astrof{\'i}sica (IRyA), Universidad Nacional Aut{\'o}noma de M{\'e}xico, Antigua Carretera a P{\'a}tzcuaro 8701, ExHda. San Jos{\'e} de la Huerta, Morelia, Michoacán, M{\'e}xico C.P. 58089}

\author{Masatoshi Imanishi}
\affiliation{National Astronomical Observatory of Japan, National Institutes of Natural Sciences, 2-21-1 Osawa, Mitaka, Tokyo 181-8588, Japan}
\affiliation{Department of Astronomy, School of Science, Graduate University for Advanced Studies (SOKENDAI), Mitaka, Tokyo 181-8588, Japan}

\author{Alvaro Labiano}
\affiliation{Telespazio UK for the European Space Agency (ESA), ESAC, Camino Bajo del Castillo s/n, 28692 Villanueva de la Ca{\~n}ada, Spain}

\author{Mason T. Leist}
\affiliation{The University of Texas at San Antonio, One UTSA Circle, San Antonio, TX 78249, USA; lulu.zhang@utsa.edu; l.l.zhangastro@gmail.com}

\author[0000-0003-4209-639X]{Nancy A. Levenson}
\affiliation{Space Telescope Science Institute, 3700 San Martin Drive Baltimore, Maryland 21218, USA}

\author{Cristina Ramos Almeida}
\affiliation{Instituto de Astrof{\'i}sica de Canarias, Calle V{\'i}a L{\'a}ctea, s/n, E-38205, La Laguna, Tenerife, Spain}
\affiliation{Departamento de Astrof{\'i}sica, Universidad de La Laguna, E-38206, La Laguna, Tenerife, Spain}

\author[0000-0001-5231-2645]{Claudio Ricci}
\affiliation{Instituto de Estudios Astrof{\'i}sicos, Facultad de Ingenier{\'i}a y Ciencias, Universidad Diego Portales, Av. Ej{\'e}rcito Libertador 441, Santiago, Chile}
\affiliation{Kavli Institute for Astronomy and Astrophysics, Peking University, Beijing 100871, People’s Republic of China}

\author{Dimitra Rigopoulou}
\affiliation{Department of Physics, University of Oxford, Keble Road, Oxford OX1 3RH, UK}
\affiliation{School of Sciences, European University Cyprus, Diogenes street, Engomi, 1516 Nicosia, Cyprus}

\author{David J. Rosario}
\affiliation{School of Mathematics, Statistics and Physics, Newcastle University, Newcastle upon Tyne, NE1 7RU, UK}

\author{Marko Stalevski}
\affiliation{Astronomical Observatory, Volgina 7, 11060 Belgrade, Serbia}
\affiliation{Sterrenkundig Observatorium, Universiteit Gent, Krijgslaan 281-S9, Gent B-9000, Belgium}

\author{Martin J. Ward}
\affiliation{Centre for Extragalactic Astronomy, Durham University, South Road, Durham DH1 3LE, UK}

\author{Donaji Esparza-Arredondo}
\affiliation{Instituto de Astrof{\'i}sica de Canarias, Calle V{\'i}a L{\'a}ctea, s/n, E-38205, La Laguna, Tenerife, Spain}
\affiliation{Departamento de Astrof{\'i}sica, Universidad de La Laguna, E-38206, La Laguna, Tenerife, Spain}

\author{Dan Delaney}
\affiliation{Department of Physics and Astronomy, University of Alaska Anchorage, Anchorage, AK 99508-4664, USA}
\affiliation{Department of Physics, University of Alaska, Fairbanks, Alaska 99775-5920, USA}

\author{Lindsay Fuller}
\affiliation{The University of Texas at San Antonio, One UTSA Circle, San Antonio, TX 78249, USA; lulu.zhang@utsa.edu; l.l.zhangastro@gmail.com}

\author{Houda Haidar}
\affiliation{School of Mathematics, Statistics and Physics, Newcastle University, Newcastle upon Tyne, NE1 7RU, UK}

\author{Sebastian H{\"o}nig}
\affiliation{School  of Physics \& Astronomy, University of Southampton, Hampshire SO17 1BJ, Southampton, UK}

\author{Takuma Izumi}
\affiliation{National Astronomical Observatory of Japan, National Institutes of Natural Sciences, 2-21-1 Osawa, Mitaka, Tokyo 181-8588, Japan}
\affiliation{Department of Astronomy, School of Science, Graduate University for Advanced Studies (SOKENDAI), Mitaka, Tokyo 181-8588, Japan}

\author{Daniel Rouan}
\affiliation{LESIA, Observatoire de Paris, Universit{\'e} PSL, CNRS, Sorbonne Universit{\'e}, Sorbonne Paris Cite{\'e}, 5 place Jules Janssen, 92195 Meudon, France}

%\author{Other GATOS Collaborators}

%\email{lulu.zhang@utsa.edu; l.l.zhangastro@gmail.com}

\begin{abstract}
%Combining the observed ionized line ratios and theoretical calculations, we provide the further evidence of potential ionized gas outflows for the six targets. 
%{\bf and are responsible for the unique ionized gas kinematics in ESO137-G034}
Utilizing JWST MIRI/MRS IFU observations of the kiloparsec scale central regions, we showcase the diversity of ionized gas distributions and kinematics in six nearby Seyfert galaxies included in the GATOS survey. Specifically, we present spatially resolved flux distribution and velocity field maps of six ionized emission lines covering a large range of ionization potentials ($15.8-97.1$ eV). Based on these maps, we showcase the evidence of ionized gas outflows in the six targets, and find some highly disturbed regions in NGC\,5728, NGC\,5506, and ESO137-G034. We propose AGN-driven radio jets plausibly play an important role in triggering these highly disturbed regions. With the outflow rates estimated based on [Ne~{\footnotesize V}] emission, we find the six targets tend to have ionized outflow rates converged to a narrower range than previous finding. These results have important implication for the outflow properties in AGN of comparable luminosity.
\end{abstract}
%For the integrated nuclear spectra, these three targets also exhibit broader ionized emission lines than the others.


\section{Introduction}

An important signature of feedback effects in galaxy evolution are gas outflows in different phases over diverse scales, which are frequently observed in nearby and distant galaxies of different types (e.g, \citealt{Muller-Sanchez.etal.2011, Bellocchi.etal.2013, Veilleux.etal.2013, Arribas.etal.2014, Cicone.etal.2014, Garcia-Burillo.etal.2014, Harrison.etal.2014, Harrison.etal.2018, Concas.etal.2019, Freeman.etal.2019, Forster-Schreiber.etal.2019, Leung.etal.2019, Garcia-Bernete.etal.2021, Veilleux.etal.2020, Ramos Almeida.etal.2022, Maksym.etal.2023, Winkel.etal.2023}). The role of such outflows in galaxy evolution is crucial as they regulate and/or even quench both star formation and black hole activity by heating up cold gas and/or expelling it from the host galaxy (e.g., \citealt{Schawinski.etal.2007, Page.etal.2012, Cheung.etal.2016, Harrison2017}, also see review by \citealt{Veilleux.etal.2005, Fabian2012, Somerville&Dave2015, Harrison&RamosAlmeida2024}). Moreover, gas outflows are primarily responsible for redistributing of dust and metals over large scales within a galaxy and even outside a galaxy in circum- and intergalactic environments (e.g., \citealt{Heckman.etal.1990, Melioli.etal.2015}).

AGN activity is required to drive the strongest outflows, some of which are powerful enough to rapidly suppress star formation in their host galaxies, particularly for the evolution of massive galaxies (e.g., \citealt{Benson.etal.2003, McCarthy.etal.2011, Cano-Diaz.etal.2012, DaviesJJ.etal.2020}). Gas outflows are also of great importance in starburst systems, especially for star-forming galaxies during the peak epoch of star-formation and black-hole growth (redshift of $\sim1-3$, e.g., \citealt{Shapley.etal.2003, Weiner.etal.2009, Rubin.etal.2014, Harrison.etal.2016, Forster-Schreiber.etal.2019}). AGN activity can affect their hosts and large-scale surroundings through outflows via different modes of feedback. Specifically, luminous, highly accreting AGN inject enough radiative energy through winds/outflows that are able to efficiently expel the surrounding gas by the ``quasar-mode'' feedback (e.g., \citealt{DiMatteo.etal.2005, Hopkins.etal.2008}). Meanwhile, low-luminosity AGN showing low-level nuclear activity exhibit increasingly prominent signatures of jet-like outflows that can heat surrounding gas through shocks, interacting with their environment mainly via the ``kinetic-mode'' feedback (e.g., \citealt{Weinberger.etal.2017, Dave.etal.2019}, and see review by \citealt{McNamara&Nulsen2007}).

Although the general theoretical framework of state-of-the-art cosmological simulations has shown the importance of outflows in reproducing observed properties of different galaxy populations, the relative role of outflow mechanisms is not fully understood, albeit with some studies on the dependence of outflow properties on galaxy properties such as host galaxy mass (e.g., \citealt{Chisholm.etal.2017}), star formation rate (e.g., \citealt{Roberts-Borsani.etal.2020}), AGN luminosity (e.g., \citealt{Fiore.etal.2017}). Even for AGN of the same luminosity, gas outflow rates can vary by up to 2 orders of magnitude, illustrating the diversity of gas outflow properties (e.g., \citealt{Baron&Netzer2019, Davies.etal.2020, Lamperti.etal.2022, Bessiere.etal.2024, Speranza.etal.2024}). \cite{Zubovas&Nardini2020} proposed this diversity results from the different duty cycles of AGN variability and outflow process, as AGN luminosity varies on a much shorter timescale than that of outflows. Additionally, \cite{Fischer.etal.2017, Fischer.etal.2018} proposed this diversity is due to the different launching directions of gas outflows, with the outflow being most intense when the AGN is sufficiently tilted for the outflow to fully interact with the galactic disk (see also \citealt{Ramos Almeida.etal.2022, Audibert.etal.2023}).

High resolution spatially resolved observations of ionic and molecular emission lines are crucial to distinguish the above scenarios (e.g., \citealt{Garcia-Burillo.etal.2014, Garcia-Burillo.etal.2019, Shimizu.etal.2019, Garcia-Bernete.etal.2021, PeraltadeArriba.etal.2023, Esposito.etal.2024}). At nuclear scales, the outflow duty cycle is more rapid so that the effect proposed by \cite{Zubovas&Nardini2020} will be mitigated, and thus outflow properties at different AGN luminosity can be reconsidered. High resolution observations directly resolve gas outflow launching directions, affording the opportunity to determine the influence of the launching direction on outflow properties. The Medium Resolution Spectrograph (MRS; \citealt{Wells.etal.2015,  Labiano.etal.2021, Argyriou.etal.2023}) on the Mid-Infrared Instrument (MIRI; \citealt{Rieke.etal.2015, Wright.etal.2015, Wright.etal.2023}) of James Webb Space Telescope (JWST; \citealt{Gardner.etal.2023, Rigby.etal.2023}) provides an unprecedented opportunity with its excellent spectral/angular resolution and sensitivity to advance our understanding of the diverse outflow properties in innermost regions of AGN (e.g., \citealt{Garcia-Bernete.etal.2022a, Garcia-Bernete.etal.2024b, U.etal.2022, Armus.etal.2023, Zhang&Ho2023}).% MRS has the sensitivity of $\sim0.1 - 10$ mJy, full width at half maximum (FWHM) angular resolution of $\sim0. 30-1.10$, and spectral resolution $R \approx 4000-1500$ over the wavelength range of $\sim5 - 28\,\mum$ (\citealt{Glasse.etal.2015, Labiano.etal.2021}).

%\afterpage{
\startlongtable
%\begin{longrotatetable}
%\centerwidetable
%\movetableright=-1in
\setlength{\tabcolsep}{6pt}
\begin{deluxetable*}{cccccccccc}
\tabletypesize{\footnotesize}
\tablecolumns{10}
\tablecaption{Properties of the Sample}
\tablehead{
\colhead{\,Galaxy\,} & \colhead{\,Type\,} & \colhead{\,$z$\,} & \colhead{\,$D_{L}$\,} & \colhead{\,\,$i_{disk}$\,\,} & \colhead{\,\,$i_{cone}$\,\,} & \colhead{\,\,log $L_{\rm bol}$\,\,} & \colhead{\,\,$\dot{M}_{\rm out}$\,\,} & \colhead{\,\,\,log $\dot{E}_{\rm kin}$\,\,\,} & \colhead{\,\,\,$\frac{L_{\rm bol}}{L_{\rm Edd}}$\,\,\,} \\
\colhead{(-)} & \colhead{(-)} & \colhead{(-)} & \colhead{(Mpc)} & \colhead{($\rm deg$)} & \colhead{($\rm deg$)} & \colhead{($\rm erg\,s^{-1}$)} & \colhead{($\rm M_{\odot}\,yr^{-1}$)} & \colhead{($\rm erg\,s^{-1}$)} & \colhead{(-)} \\
\colhead{(1)} & \colhead{(2)} & \colhead{(3)} & \colhead{(4)} & \colhead{(5)} & \colhead{(6)} & \colhead{(7)} & \colhead{(8)} & \colhead{(9)} & \colhead{(10)}}
\startdata
NGC\,5728 & SAB(r)a & 0.00932 & 39 & 43 & 49 & 44.1 & 0.09 & 40.3 & 0.05 \\
NGC\,5506 & Sa pec  & 0.00608 & 27 & 80 & 42 & 44.1 & 0.21 & 40.6 & 0.04 \\
ESO137-G034 & SAB0/a & 0.00914 & 35 & 38 & \nodata & 43.4 & 0.52 & 40.7 & 0.01\\
NGC\,7172 & Sa pec  & 0.00868 & 37 & 88 & 67 & 44.1 & 0.005 & 38.4 & 0.02 \\
MCG-05-23-016 & S0 & 0.00849 & 35 & 66 & 50 & 44.3 & 0.003 &  37.8 & 0.06 \\
NGC\,3081 & (R)SAB0/a(r) & 0.00798 & 34 & 41 & 71 & 44.1 & 0.04 &  39.0 & 0.02 \\
\enddata
\tablecomments{\footnotesize Column (1): Target name.  Column (2): Target host type from NASA/IPAC Extragalactic Database (NED), with (r) and (R) indicating inner and outer ring, respectively. Column (3): Redshift taken from NED. Column (4): Luminosity distance taken from NED using redshift independent estimates or peculiar velocity corrections (\citealt{Theureau.etal.2007}). Column (5): Galactic disk inclination. Column (6): AGN ionization cone inclination assuming the cone is aligned with the torus. Column (7)-(10): Bolometric AGN luminosity derived from X-ray luminosity, ionized gas mass outflow rate and ionized outflow kinetic energy derived from [O~{\scriptsize III}] 5007\AA\ emission, and Eddington ratio, respectively. References for Column (5): NGC\,5728 (\citealt{Shimizu.etal.2019}), NGC\,5506 (\citealt{Esposito.etal.2024}), NGC\,7172 (\citealt{Alonso-Herrero.etal.2023}), ESO137-G034, MCG-05-23-016, and NGC\,3081 (\citealt{Burtscher.etal.2021}, galaxy axis ratios therein). References for Column (6): NGC\,5728 (\citealt{Shimizu.etal.2019}), NGC\,5506 (\citealt{Sun.etal.2018}), NGC\,7172 (\citealt{Alonso-Herrero.etal.2023}), MCG-05-23-016 (\citealt{Zoghbi.etal.2017}), and NGC\,3081 (\citealt{Ramos Almeida.etal.2011}). Reference for Column (7)-(10): (\citealt{Davies.etal.2015, Davies.etal.2020, Caglar.etal.2020}).}
\label{tabinfo}
\end{deluxetable*}
%\end{longrotatetable}
%\footnote{\url{https://ned.ipac.caltech.edu/}}
%}

Leveraging JWST/MRS observations of the innermost kpc scale regions of six nearby Seyferts with comparable luminosity, we reveal the diversity of their spatially resolved ionized gas kinematics and the evidence of their ionized gas outflows. Our final goal is to shed light on the relative outflow strength in AGN with comparable luminosity. This paper is accompanied by papers on individual targets of the six (\citealt{Davies.etal.2024, Esparza-Arredondo.etal.2024, Hermosa-Munoz.etal.2024a}, and Delany et al. in preparation, Haidar et al. in preparation). See also the specific study in \cite{Lopez-Rodriguez.etal.2024} of the emission line contribution to extended MIR emission, as well as the study in \cite{Garcia-Bernete.etal.2024c} and \cite{Zhang.etal.2024} of broad PAH features, of the six targets. In this paper, Section~2 describes the targets, observations, and data processing steps. Section~\ref{sec3} provides the maps and analysis on the spatially resolved ionized gas kinematics and showcase the evidence of ionized gas outflows in the six targets. Section~\ref{sec5} discusses the possible physical mechanism responsible for some highly disturbed regions in our targets, and Section~\ref{sec4} presents a rather quantitative comparison of the ionized gas outflow strength among the six targets. Section~\ref{sec6} contains a summary with main conclusions.

\startlongtable
%\begin{longrotatetable}
%\centerwidetable
%\movetableright=-1in
\setlength{\tabcolsep}{4pt}
\begin{deluxetable}{cccccc}
\tabletypesize{\scriptsize}
\tablecolumns{6}
\tablecaption{Observational Configurations of the Sample}
\tablehead{
\colhead{Target} & \colhead{R.A.} & \colhead{Dec.} & \colhead{Mosaic} & \colhead{Dither} & \colhead{$t_{\rm Exp}$} \\
\colhead{(-)} & \colhead{(deg)} & \colhead{(deg)} & \colhead{(-)} & \colhead{(-)} & \colhead{(s)} \\
\colhead{(1)} & \colhead{(2)} & \colhead{(3)} & \colhead{(4)} & \colhead{(5)} & \colhead{(6)}}
\startdata
NGC\,5728 & 14:42:23.880 & $-$17:15:11.08 & $2\times2$ & 4-point & 1051 \\
NGC\,5506 & 14:13:14.878 & $-$03:12:27.76 & $1\times8$ & None& 275 \\
ESO137-G034 & 16:35:13.995 & $-$58:04:47.91 & $2\times1$ & 4-point & 1147 \\
NGC\,7172 & 22:02:1.889 & $-$31:52:10.47 & None & 4-point & 1121 \\
MCG-05-23-016 & 09:47:40.135 & $-$30:56:56.00 & None & 4-point & 1121 \\
NGC\,3081 & 09:59:29.534 & $-$22:49:34.78 & $2\times2$ & 4-point & 1147 \\
\enddata
\tablecomments{\scriptsize Column (1): Target name. Column (2): Right ascension of target. Column (3): Declination of target. Column (4): Mosaic pattern. (5): Dither strategy. Column (6): Total exposure time of each subband.}
\label{tabobs}
\end{deluxetable}
%\end{longrotatetable}

\section{Data and Line Properties}\label{sec2}

\subsection{Targets and Observations}\label{sec2.1}
%\footnote{\href{https://stsci.app.box.com/s/vq100k7snieftecjgbjji9r35junqjuw}{JWST Mid Infrared Instrument.pdf}}
%\footnote{\href{https://jwst-docs.stsci.edu/jwst-science-calibration-pipeline-overview/stages-of-jwst-data-processing}{jwst-science-calibration-pipeline-overview}}
%\footnote{\url{https://mast.stsci.edu/portal/Mashup/Clients/Mast/Portal.html}}
This paper is part of a series studying six type 1.9/2 Seyfert galaxies ($L_{\rm bol} \approx 10^{43.4} - 10^{44.3}\, {\rm erg\,s^{-1}}$) with MRS integral field unit (IFU) spectral observations obtained by the JWST cycle 1 GO program \#1670 (PI: Shimizu, T. Taro). The six targets are part of the Galactic Activity, Torus, and Outflow Survey (GATOS; \citealt{Garcia-Burillo.etal.2021, Alonso-Herrero.etal.2021, Garcia-Bernete.etal.2024a})\footnote{\url{https://gatos.myportfolio.com}} and some of their basic properties are summarized in Table~\ref{tabinfo}. The full sample of GATOS is selected from the 70 Month Swift-BAT Allsky Hard X-ray Survey (\citealt{Baumgartner.etal.2013}), ensuring a nearly complete selection of AGN with the luminosities $L_{14-150\,\rm keV} > 10^{42}\,\rm erg\,s^{-1}$ at distances of $10-40$ Mpc. The sample is largely unbiased to obscuration/absorption even up to column densities of $N_{\rm H} \sim 10^{24}\,\rm cm^{-2}$. The AGN luminosity and absorbing column density can be obtained by the analysis of the X-ray data (\citealt{Ricci.etal.2017}). 

\begin{figure*}[!ht]
\center{\includegraphics[width=0.625\linewidth]{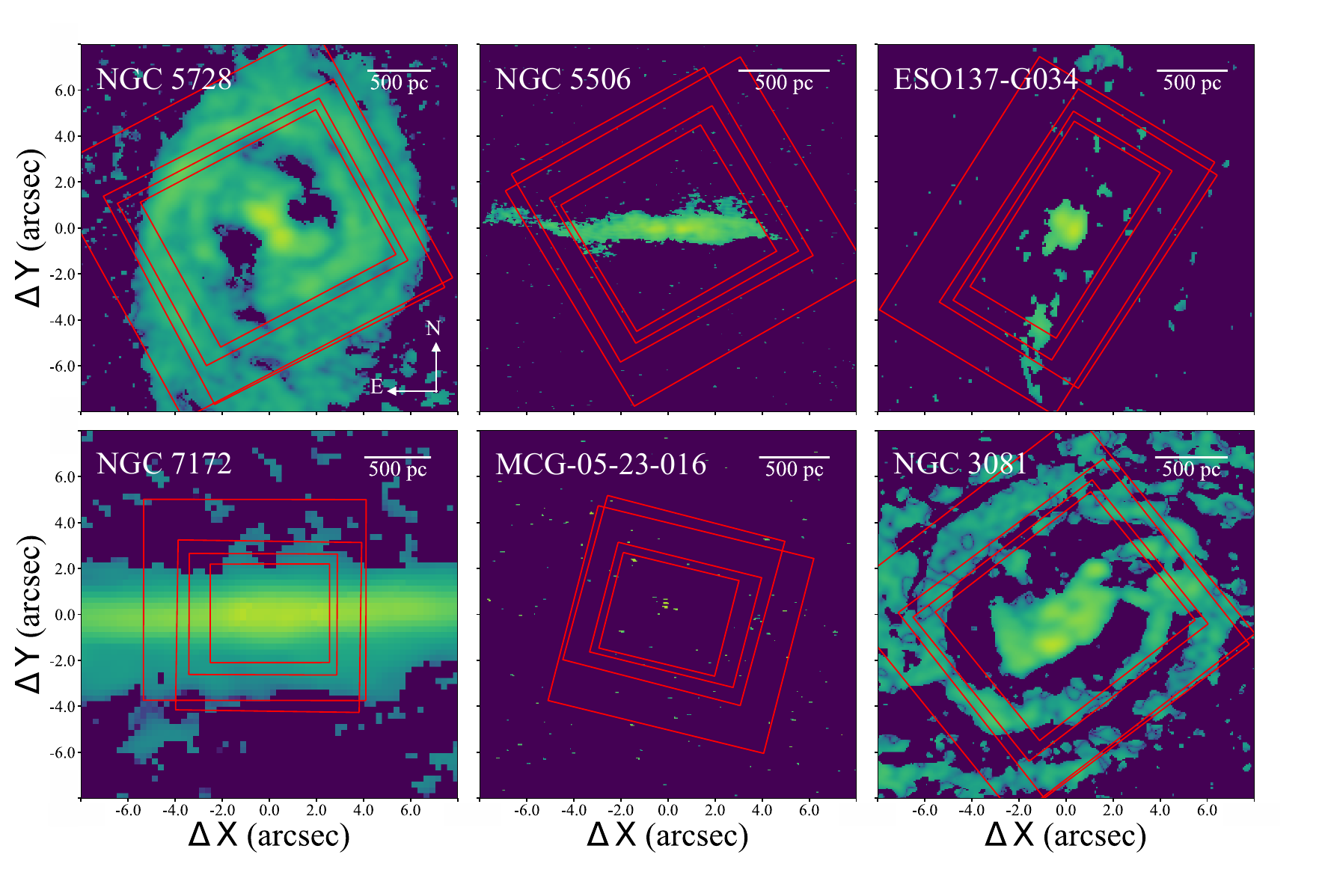}}
\caption{Illustration of channel 1 to 4 (red rectangles from small to large) MIRI/MRS coverages of the six Seyfert galaxies, with ALMA CO(2-1) maps (see acknowledgements) as the background. A scale bar of 500 pc is in the top-right of each panel, and a compass is in the top-left panel and is the same for all panels with north is up and east is to the left .}\label{figc}
\end{figure*}

The MIRI/MRS observations were carried out using the full set of four IFUs (channels 1 to 4), covering $4.9-7.6\,\mum$, $7.51-11.71\,\mum$, $11.55-17.98\,\mum$, and $17.7-27.9\,\mum$, respectively. These four channels are observed simultaneously, but each exposure can only cover one of the three grating settings (short, medium, and long sub-bands). The observational configuration and total science exposure time of each sampled Seyfert galaxy are summarized in Table~\ref{tabobs}. For background observations, a 2-point dither pattern is taken for all targets, in a blank region of sky a few arcminutes away from the targets.

We primarily follow the standard JWST MIRI/MRS pipeline (release 1.11.4) to reduce the raw data (e.g., \citealt{Labiano.etal.2016, Bushouse.etal.2023}), using the same configuration (the calibration context 1130) of the pipeline stages as is in \cite{Garcia-Bernete.etal.2022a} and \cite{Pereira-Santaella.etal.2022}. Residual fringes remain with the standard fringe removal, which could have a significant influence on weak spectral features (\citealt{Argyriou.etal.2020, Gasman.etal.2023}). Therefore, we apply an extra JWST pipeline step (i.e., residual\_fringe) not implemented in the standard JWST pipeline to correct the low-frequency fringe residuals (\citealt{Law.etal.2023}), before performing the standard process to generate the 3D spectral data cubes. Moreover, some hot and cold pixels are not identified by the current pipeline version, so we also added an extra step before creating the data cubes to mask them. The data reduction and extra steps are described in \cite{Garcia-Bernete.etal.2024a}. 

\subsection{Data Analysis}\label{sec2.2}

\begin{figure*}[!ht]
\center{\includegraphics[width=0.65\linewidth]{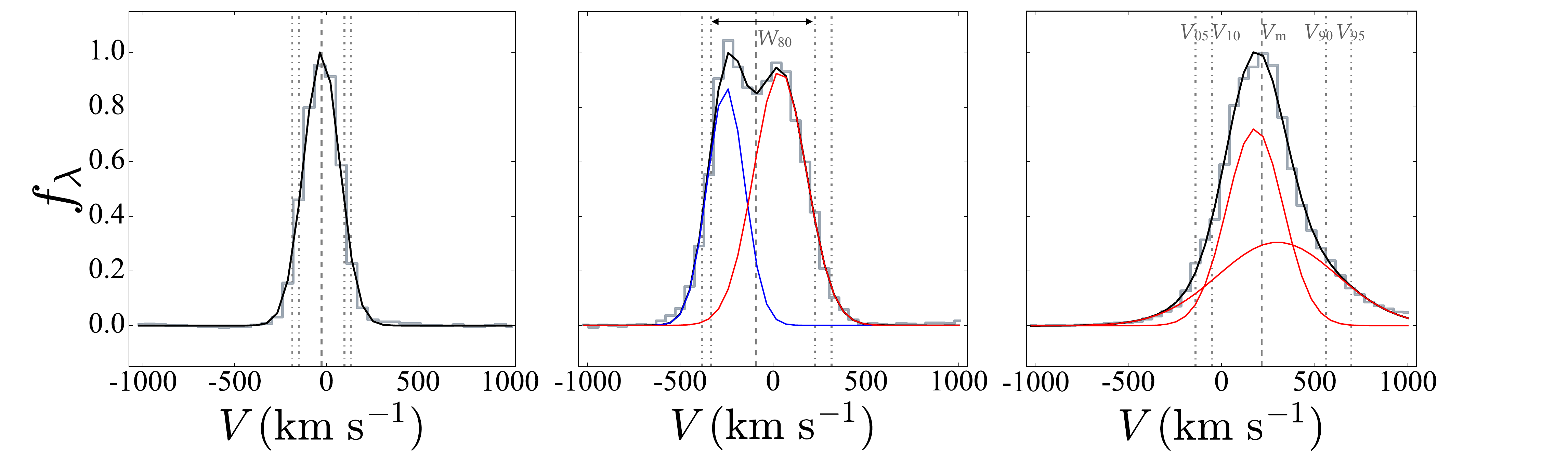}}
\caption{Illustration of emission line fitting for different observed line profiles, where gray histograms and black curves represent (with the local linear continuum subtracted) the observed and best-fitted emission line profiles, respectively. The blue and red curves further indicate separate Gaussian profiles if the emission line is better fitted with a double-Gaussian profile (blue- and/or red-shifted). The vertical lines in each panel are, from left to right, $v_{05}$, $v_{10}$, $v_{\rm m}$, $v_{90}$, and $v_{95}$, respectively, and the $W_{80}$ is also indicated in the middle panel.}\label{fig0}
\end{figure*}

\cite{Zhang.etal.2021} developed a strategy for spatially resolved analysis of mid-infrared (mid-IR) properties of nearby galaxies, based on the mapping-mode observations taken with the Infrared Spectrograph (IRS; \citealt{Houck.etal.2004}) on Spitzer Space Telescope (\citealt{Werner.etal.2004}). This strategy, as summarized below, has already been applied to investigate the interaction between AGN and starburst activity in the central $\sim$1.5 kpc$\times$1.3 kpc region of type 1 Seyfert galaxy NGC 7469 on a $\sim$100 pc scale, based on JWST MIRI/MRS IFU observations (\citealt{Zhang&Ho2023}). We adopt the same strategy to study ionized gas kinematics in central kpc-scale regions of the six Seyfert galaxies (see Figure~\ref{figc}).

A key step of this strategy for extracting spatially resolved diagnostics on a common physical scale is to have the same angular resolution for all the slices within MIRI/MRS data cubes. To this end, we need to convolve all the slices within different data cubes of each sampled Seyfert galaxy to have the same angular resolution as the slice of the widest point-spread function (PSF). We focus on the first three channels (i.e., channel 1, 2, and 3; $\lambda\approx5-18\, \mum$) to have as higher full width at half maximum (FWHM) angular resolution as possible (i.e., $\rm FWHM \approx 0\farcs3 - 0\farcs7$), while covering all the emission lines relevant for this study. Based on the measured PSF FWHMs of JWST/MRS observations as detailed in \cite{Zhang&Ho2023}, we use a two dimension Gaussian function to mimic the PSF of each slice and then construct the convolution kernels (\citealt{Aniano.etal.2011}). After convolving all slices to the same angular resolution, of $0\farcs7$, we reproject all spectral data cubes into the same coordinate frame with a pixel size of 0\farcs35 (half of the angular resolution, $\sim 45 - 70$ pc at distances of the targets), for further spectrum extraction and emission line fitting.

\subsection{Emission Line Fitting}\label{sec2.3}

Mid-IR spectra of galaxies exhibit abundant ionic fine-structure lines and molecular hydrogen rotational lines that are well resolved with the spectral resolution of JWST/MRS. In general the relative strength and broadening of emission lines provide valuable diagnostics of galaxy properties (e.g., \citealt{Pereira-Santaella.etal.2010, Pereira-Santaella.etal.2017, DAgostino.etal.2019, Sajina.etal.2022, Feltre.etal.2023}). This paper focuses on six ionized emission lines covering a large range of ionization potentials (i.e., [Ar~{\footnotesize II}] 6.985~$\mum$, [Ne~{\footnotesize II}] 12.814~$\mum$, [Ar~{\footnotesize III}] 8.991~$\mum$, [S~{\footnotesize IV}] 10.511~$\mum$, [Ne~{\footnotesize III}] 15.555~$\mum$, and [Ne~{\footnotesize V}] 14.322~$\mum$ lines with ionization potentials ranging from 15.8 to 97.1 eV, see Table~\ref{tabnucs}). %Some of the companion papers focus on diagnostics based on ionized emission lines as well (i.e., \citealt{Hermosa-Munoz.etal.2024a}, and Haidar et al. in prep), while the others focus more on molecular hydrogen emission lines (i.e., \citealt{Davies.etal.2024}, and Delany et al. in preparation, Esparza-Arredondo et al. in preparation). 
In this work, we fit each ionic emission line using a single- and then a double-Gaussian profile, plus a local linear continuum. The fitting is implemented with the Levenberg-Marquardt least-squares minimization algorithm. Based on the reduced $\chi^{2}$ values of the two sets of fitting, we calculate the $p$-value using a statistical $F$-test and only adopt the double-Gaussian fitting result if $p<0.05$ (see also \citealt{Hao.etal.2005}). Stronger constraints tend to include only one Gaussian component in the best-fit profile for some low signal-to-noise spaxels, but will not affect our conclusions. To get more robust statistics, each emission line is perturbed with random noise at its uncertainty level and then the fitting is repeated 100 times. The median and standard deviation of those 100 fits are taken as the final estimate and corresponding uncertainty of each emission line, respectively.

\begin{figure*}[!ht]
\center{\includegraphics[width=0.625\linewidth]{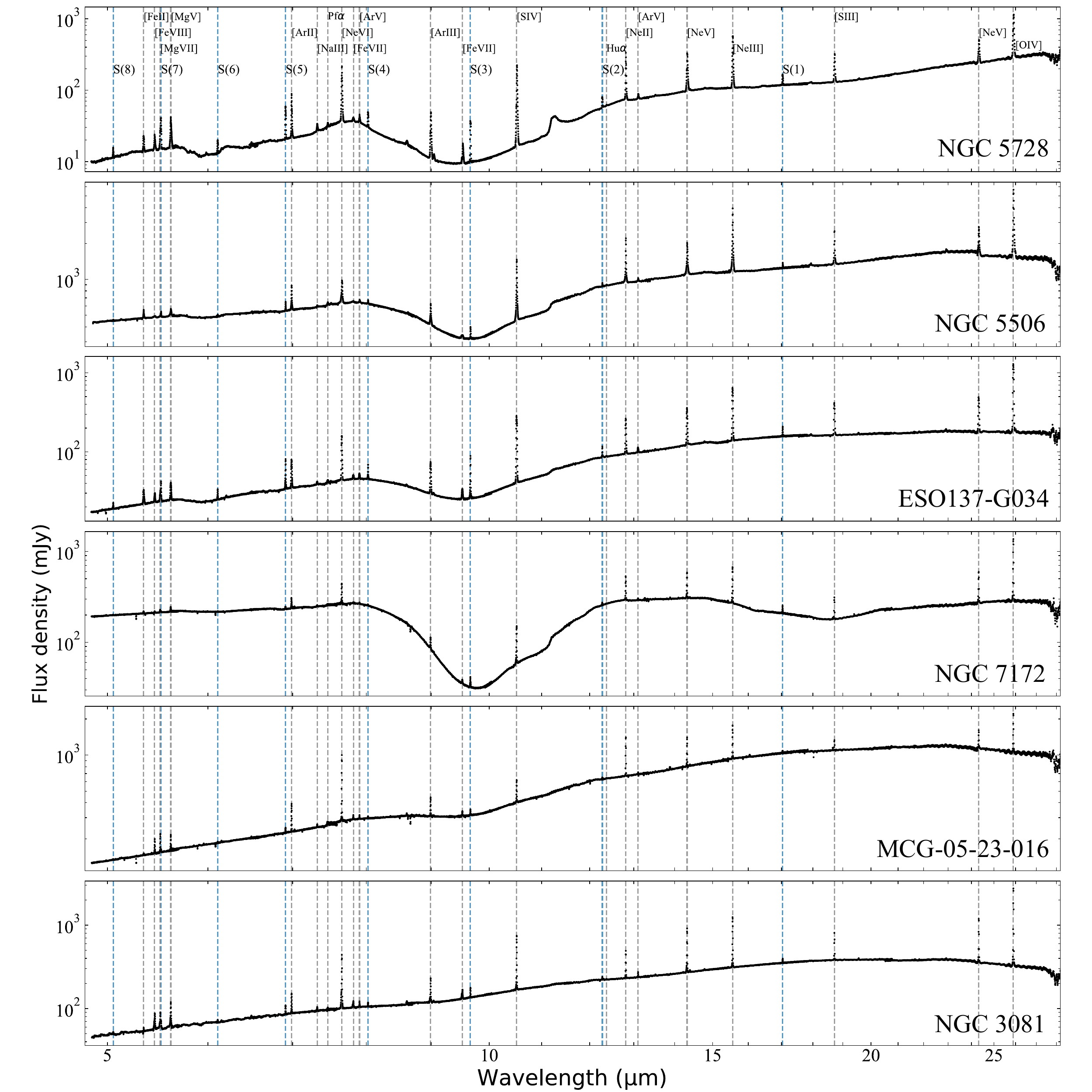}}
\caption{Mid-IR spectra extracted from the central $r = 0\farcs75 $ aperture (centered on the continuum peak) for each sampled Seyfert galaxy, where 21 prominent ionized and 8 molecular lines as listed in Table~\ref{tabnucs} are marked in the top panel (zoom in to see more details).}\label{figsn}
\end{figure*}

\begin{figure*}[!ht]
\center{\includegraphics[width=0.6\linewidth]{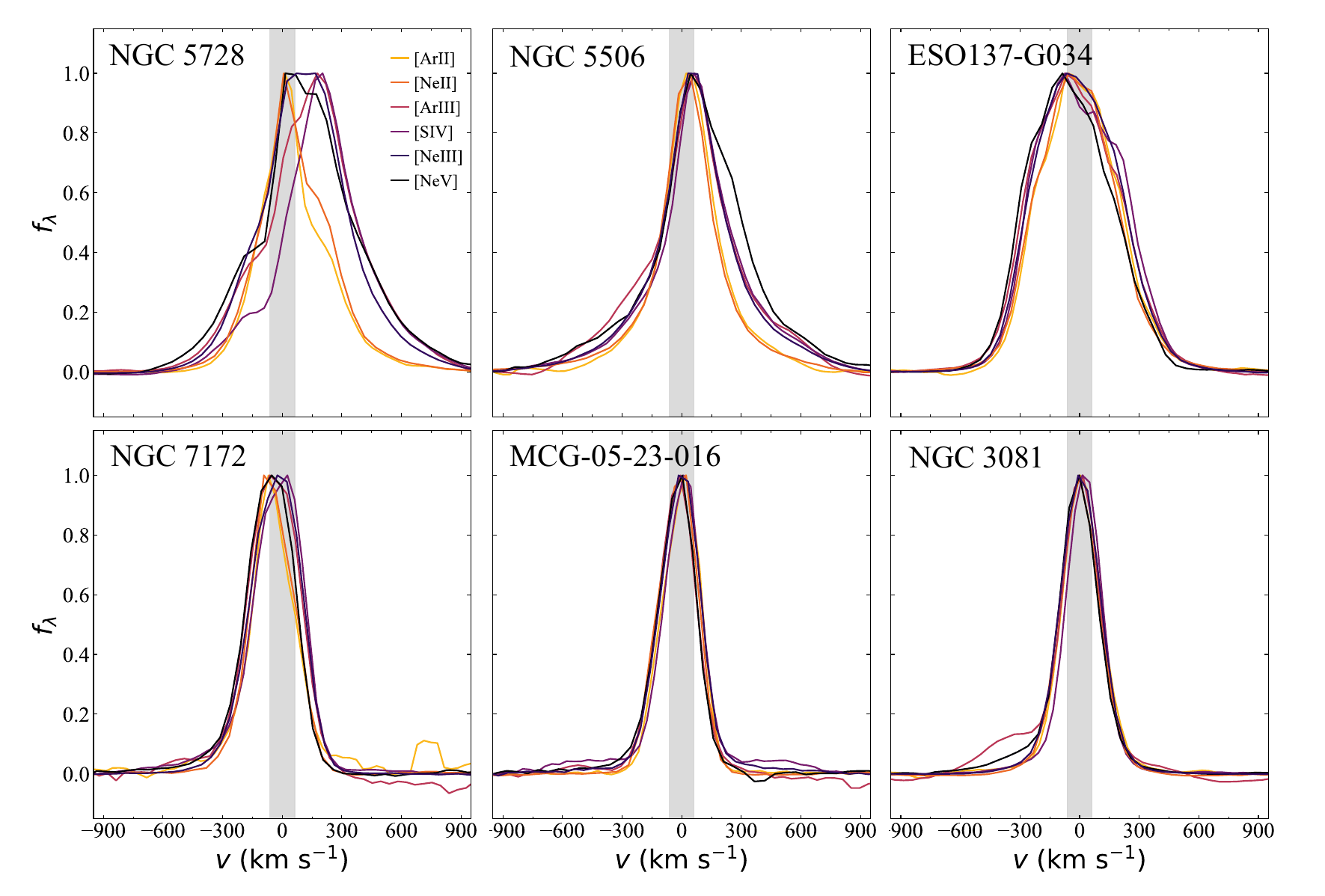}}
\caption{Normalized profiles of [Ar~{\footnotesize II}] 6.985~$\mum$, [Ne~{\footnotesize II}] 12.814~$\mum$, [Ar~{\footnotesize III}] 8.991~$\mum$, [S~{\footnotesize IV}] 10.511~$\mum$, [Ne~{\footnotesize III}] 15.555~$\mum$, and [Ne~{\footnotesize V}] 14.322~$\mum$ emission lines from spectra in Figure~\ref{figsn}, with the darker color indicating the higher ionization potential. The grey shaded region in each panel indicates the FWHM of the instrumental broadening at the wavelength of [Ne~{\footnotesize III}], which is the largest among the six lines. Note that these line profiles are in the rest frame converted according to the redshift listed in Table~\ref{tabinfo}.}\label{figln}
\end{figure*}
%High ionization potential lines (e.g., [Ne~{\footnotesize III}], [Ne~{\footnotesize V}]) exhibit stronger outflow signatures with broader profiles and/or more significant velocity shifts, comparing with lines of lower ionization potentials (e.g., [Ar~{\footnotesize II}], [Ne~{\footnotesize II}]).
%Targets in the top row display stronger signatures of disturbed ionized gas with broader emission line profiles (with $W_{80} \gtrsim 600\,{\rm km\,s^{-1}}$) than targets in the bottom row ($W_{80} \approx 300\,{\rm km\,s^{-1}} $)

\subsection{Non-Parametric Measurement}\label{sec2.4}

As shown in Figure~\ref{fig0}, emission lines of certain spaxels of our targets exhibit a double-peaked profile. Such double-peaked profiles are caused by outflows with disturbed velocity field as will be detailed later (e.g. \citealt{Fischer.etal.2011, Bae&Woo2016}). In such cases, certain bias will occur when assigning the broader component of the best-fit result to outflow-dominated motions. Therefore, to provide consistent measurements for all the six targets, in this work we adopt the non-parametric methodology to study the kinematics of the six targets (\citealt{Rupke&Veilleux2011, Liu.etal.2013, Harrison.etal.2014, McElroy.etal.2015, HervellaSeoane.etal.2023}). Similar to previous studies leveraging non-parametric measurements (e.g., \citealt{Liu.etal.2013, Harrison.etal.2014, McElroy.etal.2015, Speranza.etal.2024}), we derive the non-parametric velocities (Figure~\ref{fig0}) based on the best-fit profile of each emission line to address low signal-to-noise regions. This work focuses on $v_{\rm m}$, the velocity that bisects the area under the emission line profile, and  $W_{80}$, the line width that contains the central 80 per cent of the flux as $W_{80} = v_{90} - v_{10}$, where $v_{10}$ and $v_{90}$ are velocities corresponding to 10 and 90 per cent of the line flux, respectively.

The non-parametric methodology diagnoses the flux weighted motion, i.e., the primary motion, of the gas component. Thus, this methodology is ood at distinguishing line emission that is sensitive to gas outflows from that to gas rotation. Meanwhile, for more general cases (as is in companion papers on individual targets), the parametric methodology is also widely used. The parametric methodology considers individual Gaussian components in the best-fit profile separately, and provides a further separation of outflow-dominated components from rotation-dominated ones for emission lines that are dominated by gas rotation (e.g., \citealt{Bellocchi.etal.2019, Leung.etal.2019, Cazzoli.etal.2020, Cazzoli.etal.2022, PeraltadeArriba.etal.2023, Venturi.etal.2023, Hermosa-Munoz.etal.2024b}). Therefore, a brief discussion on the ionized gas kinematics in central regions of the six targets based on the parametric methodology is also provided in Section~\ref{secA} of the appendix. 

\startlongtable
\begin{longrotatetable}
%\centerwidetable
%\movetableright=-1in
\setlength{\tabcolsep}{2pt}
\begin{deluxetable*}{ccccccccccccccc}
\tabletypesize{\scriptsize}
\tablecolumns{15}
\tablecaption{Measurements for Emission Lines Furnished on Spectra Extracted from Central Apertures}
\tablehead{
\colhead{} & \colhead{} & \colhead{} & \colhead{NGC\,5728} & \colhead{} & \colhead{NGC\,5506} & \colhead{} & \colhead{ESO137-G034} & \colhead{} & \colhead{NGC\,7172} & \colhead{} & \colhead{MCG-05-23-016} & \colhead{} & \colhead{NGC\,3081} & \colhead{} \\
\colhead{Line} & \colhead{$\lambda$} & \colhead{IP} & \colhead{log flux} & \colhead{$W_{80}$} & \colhead{log flux} & \colhead{$W_{80}$} & \colhead{log flux} & \colhead{$W_{80}$} & \colhead{log flux} & \colhead{$W_{80}$} & \colhead{log flux} & \colhead{$W_{80}$} & \colhead{log flux} & \colhead{$W_{80}$} \\
\colhead{(-)} & \colhead{($\mu$m)} & \colhead{(eV)} & \colhead{[$\rm erg/s/cm^{2}$]} & \colhead{($\rm km/s$)} & \colhead{[$\rm erg/s/cm^{2}$]} & \colhead{($\rm km/s$)} & \colhead{[$\rm erg/s/cm^{2}$]} & \colhead{($\rm km/s$)} & \colhead{[$\rm erg/s/cm^{2}$]} & \colhead{($\rm km/s$)} & \colhead{[$\rm erg/s/cm^{2}$]} & \colhead{($\rm km/s$)} & \colhead{[$\rm erg/s/cm^{2}$]} & \colhead{($\rm km/s$)} \\
\colhead{(1)} & \colhead{(2)} & \colhead{(3)} & \colhead{(4)} & \colhead{(5)} & \colhead{(6)} & \colhead{(7)} & \colhead{(8)} & \colhead{(9)} & \colhead{(10)} & \colhead{(11)} & \colhead{(12)} & \colhead{(13)} & \colhead{(14)} & \colhead{(15)} }
\startdata
${\rm H_{2}}\ S(8)$ & 5.053 & -- & $-$14.54(0.02) & 366.2(13.0) & $-$14.61(0.35) & 219.2(2763.0) & $-$14.69(0.18) & 434.1(2127.0) & $-$14.91(0.36) & 236.2(1146.0) & $-$15.02(0.05) & 229.1(1022.0) & $-$14.87(0.03) & 247.3(12.3) \\
$\rm [Fe\ II]$ & 5.34 & 7.9 & $-$14.05(0.01) & 535.4(11.3) & $-$13.23(0.01) & 411.6(11.8) & $-$13.89(0.01) & 634.9(14.0) & $-$14.18(0.03) & 312.8(20.4) & $-$14.51(0.03) & 303.6(16.1) & $-$14.66(0.02) & 236.9(14.3) \\
$\rm [Fe\ VIII]$ & 5.447 & 124 & $-$14.03(0.02) & 664.9(11.9) & $-$13.60(0.06) & 1196.0(56.8) & $-$14.17(0.06) & 510.8(6.5) & $-$15.17(1.09) & 241.5(205.0) & $-$13.64(0.08) & 275.7(5.2) & $-$13.60(0.01) & 561.3(6.5) \\
$\rm [Mg\ VII]$ & 5.503 & 186.5 & $-$13.97(0.27) & 584.4(326.0) & $-$13.61(0.14) & 1186.0(272.5) & $-$13.88(0.22) & 732.6(4242.0) & $-$14.25(0.23) & 331.2(935.8) & $-$13.54(0.18) & 342.6(1404.0) & $-$13.58(0.11) & 778.6(612.6) \\
${\rm H_{2}}\ S(7)$ & 5.511 & -- & $-$13.78(0.16) & 399.0(1387.0) & $-$13.67(0.11) & 260.6(1246.0) & $-$14.36(0.20) & 207.3(1412.0) & $-$14.22(0.14) & 464.6(139.1) & $-$14.33(0.57) & 689.7(3282.0) & $-$14.49(1.20) & 249.5(484.6) \\
$\rm [Mg\ V]$ & 5.61 & 109.2 & $-$13.57(0.04) & 750.1(6.4) & $-$12.95(0.03) & 1470.0(229.5) & $-$13.80(0.06) & 535.1(5.6) & $-$13.81(0.02) & 336.0(12.1) & $-$13.61(0.09) & 283.3(12.4) & $-$13.43(0.03) & 492.2(17.4) \\
${\rm H_{2}}\ S(6)$ & 6.109 & --& $-$14.39(0.12) & 370.0(9.2) & $-$14.29(0.01) & 207.9(5.4) & $-$14.43(0.19) & 302.0(4.4) & $-$14.83(0.09) & 315.4(83.0) & $-$14.82(0.03) & 207.1(11.9) & $-$14.78(0.02) & 228.4(8.6) \\
${\rm H_{2}}\ S(5)$ & 6.91 & -- & $-$13.63(0.09) & 398.3(5.1) & $-$13.46(0.00) & 238.4(2.8) & $-$13.74(0.01) & 332.8(4.6) & $-$14.16(0.01) & 327.0(7.7) & $-$14.12(0.02) & 221.1(9.0) & $-$14.03(0.02) & 257.2(7.7) \\
$\rm [Ar\ II]$ & 6.985 & 15.8& $-$13.49(0.02) & 451.1(5.9) & $-$12.77(0.01) & 471.3(3.4) & $-$13.51(0.05) & 485.7(3.8) & $-$13.59(0.15) & 283.3(91.2) & $-$13.27(0.12) & 238.6(3.0) & $-$13.65(0.11) & 263.5(3.2) \\
$\rm [Na\ III]$ & 7.318 & 47.3 & $-$14.27(0.01) & 614.3(17.4) & $-$13.96(0.02) & 456.3(45.8) & $-$14.56(0.01) & 592.1(14.5) & $-$14.34(0.02) & 417.9(27.3) & $-$14.31(0.01) & 257.5(9.4) & $-$14.39(0.01) & 314.0(11.7) \\
$\rm Pf{\alpha}$ & 7.46 & 13.6 & $-$14.60(0.02) & 562.5(27.9) & $-$13.60(0.02) & 678.0(30.5) & $-$14.50(0.02) & 684.9(21.8) & $-$14.54(0.20) & 341.3(237.3) & $-$14.03(0.07) & 591.4(94.8) & $-$14.75(0.05) & 246.4(26.8) \\
$\rm [Ne\ VI]$ & 7.652 & 126.2 & $-$12.96(0.06) & 623.6(9.8) & $-$12.52(0.02) & 945.4(32.0) & $-$13.13(0.17) & 492.1(13.0) & $-$13.16(0.01) & 300.0(4.9) & $-$12.67(0.01) & 241.9(4.3) & $-$12.87(0.01) & 445.7(5.7) \\
$\rm [Fe\ VII]$ & 7.815 & 99.1 & $-$14.31(0.01) & 822.3(16.2) & $-$13.48(0.03) & 1849.0(115.8) & $-$14.37(0.08) & 567.3(15.7) & \nodata & \nodata & $-$14.09(0.02) & 271.4(16.7) & $-$13.91(0.13) & 563.7(15.6) \\
$\rm [Ar\ V]$ & 7.902 & 40.7 & $-$14.18(0.14) & 697.8(404.3) & $-$13.56(0.03) & 883.3(55.6) & $-$14.28(0.10) & 518.4(93.5) & $-$14.54(0.03) & 268.7(14.3) & $-$14.08(0.04) & 442.4(140.1) & $-$14.20(0.03) & 392.2(44.0) \\
${\rm H_{2}}\ S(4)$ & 8.025 & -- & $-$13.97(0.09) & 414.0(7.3) & $-$13.85(0.01) & 261.3(12.7) & $-$14.08(0.01) & 331.8(3.5) & $-$14.42(0.03) & 409.6(33.2) & $-$14.44(0.02) & 254.0(15.0) & $-$14.39(0.02) & 250.4(9.7) \\
$\rm [Ar\ III]$ & 8.991 & 27.6 & $-$13.66(0.02) & 636.4(7.0) & $-$12.96(0.01) & 697.5(8.3) & $-$13.58(0.04) & 528.9(2.6) & $-$14.00(0.18) & 310.4(10.6) & $-$13.43(0.06) & 270.0(7.0) & $-$13.45(0.01) & 427.0(9.9) \\
$\rm [Fe\ VII]$ & 9.527 & 99.1 & $-$14.31(0.12) & 784.8(43.0) & $-$13.70(0.02) & 961.5(13.3) & $-$14.24(0.08) & 593.1(6.6) & $-$14.94(0.03) & 355.6(24.7) & $-$14.03(0.03) & 305.8(33.6) & $-$13.73(0.02) & 557.6(4.7) \\
${\rm H_{2}}\ S(3)$ & 9.665 & -- & $-$13.86(0.07) & 433.6(7.4) & $-$13.70(0.01) & 245.2(6.7) & $-$13.76(0.01) & 330.6(3.5) & $-$14.60(0.02) & 314.3(19.5) & $-$14.02(0.01) & 249.3(7.0) & $-$13.97(0.01) & 241.9(6.6) \\
$\rm [S\ IV]$ & 10.511 & 34.8 & $-$13.03(0.01) & 641.6(4.8) & $-$12.38(0.01) & 655.9(7.4) & $-$12.90(0.03) & 524.6(2.0) & $-$13.54(0.07) & 347.5(9.4) & $-$13.28(0.03) & 296.7(217.5) & $-$12.89(0.00) & 248.7(1.1) \\
${\rm H_{2}}\ S(2)$ & 12.279 & -- & $-$14.06(0.15) & 416.3(26.6) & $-$13.92(0.02) & 245.3(13.0) & $-$14.22(0.01) & 242.2(6.5) & $-$14.45(0.03) & 269.4(16.3) & $-$14.43(0.04) & 221.7(17.7) & $-$14.43(0.02) & 234.5(14.1) \\
$\rm Hu{\alpha}$ & 12.37 & 13.6 & $-$15.22 (0.03) & 518.8 (37.3) & $-$14.28 (0.04) & 502.9 (55.0) & $-$14.96 (0.02) & 685.1 (35.4) & $-$14.91 (0.06) & 641.6 (89.5) & $-$14.63 (0.09) & 414.3 (120.6) & $-$15.15 (0.43) & 313.3 (387.9) \\
$\rm [Ne\ II]$ & 12.814 & 21.6 & $-$13.13(0.15) & 465.1(13.1) & $-$12.45(0.01) & 523.9(11.1) & $-$13.20(0.20) & 481.8(8.0) & $-$13.25(0.22) & 333.9(29.4) & $-$12.85(0.19) & 241.2(3.6) & $-$13.30(0.03) & 267.2(1.7) \\
$\rm [Ar\ V]$ & 13.102 & 40.7 & $-$14.22(0.02) & 758.4(17.8) & $-$13.42(0.09) & 1414.0(299.2) & $-$14.21(0.07) & 522.5(11.4) & $-$14.58(0.17) & 254.9(251.9) & $-$14.42(0.10) & 181.9(20.0) & $-$14.24(0.01) & 208.5(3.0) \\
$\rm [Ne\ V]$ & 14.322 & 97.1  & $-$13.04(0.02) & 695.3(9.0) & $-$12.43(0.01) & 896.6(21.3) & $-$13.06(0.04) & 517.7(3.4) & $-$13.22(0.19) & 308.4(9.2) & $-$13.01(0.12) & 248.8(4.5) & $-$12.94(0.01) & 292.1(3.5) \\
$\rm [Ne\ III]$ & 15.555 & 41.0 & $-$12.79(0.06) & 596.7(5.2) & $-$12.10(0.01) & 655.5(6.5) & $-$12.76(0.02) & 512.7(1.6) & $-$13.08(0.17) & 316.2(4.9) & $-$12.87(0.15) & 251.5(3.7) & $-$12.80(0.00) & 275.5(0.7) \\
${\rm H_{2}}\ S(1)$ & 17.035 & -- & $-$13.92(0.13) & 399.6(13.7) & $-$13.78(0.03) & 231.5(16.4) & $-$14.16(0.01) & 236.4(4.1) & $-$14.25(0.01) & 231.7(10.8) & $-$14.16(0.04) & 241.8(25.6) & $-$14.19(0.02) & 251.6(10.0) \\
$\rm [S\ III]$ & 18.71 & 23.3 & $-$13.25(0.01) & 634.6(9.7) & $-$12.55(0.01) & 574.2(10.1) & $-$13.12(0.04) & 551.1(3.3) & $-$13.66(0.15) & 302.9(71.8) & $-$13.51(0.05) & 241.8(21.9) & $-$13.24(0.01) & 258.4(13.9) \\
$\rm [Ne\ V]$ & 24.32 & 97.1 & $-$13.16(0.02) & 773.0(13.1) & $-$12.67(0.05) & 618.3(30.8) & $-$13.11(0.07) & 570.1(4.7) & $-$13.33(0.02) & 364.1(9.2) & $-$13.18(0.05) & 439.2(120.6) & $-$13.03(0.01) & 289.6(3.8) \\
$\rm [O\ IV]$ & 25.89 & 54.9 & $-$12.68(0.01) & 751.6(10.3) & $-$12.10(0.01) & 648.3(10.7) & $-$12.59(0.05) & 562.9(2.4) & $-$12.87(0.02) & 332.8(3.4) & $-$12.88(0.11) & 326.0(24.7) & $-$12.61(0.01) & 275.8(1.8) \\
\enddata
\tablecomments{\scriptsize Column (1): Line name. Column (2): Rest wavelength. Column (3): Ionization potential. Column (4-15): The flux and $W_{80}$ measurements (with the uncertainty in parentheses) of each emission line on the nuclear spectrum extracted from a $r = 0\farcs75$ aperture for each sampled Seyfert galaxy.}
\label{tabnucs}
\end{deluxetable*}
\end{longrotatetable}

\section{Characteristics of Ionized Gas Kinematics and Evidence of Ionized Gas Outflows}\label{sec3}

\subsection{Extracted Spectra and Emission Line Measurements}\label{sec3.0}

To illustrate the rich information captured by JWST MIRI/MRS spectra, Figure~\ref{figsn} presents spectra extracted from $r = 0\farcs75$ apertures centered on the IR continuum peak (as the location of AGN) for each target. Channel 1, 2, and 3 spectra are extracted from the convolved spectral data cubes (Section~\ref{sec2.2}), and channel 4 spectra are extracted from spectral data cubes without convolution. The extraction aperture radius (i.e., 0\farcs75, $\sim 100 - 150$ pc at distances of the targets) for the four channels is large enough and hence no aperture correction was applied (as confirmed by the smooth transition from channel 3 to channel 4 spectra). These spectra exhibit diverse properties in terms of continuum shape and emission line intensity.

For example, the nuclear spectrum of NGC\,7172 is different from others, which is attributable to the strong silicate absorption around 9.7 and 18 $\mum$ of a dust lane and/or hot dust around its nucleus (\citealt{Smajic.etal.2012, Alonso-Herrero.etal.2021}). More importantly, spectra in Figure~\ref{figsn} exhibit more narrow emission lines and the lacking of PAH features (see also \citealt{Garcia-Bernete.etal.2024a, Garcia-Bernete.etal.2024c}), compared to Spitzer/IRS spectra (\citealt{Asmus.etal.2014, Garcia-Bernete.etal.2016}). Table~\ref{tabnucs} provides the measured flux and $W_{80}$ for 21 prominent ionized and 8 molecular lines from these spectra based on the non-parametric methodology (Sections~\ref{sec2.3}~\&~\ref{sec2.4}). Specific measurements of relatively broad PAH features, which are diluted by the underlying continuum and could be destroyed by AGN (e.g., \citealt{Garcia-Bernete.etal.2022a, Garcia-Bernete.etal.2022b, Zhang.etal.2022, Ramos Almeida.etal.2023}), will be presented and studied in \citealt{Garcia-Bernete.etal.2024c} and \citealt{Zhang.etal.2024}.

In this paper we focus on spatially resolved ionized gas kinematics targeting [Ar~{\footnotesize II}], [Ne~{\footnotesize II}], [Ar~{\footnotesize III}], [S~{\footnotesize IV}], [Ne~{\footnotesize III}], and [Ne~{\footnotesize V}] emission lines. Before the detailed analysis of the spatially resolved maps of these emission lines, we first look at the normalized profiles (with local continuum subtracted) of these six ionic emission lines from the above spectra. As shown in Figure~\ref{figln}, we find that NGC\,5728, NGC\,5506, and ESO137-G034 appear to display stronger signatures of disturbed ionized gas with broader emission line profiles ($W_{80} \gtrsim \rm 500\, km\,s^{-1}$) than NGC\,7172, MCG-05-23-016, and NGC\,3081 ($W_{80} \lesssim \rm 400\, km\,s^{-1}$). The widths of the observed line profiles include the contribution from instrumental broadening, which impacts the [Ne~{\scriptsize III}] 15.555~$\mum$ line the most as it has the longest wavelength among the six emission lines (\citealt{Labiano.etal.2021}). Taking such effect into consideration for [Ne~{\footnotesize III}] line, will reduce $W_{80}$ by less than 15\% (10\%) for measured $W_{80}$ with values larger than $250\,\rm km\,s^{-1}$ ($300\,\rm km\,s^{-1})$, assuming a Gaussian profile.

High ionization lines exhibit stronger signatures of disturbed kinematics with relatively broader profiles and/or larger velocity shifts, compared to low ionization lines (see also \citealt{Armus.etal.2023, Hermosa-Munoz.etal.2024a}). Additionally, for the six targets, all H$_2$ rotational lines have $W_{80} \lesssim 300\, \rm km\,s^{-1}$, except for NGC\,5728 with $W_{80}$ of H$_2$ lines $\sim 400\, \rm km\,s^{-1}$ (Table~\ref{tabnucs}). The relatively narrower H$_2$ lines against ionized emission lines are consistent with the result that the broadening of H$_2$ lines is more dominated by gravitational potential of a galaxy. Specific analysis of the H$_2$ lines of these targets is beyond the scope of this paper and is left to dedicated works in this series (i.e., \citealt{Davies.etal.2024, Esparza-Arredondo.etal.2024}, and particularly Delany et al. in preparation) % This value will be used later for the calculation of spatially resolved outflow rates.

 \begin{figure*}[!ht]
\center{\includegraphics[width=0.8\linewidth]{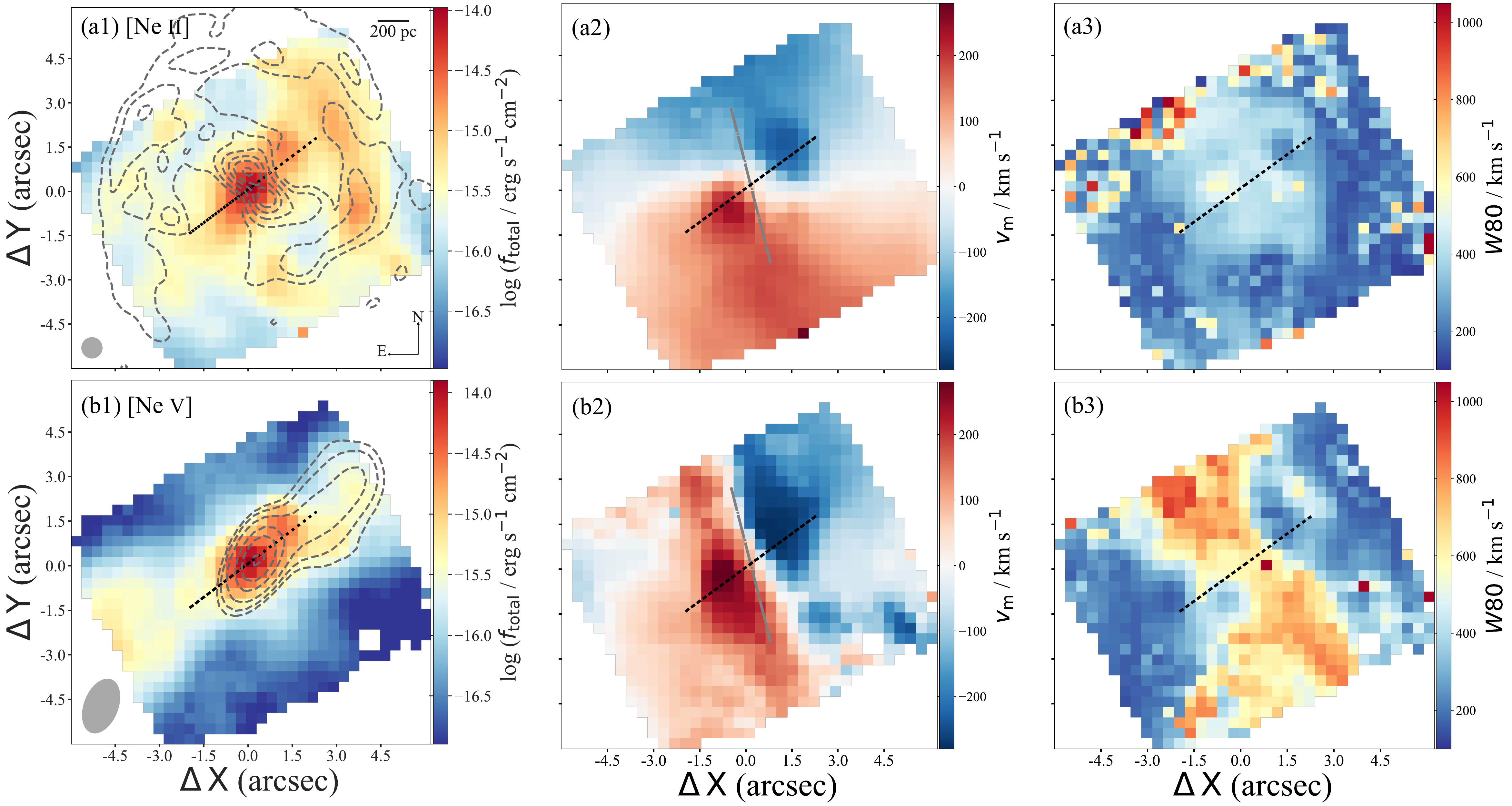}}
\caption{The distributions in NGC\,5728 central region for $f_{\rm total}$ (left panels), $v_{\rm m}$ (middle panels), and $W_{80}$ (right panels) of [Ne~{\footnotesize II}] 12.814~$\mum$ (top) and [Ne~{\footnotesize V}] 14.322~$\mum$ (bottom) emission lines. For all panels, the origin of coordinates is on the infrared continuum peak as location of the AGN. Panel (a1) also features a filled gray circle to indicate the angular resolution (i.e., 0\farcs7$\times0\farcs7$) of all colored maps, a compass with north is up and east is to the left, and a scale bar of 200 pc; they are the same for other panels in this figure. The gray dashed contours in panel (a1) indicate the ALMA CO(2-1) emission map as in Figure~\ref{figc}. The CO emission map here and hereafter is, unless specifically noted, convolved to the same angular resolution as the emission line maps, from 0.075, 0.15, 0.3, 0.5, 0.7, to 0.95 of the peak CO intensity. The gray dashed contours in panel (b1) indicate the VLA 4.9 GHz (6 cm) radio emission map (see acknowledgements), from 0.15, 0.2, 0.3, 0.5, 0.7, to 0.975 of the peak radio intensity unless specifically noted, with the gray ellipse indicating the beam size of the radio map. The black dashed line in each panel indicates the measured AGN ionization axis with the position angle (PA) of 127\degree (see Section~\ref{sec3.1.1}). The gray dotted-dashed lines in panels (a2, b2) indicate the kinematic major axis of the rotating gas disk in NGC\,5728 with the PA of 194\degree, which is fitted from CO(2-1) velocity field by \cite{Shimizu.etal.2019}.}\label{NGC5728_Plus}
\end{figure*}

\begin{figure*}[!ht]
\center{\includegraphics[width=0.8\linewidth]{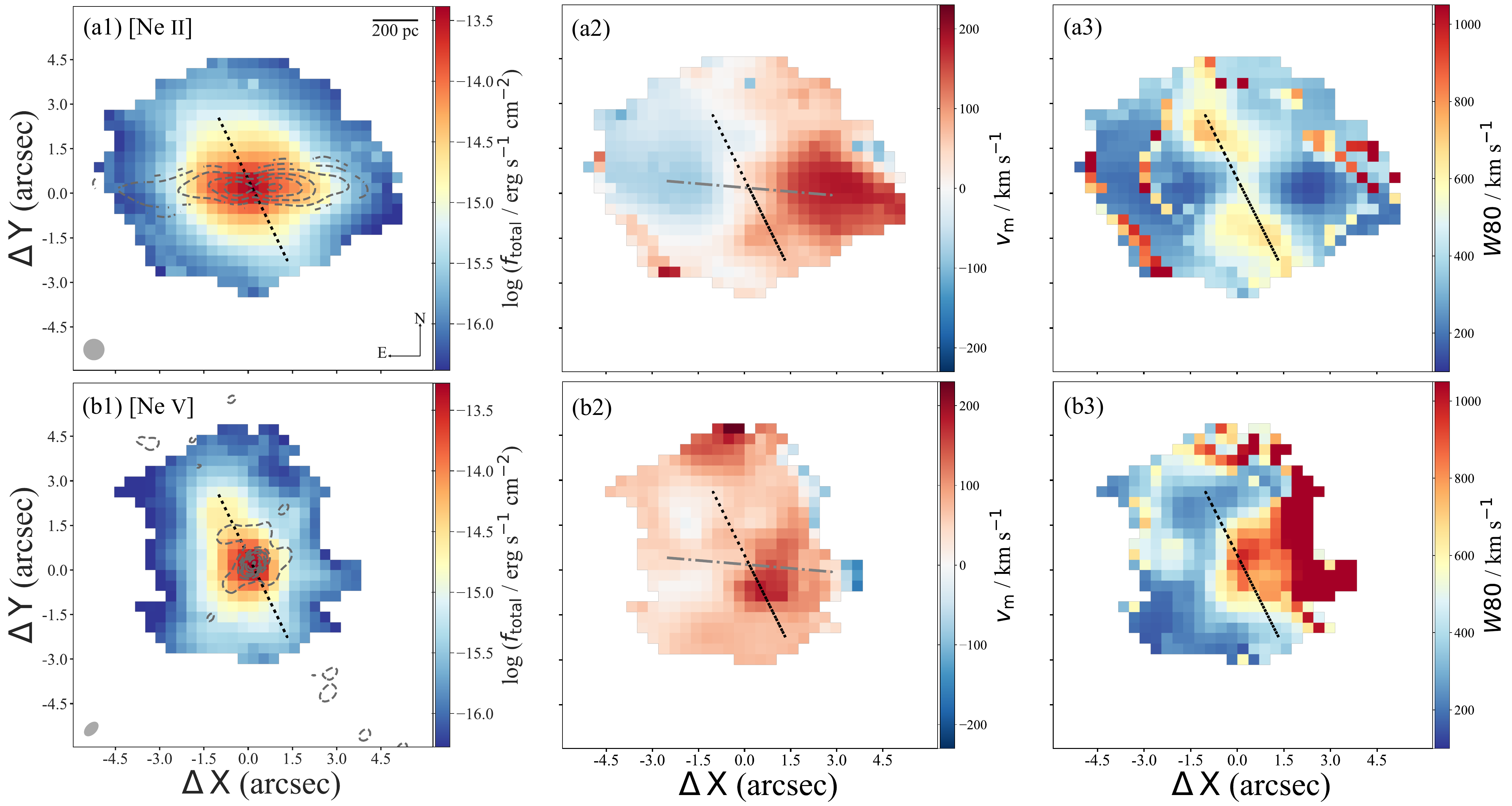}}
\caption{The same as Figure~\ref{NGC5728_Plus} but for NGC\,5506. The contours of the VLA 4.9 GHz radio emission map are from 0.01, 0.075, 0.15, 0.2, 0.3, 0.5, 0.7, to 0.975 of the peak radio intensity to highlight the diffuse wing-like radio emission. The PA of the measured AGN ionization axis (black dashed line in each panel) is 26\degree. The PA of the kinematic major axis (gray dotted-dashed lines in panels a2 \& b2) of the rotating gas disk in NGC\,5506 is 265\degree, which is fitted from CO(3-2) data cube by \cite{Esposito.etal.2024}.}\label{NGC5506_Plus}
\end{figure*}

\begin{figure*}[!ht]
\center{\includegraphics[width=0.8\linewidth]{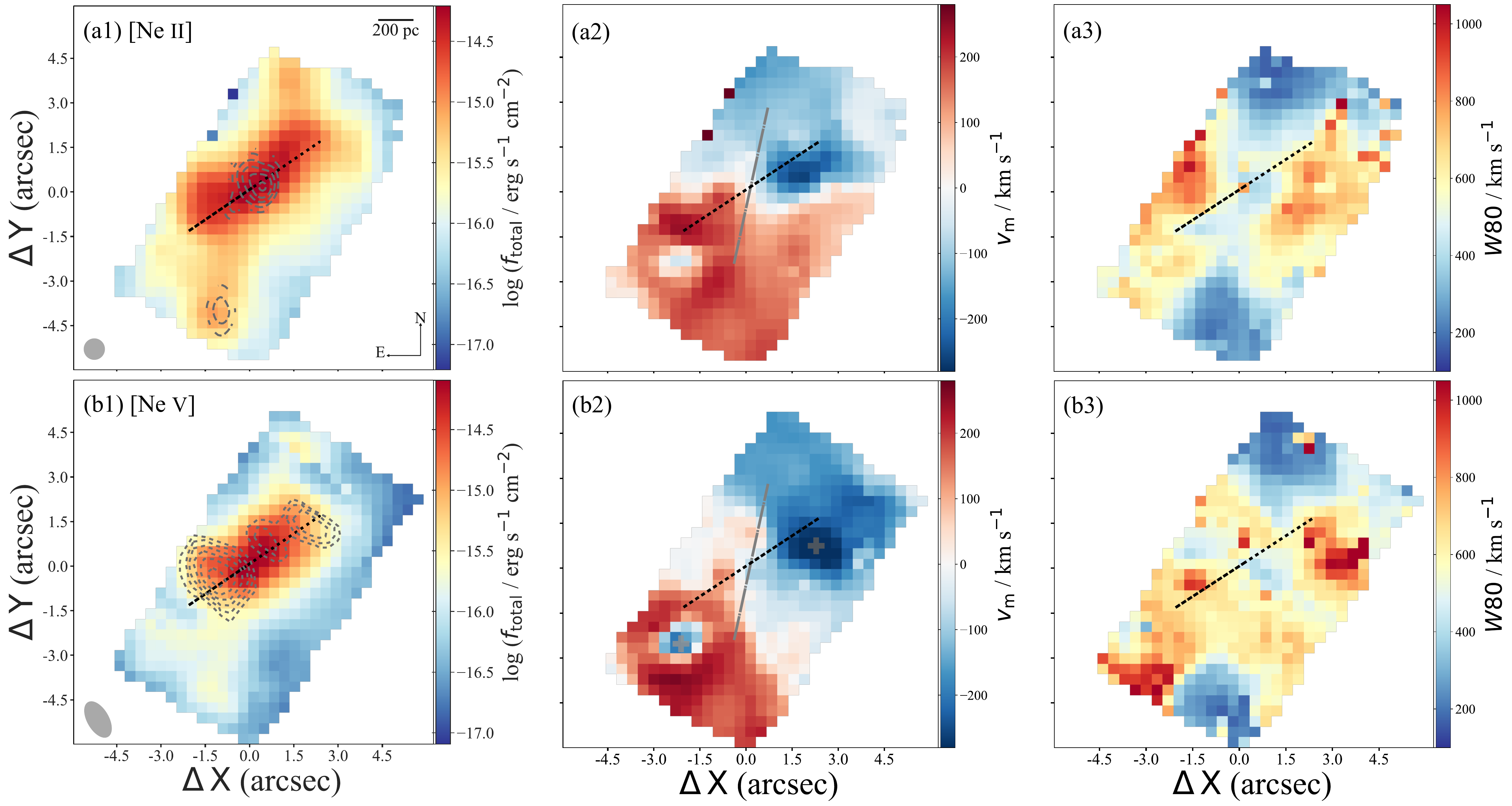}}
\caption{The same as Figure~\ref{NGC5728_Plus} but for ESO137-G034. The gray dashed contours in panel (b1) indicate 8.6 GHz (3.5 cm) radio emission map extracted from \cite{Morganti.etal.1999}. The observation is taken by the Australia Telescope Compact Array (ATCA) and the contours are from $-$0.5, $-$0.4, 0.4, 0.5, 0.7, 1, 1.3, to 1.8 mJy beam$^{-1}$. The PA of the measured AGN ionization axis (black dashed line in each panel) is 124\degree. The PA of the kinematic major axis (gray dotted-dashed lines in panels a2 \& b2) of the rotating gas disk is 168\degree, which is measured from the $v_{\rm m}$ distribution of [Ne~{\footnotesize II}] emission line in panel (a2) after masking the two kinematically distinct regions as marked by gray plus signs in panel (b2) (see the method in \citealt{Krajnovic.etal.2006}).}\label{ESO137-G034_Plus}
\end{figure*}

\begin{figure*}[!ht]
\center{\includegraphics[width=0.8\linewidth]{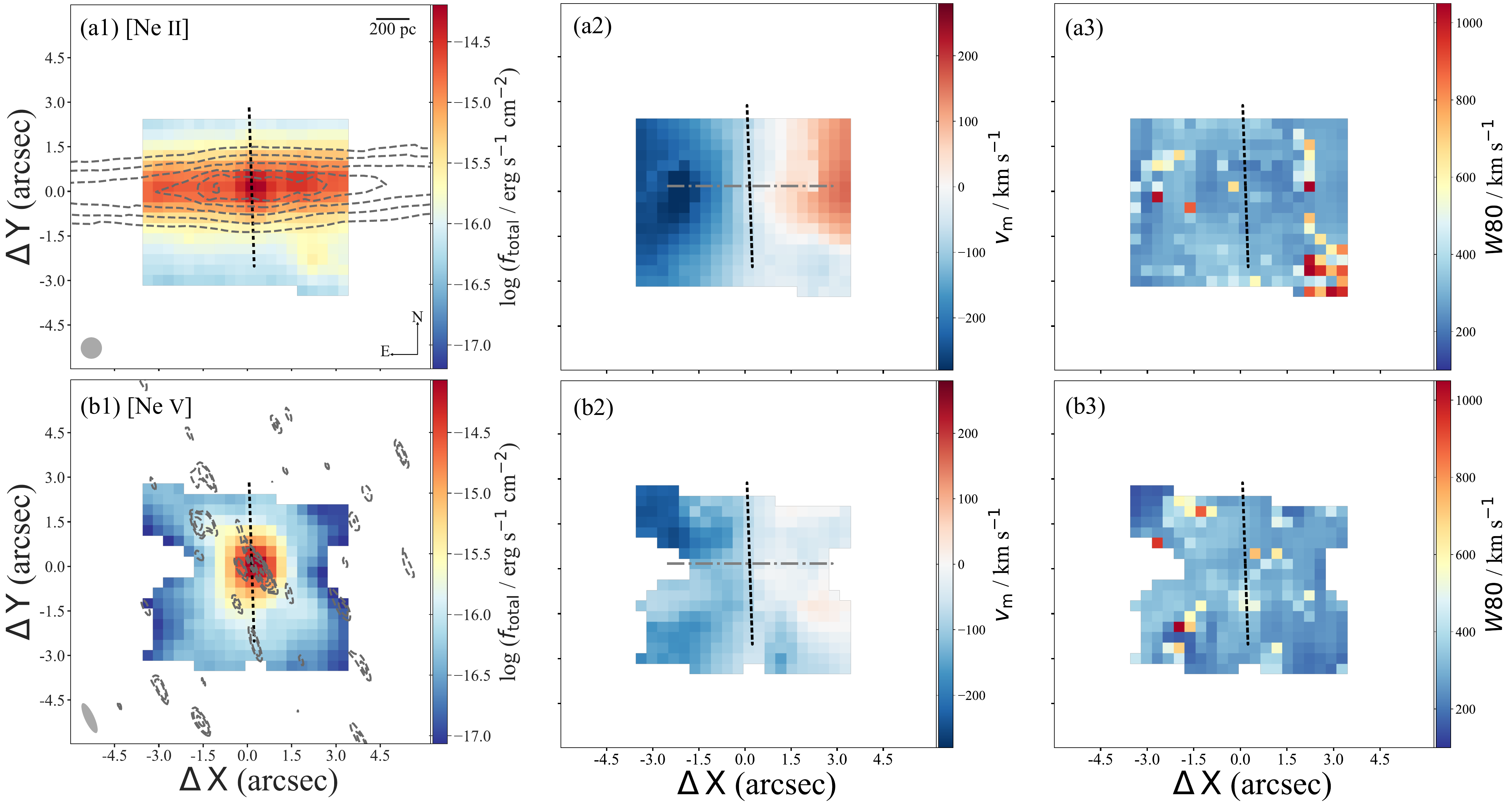}}
\caption{The same as Figure~\ref{NGC5728_Plus} but for NGC\,7172. The CO(2-1) emission map here has a little worse angular resolution comparing with the emission line maps, and hence no convolution of the CO emission map is performed. The PA of the measured AGN ionization axis (black dashed line in each panel) is 2\degree. The PA of the kinematic major axis (gray dotted-dashed lines in panels a2 \& b2) of the rotating gas disk in NGC\,7172 is 270\degree, which is fitted from CO(3-2) data cube by \cite{Alonso-Herrero.etal.2023}.}\label{NGC7172_Plus}
\end{figure*}

\begin{figure*}[!ht]
\center{\includegraphics[width=0.8\linewidth]{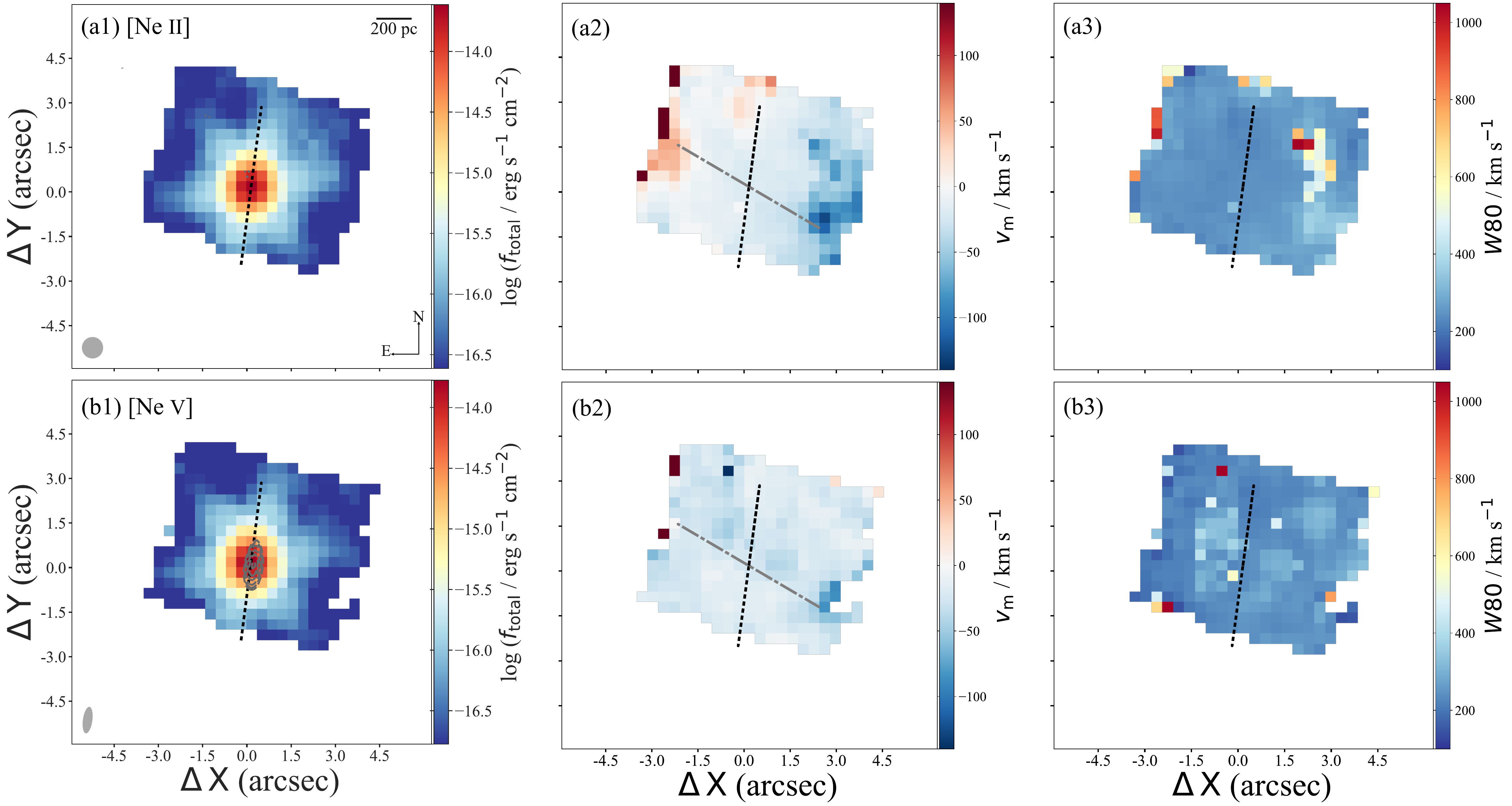}}
\caption{The same as Figure~\ref{NGC5728_Plus} but for MCG-05-23-016. The PA of the measured AGN ionization axis (black dashed line in each panel) is 172\degree. The PA of the kinematic major axis (gray dotted-dashed lines in panels a2 \& b2) of the rotating gas disk in MCG-05-23-016 is 59\degree, which is fitted from CO(2-1) data cube by \cite{Esparza-Arredondo.etal.2024}.}\label{MCG-05-23-016_Plus}
\end{figure*}

\begin{figure*}[!ht]
\center{\includegraphics[width=0.8\linewidth]{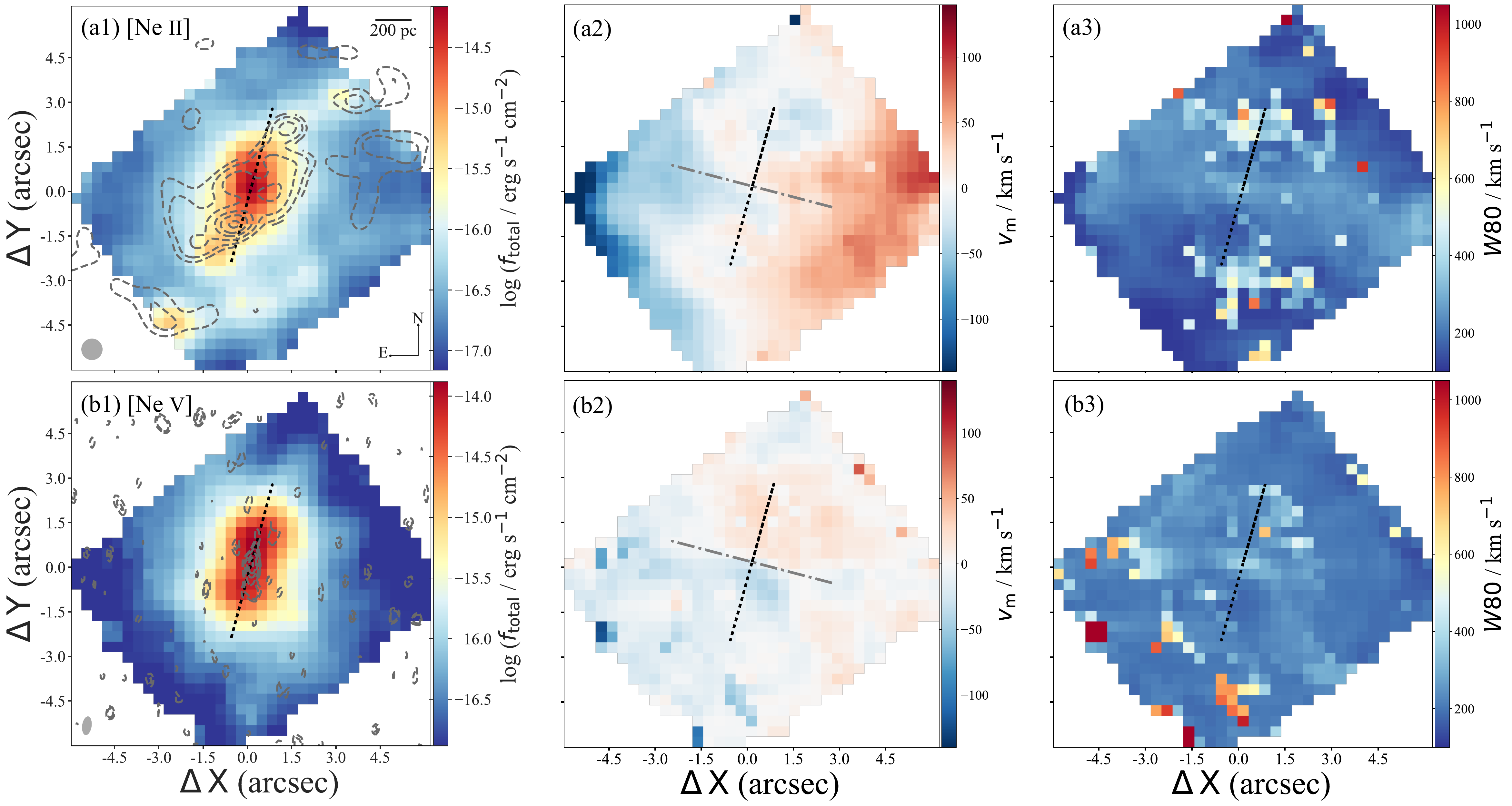}}
\caption{The same as Figure~\ref{NGC5728_Plus} but for NGC\,3081. The PA of the measured AGN ionization axis (black dashed line in each panel) is 165\degree. The PA of the kinematic major axis (gray dotted-dashed lines in panels a2 \& b2) of rotating gas disk is 255\degree, which is fitted from [N~{\footnotesize II}] velocity field by \cite{Ruschel-Dutra.etal.2021}.}\label{NGC3081_Plus}
\end{figure*}

 \subsection{Spatially Resolved Ionized Gas Distributions and Kinematics}\label{sec3.1}

Based on non-parametric measurements for individual 0\farcs35$\times$0\farcs35 spaxels in central regions of our targets, this section presents flux distributions and kinematics of the six emission lines covering a large range of ionization potentials. Based on the rather qualitative analysis as detailed below, we (1) showcase the evidence of ionized gas outflows along their AGN ionization cones in NGC\,5728, NGC\,5506, NGC\,3081, and likely in ESO137-G034, and (2) find some intriguing highly disturbed regions perpendicular to their AGN ionization cones in NGC\,5728, NGC\,5506, and ESO137-G034. In addition, we find that NGC\,7172 and MCG-05-23-016 also display the evidence of outflows, especially along the direction of their AGN ionization cones.

\subsubsection{Flux Distributions}\label{sec3.1.1}

Figure~\ref{NGC5728_Plus}(a1,b1)$-$\ref{NGC3081_Plus}(a1, b1) present, using [Ne~{\footnotesize II}] and [Ne~{\footnotesize V}] lines as examples, distributions of the total flux (i.e., the integral of the best-fit emission line profile for each spaxel, denoted as $f_{\rm total}$ hereafter) of these six lines (Figure~\ref{figa1}-\ref{figf1}).\footnote{Among these six emission lines, [Ne~{\scriptsize V}] line has the highest (97.1 eV) ionization potential, while [Ne~{\scriptsize II}] line has the lowest ionization potential (21.6 eV) except for [Ar~{\scriptsize II}] line (15.8 eV). [Ne~{\footnotesize II}] line is used for the illustration here and hereafter as this line is in the MRS sub-channel with a field of view that is much larger than the one of [Ar~{\scriptsize II}] line and is also more comparable to the one of [Ne~{\scriptsize V}] line.} Specifically, emission lines with relatively low ionization potentials (e.g.,  [Ar~{\footnotesize II}], [Ne~{\footnotesize II}]) exhibit more extended $f_{\rm total}$ distributions with more sub-structures. Meanwhile, emission lines with higher ionization potentials (e.g., [Ne~{\footnotesize V}]) tend to exhibit more concentrated $f_{\rm total}$ distributions with certain orientation dependence. This behavior is similar for all targets except for MCG-05-23-016, which only exhibits slight extension along the AGN ionization axis (as measured below) for all six lines. See Figure~\ref{figa1}-\ref{figf1} in the appendix for $f_{\rm total}$ distributions of all six lines panel by panel and ordered according to their ionization potentials.

Different excitation sources are able to explain these different flux distributions (e.g., \citealt{Sajina.etal.2022}). Specifically, both the star-formation and AGN activity can contribute to the excitation of relatively low ionization lines. Consistent with this, the $f_{\rm total}$ distributions of low ionization lines (e.g., [Ar~{\footnotesize II}], [Ne~{\footnotesize II}]) exhibit some extended structures, following the distributions of star-forming cold molecular gas traced by CO emission (gray dashed contours in Figure~\ref{NGC5728_Plus}a1-\ref{NGC3081_Plus}a1). Figure~\ref{NGC5728_Plus}(a1)-\ref{NGC3081_Plus}(a1) also exhibit that AGN ionization contributes to these low ionization lines around the nucleus as well, and we will discuss this again in Section~\ref{sec3.1.2}. Meanwhile, for high ionization lines (e.g., [Ne~{\footnotesize V}]), AGN ionization with a certain orientation dependence dominates their excitation as star formation activity cannot generate such high energy photons ($\sim 100$ eV).

\begin{figure*}[!ht]
%\figurenum{AI}
\center{\includegraphics[width=0.8\linewidth]{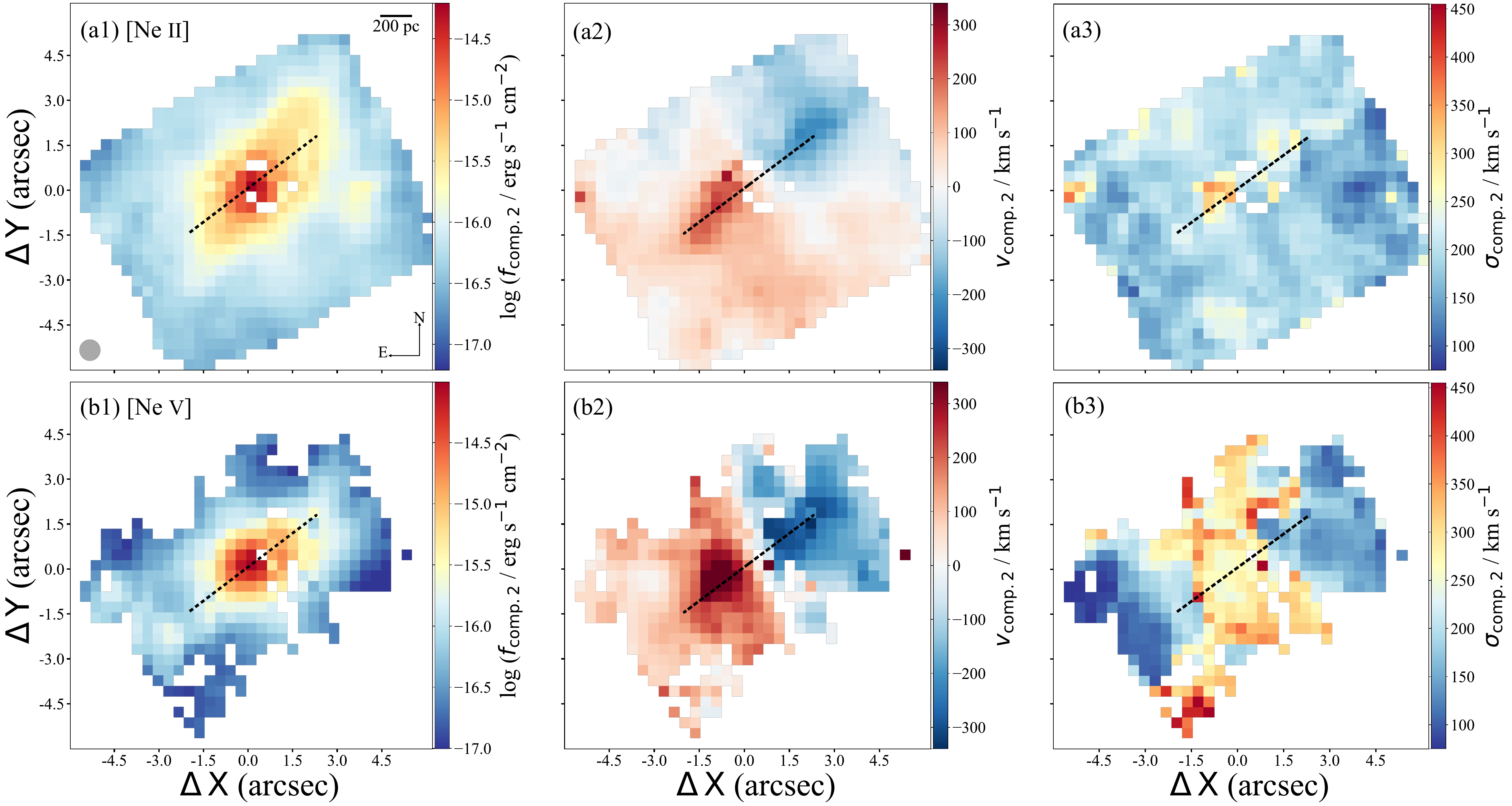}}
\caption{The distributions in NGC 5728 central region for flux (left panels), velocity (middle panels), and velocity dispersion (right panels) of [Ne~{\footnotesize II}] 12.814~$\mum$ (top) and [Ne~{\footnotesize V}] 14.322~$\mum$ (bottom) emission lines. The black dashed line in each panel indicates the measured PA of the AGN ionization axis. Panel (a1) also features a filled gray circle to indicate the angular resolution (i.e., 0\farcs7$\times0\farcs7$) of all colored maps, a compass with north is up and east is to the left, and a scale bar of 200 pc; they are the same for other panels in this figure.}\label{NGC5728_Plus_c2}
\end{figure*}
%Note that the flux distributions and velocity fields of the broad component of [Ne~{\footnotesize II}] and [Ne~{\footnotesize V}] emission lines are respectively consistent with each other, and similar to $f_{\rm total}$ and $v_{m}$ distributions of [Ne~{\footnotesize V}] emission line as shown in Figure~\ref{NGC5728_Plus}(b1, b2). In addition, the velocity dispersion field of the broad component of [Ne~{\footnotesize V}] emission line in panel (b3) exhibits the same orientation of emission line broadened regions as revealed by the $W80$ map in Figure~\ref{NGC5728_Plus}(b3).

\begin{figure*}[!ht]
%\figurenum{AII}
\center{\includegraphics[width=0.8\linewidth]{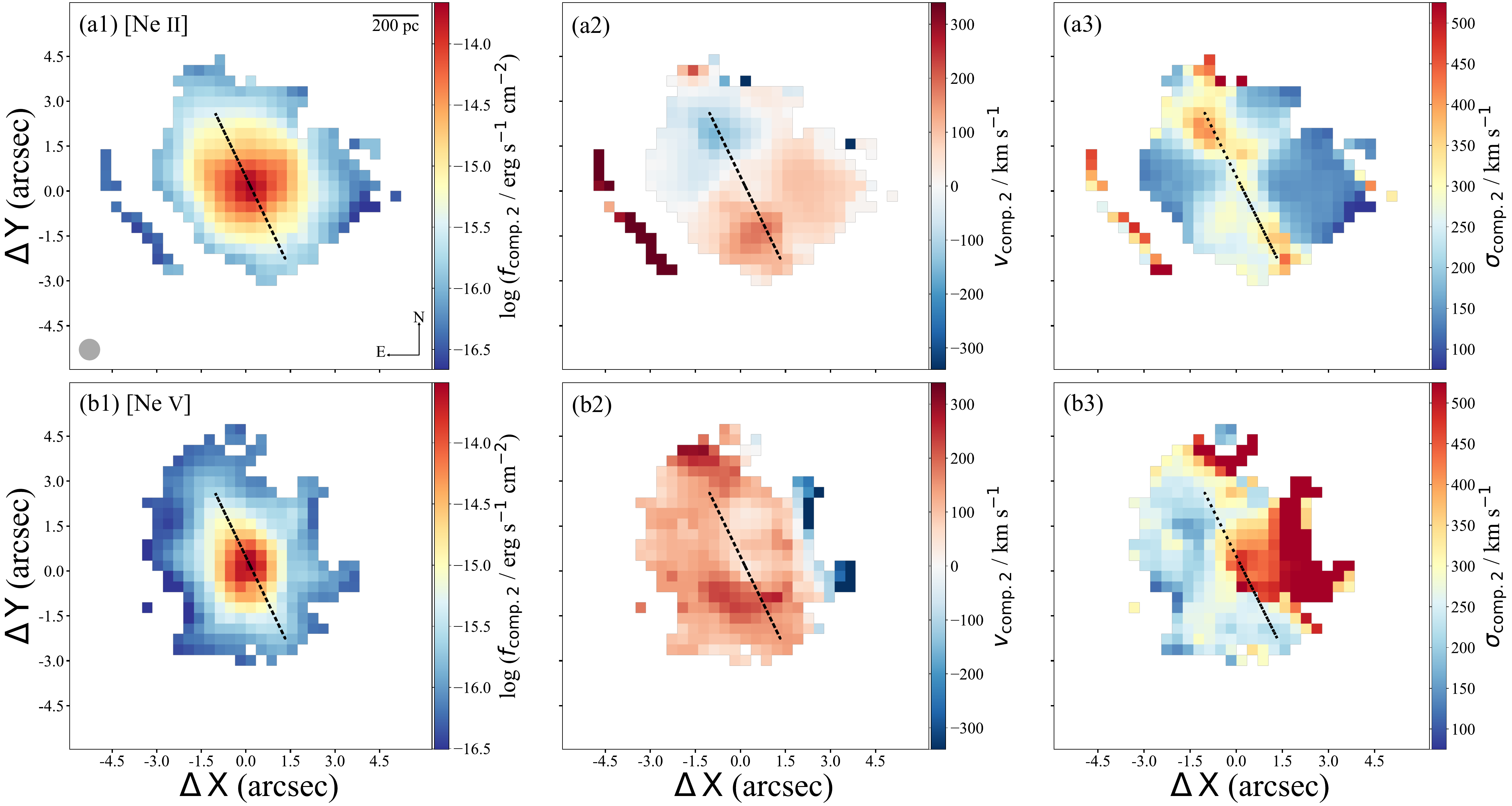}}
\caption{The same as Figure~\ref{NGC5728_Plus_c2} but for NGC\,5506.}\label{NGC5506_Plus_c2}
\end{figure*}

\begin{figure*}[!ht]
%\figurenum{AIII}
\center{\includegraphics[width=0.8\linewidth]{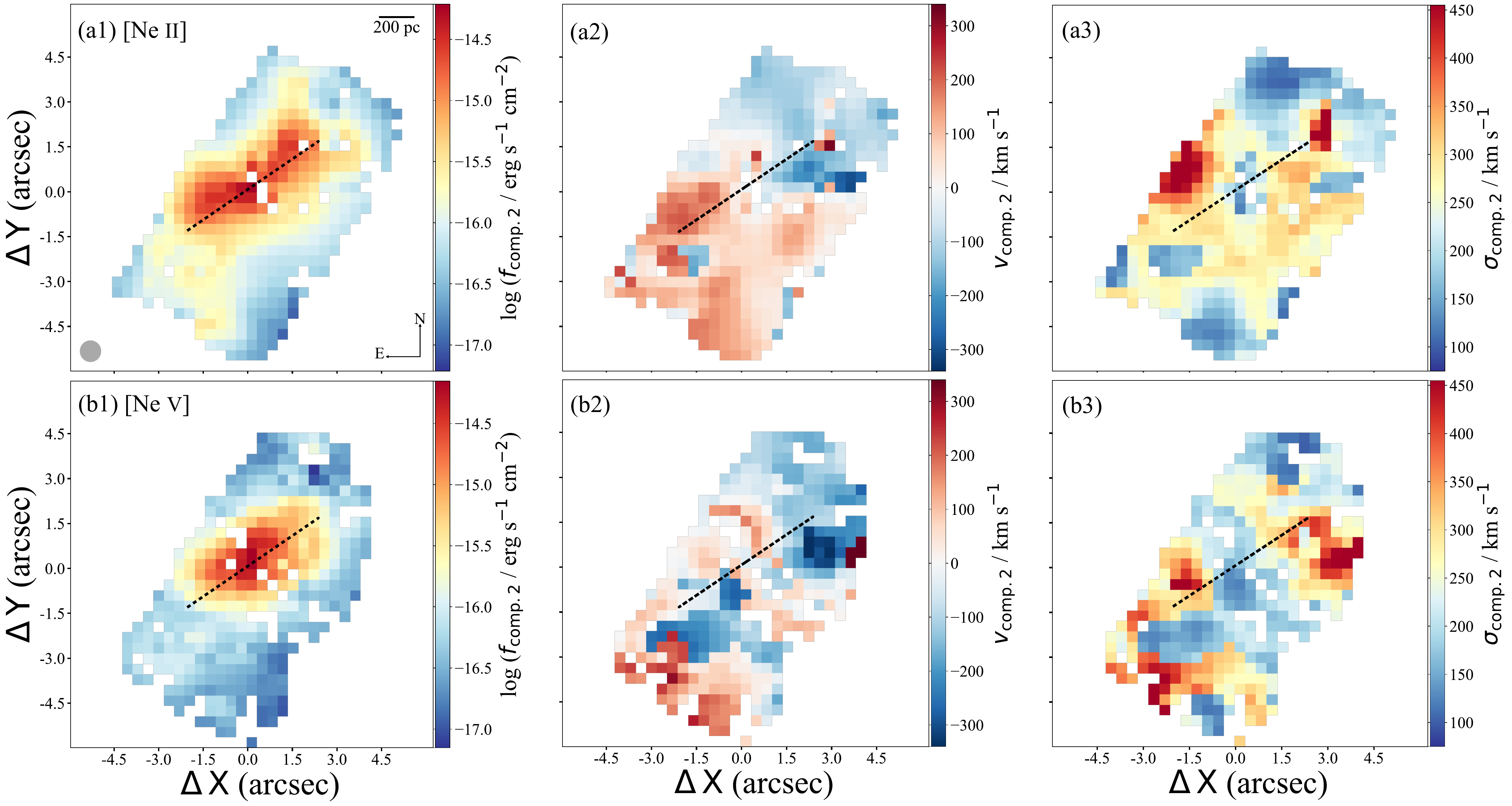}}
\caption{The same as Figure~\ref{NGC5728_Plus_c2} but for ESO137-G034. Note that velocity and velocity dispersion fields in panel (b2) and (b3) are more disturbed around the KDRs comparing with $v_{m}$ and $W_{80}$ distributions of [Ne~{\footnotesize V}] emission line as shown in Figure~\ref{ESO137-G034_Plus}(b2, b3).}\label{ESO137-G034_Plus_c2}
\end{figure*}

\begin{figure*}[!ht]
%\figurenum{AIV}
\center{\includegraphics[width=0.8\linewidth]{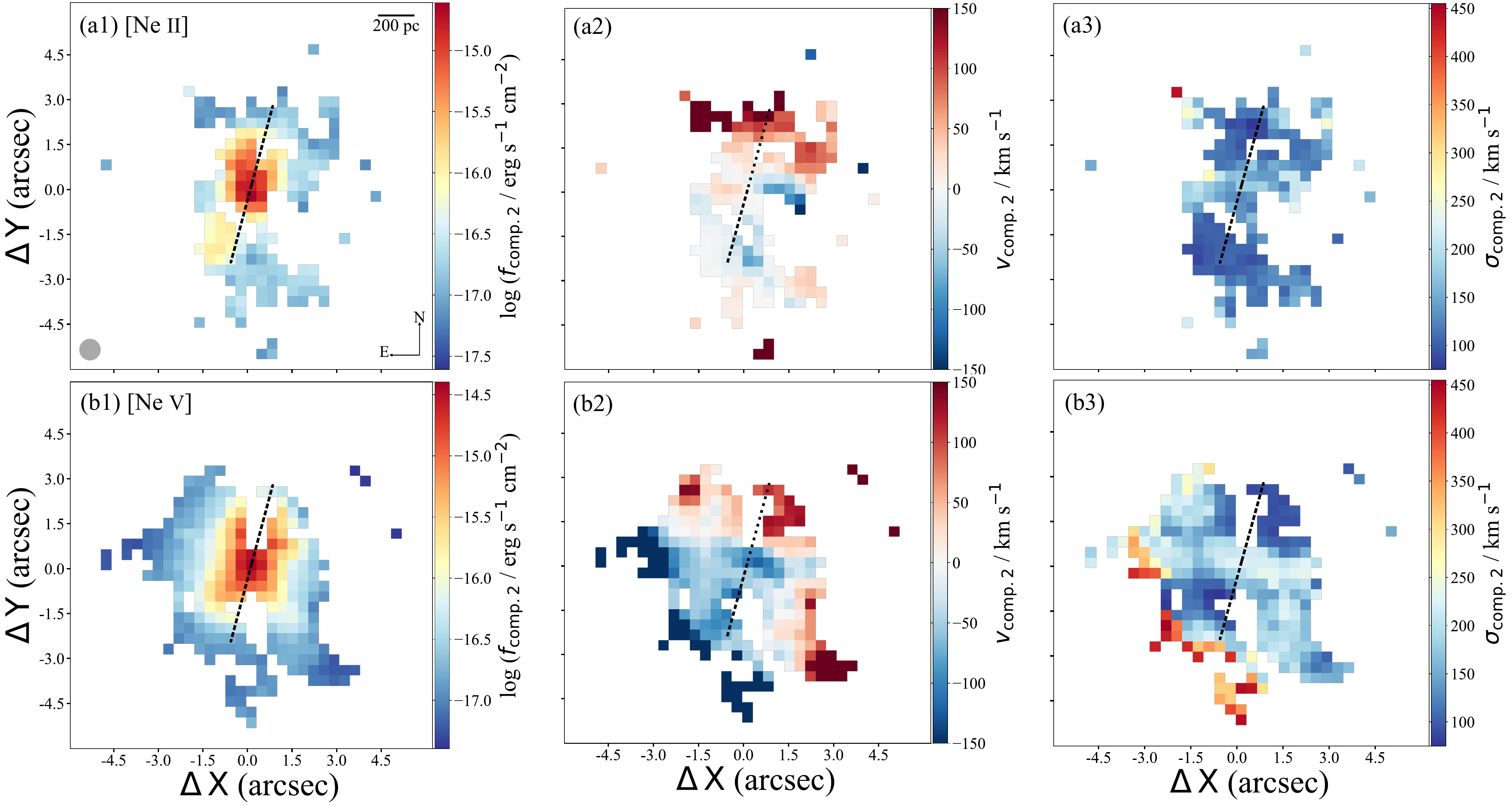}}
\caption{The same as Figure~\ref{NGC5728_Plus_c2} but for NGC\,3081.}\label{NGC3081_Plus_c2}
\end{figure*}

As indicated by black dashed lines in Figure~\ref{NGC5728_Plus}(b1)-\ref{NGC3081_Plus}(b1), the AGN ionization axis of each target is measured as the major axis of a elliptical source detected from [Ne~{\footnotesize V}] $f_{\rm total}$ maps using {\tt Source Extractor} (\citealt{Bertin&Arnouts1996, Barbary.etal.2017}). Note that the AGN ionization axis measured here of each target is basically aligned with their AGN ionization cone determined by optical emission lines (e.g., \citealt{Fischer.etal.2013, Thomas.etal.2017, Shimizu.etal.2019,Ruschel-Dutra.etal.2021, Esposito.etal.2024}), and/or radio jets (see gray dashed contours in Figure~\ref{NGC5728_Plus}b1-\ref{NGC3081_Plus}b1 and, e.g., \citealt{Morganti.etal.1999, Nagar.etal.1999, Thean.etal.2000, Orienti&Prieto2010, Durre&Mould2018} for the radio emission maps). AGN ionization cones are in general accompanied by outflows (e.g., \citealt{King&Pounds2015, Harrison&RamosAlmeida2024}),  as is further discussed in the following subsection for our 6 targets.
%which indicate the AGN ionization axis of each target, are consistent with orientations of the AGN ionization cone that determined by optical observations (NGC\,5728, NGC\,5506, NGC\,3081, and NGC\,7172; \citealt{Fischer.etal.2013, Thomas.etal.2017, Ruschel-Dutra.etal.2021}) or radio observations (NGC\,5728, NGC\,3081, NGC\,7172, MCG-05-23-016, NGC\,5506, and ESO137-G034; \citealt{Morganti.etal.1999, Nagar.etal.1999, Thean.etal.2000, Orienti&Prieto2010, Durre&Mould2018}).

\subsubsection{Velocity Fields}\label{sec3.1.2}

In addition to the flux distribution, the velocity field also helps reveal the physical condition of ionized gas and the existence of ionized gas outflow in each target. Figure~\ref{NGC5728_Plus}(a2,b2)$-$\ref{NGC3081_Plus}(a2,b2) present the $v_{m}$ distributions of [Ne~{\footnotesize II}] and [Ne~{\footnotesize V}] emission lines for the central region of each target (see Figure~\ref{figa2}-\ref{figf2} for complete $v_{m}$ distributions of the six lines). As aforementioned, the $v_{\rm m}$ distribution traces the primary velocity field of each emission line.

%{\bf In conclusion here, based on the rather qualitative analysis of flux distributions, velocity fields, and $W_{80}$ maps of six ionized emission lines covering a large range of ionization potentials, we (1) discuss the evidence of ionized gas outflows along their AGN ionization cones in NGC\,5728, NGC\,5506, NGC\,3081, and likely in ESO137-G034, and (2) find some intriguing highly disturbed regions perpendicular to their AGN ionization cones in NGC\,5728, NGC\,5506, and ESO137-G034. In addition, we find that NGC\,7172 and MCG-05-23-016 also display some evidence of outflows around their AGN.}% In terms of these findings, the following Section~\ref{sec4} and Section~\ref{sec5} provide, respectively, further quantifications on the outflow strength of the six targets, and specific discussion on these intriguing highly disturbed regions in NGC\,5728, NGC\,5506, and ESO137-G034.

For the six targets, their $v_{\rm m}$ distributions of low ionization lines (e.g., [Ar~{\footnotesize II}], [Ne~{\footnotesize II}]) all exhibit regular patterns, each with approximately centrally symmetric approaching and receding sides aligned with the kinematic major axis of the rotating gas disk (i.e., gray dashed lines in middle panels of Figure~\ref{NGC5728_Plus}-\ref{NGC3081_Plus} and see references in their captions). Such $v_{\rm m}$ distributions indicate disk-rotation-dominated motions of low ionization lines. The $v_{\rm m}$ distributions of AGN ionization dominated high ionization lines, e.g., [Ne~{\footnotesize V}], are all different from those of low ionization lines, with some of them even showing obvious twists of their kinematic axes (see Figure~\ref{NGC5728_Plus}a2,b2$-$\ref{NGC3081_Plus}a2,b2). Such twisting is more significant for NGC\,5728, NGC\,5506, NGC\,3081, and slight for ESO137-G034 (see further note on ESO137-G034 at the end of this subsection), while the $v_{\rm m}$ distributions of high ionization lines in NGC\,7172 and MCG-05-23-016 exhibit more perturbations. The contrasting $v_{\rm m}$ distributions highlight that emission lines of different ionization potentials are tracing different gas motions.

\begin{figure*}[!ht]
\center{\includegraphics[width=0.675\linewidth]{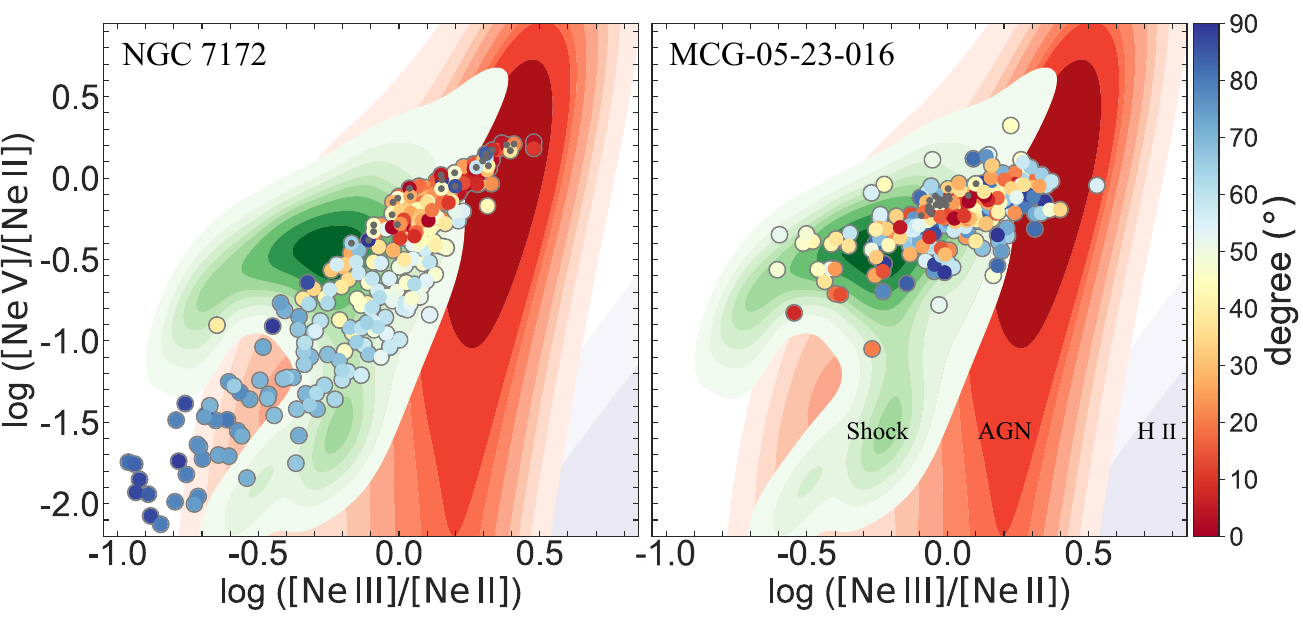}}
\caption{The diagnostic diagram of ionized emission line ratios [Ne~{\footnotesize V}]/[Ne~{\footnotesize II}] versus [Ne~{\footnotesize III}]/[Ne~{\footnotesize II}] for spatially resolved spaxels in NGC\,7172 (left) and MCG-05-23-016 (right), color coded according to the orientation of each spaxel relative to the AGN ionization axis (0$\degree$and 90$\degree$ mean parallel and perpendicular to the AGN ionization axis, respectively). The data points with a gray dot pertain to spaxels within the central $r=1\arcsec$ aperture. The greenish, reddish, and purplish contours (from the left to the right) in each panel are the model results for fast radiative shocks, AGN, and H~{\footnotesize II} regions, respectively. The fast radiative shock models (including the shock precursor) are calculated using the {\tt MAPPINGS~V} (\citealt{Sutherland&Dopita2017}) by \cite{Alarie&Morisset2019}, while the H~{\footnotesize II} and AGN models are calculated using {\tt C{\scriptsize LOUDY}} (\citealt{Ferland.etal.2017, Chatzikos.etal.2023}) by \cite{Morisset.etal.2015} and \cite{Pereira-Santaella.etal.2024}, respectively.}\label{LineD}
\end{figure*}

The kinematic major axes of the $v_{\rm m}$ distributions for [Ne~{\footnotesize V}] emission line, as the representative of high ionization lines, tend to be aligned with the AGN ionization cones (i.e., black dashed lines in Figure~\ref{NGC5728_Plus}a1\&b1$-$\ref{NGC3081_Plus}a1\&b1). This trend is also more significant for NGC\,5728, NGC\,5506, NGC\,3081, and ESO137-G034 (see Figure~\ref{NGC5728_Plus}b2$-$\ref{ESO137-G034_Plus}b2, and \ref{NGC3081_Plus}b2). For their $v_{\rm m}$ distributions, the kinematic major axis gradually twists from those of [Ar~{\footnotesize II}] and [Ne~{\footnotesize II}] to that of [Ne~{\footnotesize V}] with an angle ranging from $\sim 45 - 90\degree$ (see Figure~\ref{figa2}, \ref{figb2}, \ref{figc2}, and \ref{figf2} for the more complete trend of such twisting in these four targets). Given the consistent orientations of the [Ne~{\footnotesize V}] $v_{\rm m}$ distributions and the AGN ionization cones, a reasonable explanation for such twisting of these targets is the increasing contribution of AGN-driven outflows in the primary motion of ionized gas. Namely, while $v_{\rm m}$ distributions of low ionization lines (e.g., [Ar~{\footnotesize II}], [Ne~{\footnotesize II}]) trace disk-rotation motions, $v_{\rm m}$ distributions of high ionization lines (e.g., [Ne~{\footnotesize V}]) are increasingly dominated by outflow motions along the AGN ionization cone.%given the alignment between the $v_{\rm m}$ distribution of [Ne~{\footnotesize V}] emission line and the AGN ionization cone
%\footnote{\bf As will be mentioned later, dust extinction on [Ne~{\scriptsize II}] 12.814~$\mum$ and [Ne~{\scriptsize V}] 14.322~$\mum$ emission lines is comparable and small for these targets, and hence should be the reason for such twisting.}

More importantly, for NGC 5728, NGC 5506, ESO137-G034, and NGC\,3081, their velocity distributions of the relatively broader components (i.e., $v_{\rm comp. 2}$), which are in general purely contributed by gas outflows, in [Ne~{\footnotesize II}] and [Ne~{\footnotesize V}] emission lines are also aligned with the AGN ionization cones (see Figure~\ref{NGC5728_Plus_c2}$-$\ref{NGC3081_Plus_c2} and the brief discussion in Appendix~\ref{secA}). Moreover, their $v_{m}$ distributions of the [Ne~{\footnotesize V}] emission line are qualitatively similar to these velocity distributions of the relatively broader components in [Ne~{\footnotesize II}] and [Ne~{\footnotesize V}] emission lines. This result supports the explanation that the twisting between the $v_{\rm m}$ distributions of different emission lines in the four targets is due to the increasing contribution of AGN-driven outflows in the primary motion of ionized gas. Under such situation, the low outflow strengths of NGC\,7172 and MCG-05-23-016 are able to explain the lacking of significant twisting of their [Ne~{\footnotesize V}] $v_{\rm m}$ distributions against those of [Ne~{\footnotesize II}], as such twisting is intrinsically because of collimating motions of highly ionized gas in strong outflows.

Although NGC\,7172 and MCG-05-23-016 do not clearly exhibit such twisting between $v_{\rm m}$ distributions of different emission lines, they still display the evidence of outflows around their AGN. The observed MIR emission line ratios along with theoretical calculations indicate the existence of fast radiative shocks ($v \approx 100 -1000\,{\rm km\,s^{-1}}$) associated with outflows in NGC\,7172 and MCG-05-23-016 (\citealt{Hermosa-Munoz.etal.2024a, Zhang.etal.2024}). As further checked in Figure~\ref{LineD}, the evidence of fast radiative shocks is still there for individual spaxels within the central region of NGC\,7172 and MCG-05-23-016. This result is more evident for spaxels located along the direction of their AGN ionization axes (i.e., the reddish and yellowish points in Figure~\ref{LineD}). Namely, while the signature of relatively stronger collimating outflows (i.e., the twisting of velocity fields) in NGC\,7172 and MCG-05-23-016 is lacking, these two targets still exhibit the evidence of outflows along their AGN ionization axes given the existence of fast radiative shocks. For the other four targets, we have checked that while some spaxels in the periphery, especially for NGC\,3081, exhibit characteristics of fast radiative shocks, their central regions are dominated by the AGN excitation, especially for ESO137-G034.
% This result hints that NGC\,7172 and MCG-05-23-016 have stronger interactions between outflows and the ambient environments while more conclusive evidence requires further study.

Furthermore, with the same JWST/MRS data set, \cite{Hermosa-Munoz.etal.2024a} and \cite{Esparza-Arredondo.etal.2024} provide more specific studies for NGC\,7172 and MCG-05-23-016, respectively. Specifically, \cite{Hermosa-Munoz.etal.2024a} found from [Ne~{\footnotesize V}] and [Ne~{\footnotesize VI}] emission lines a biconical ionized gas outflow emerging north-south from the nucleus, extending at least $\sim 2\farcs5$ N and $3\farcs8$ S (projected distance of $\sim$ 450 and 680 pc, respectively). Most of the emission arising in the northern part of the cone was not previously detected due to obscuration (e.g, \citealt{Alonso-Herrero.etal.2023}). Moreover, \cite{Hermosa-Munoz.etal.2024a} revealed that NGC\,7172 is likely a case of weak coupling between outflow and the host given the kinematic properties and geometry of the outflow. This can further explain the lacking of significant collimating outflows along the AGN ionization axis in NGC\,7172. Meanwhile, the intensity maps of [Ne~{\footnotesize V}] and [Ne~{\footnotesize VI}] emission lines of MCG-05-23-016 exhibit point-like distributions. Nevertheless, \cite{Esparza-Arredondo.etal.2024} observed clumps of more extended warm molecular gas traced by the H$_2$ $S(3)$, $S(4)$, and $S(5)$ lines with velocity dispersions of up to $\sim$160 $\rm km\,s^{-1}$, in regions where cold molecular gas traced by CO(2-1) emission is absent. They found one of these clumps, located at $\sim$350 pc northwest of the nucleus, shows kinematics that are consistent with outflowing gas, while this clump is likely associated with star formation activity.
%Moreover, they found evidence for positive feedback in two distinct outflowing clumps at projected distances of $\sim 3\farcs1$ and $\sim 4\farcs3$ (i.e. $\sim$ 560 and 780 pc) south-west from the AGN. 

In addition, \cite{Davies.etal.2024} provided a detailed discussion on the complex geometry of circumnuclear region of NGC 5728 based on the data set presented here. They illustrated that the AGN ionization cone (as shown in Figure~\ref{NGC5728_Plus}) and the corresponding gas outflow in NGC\,5728 strongly intersect with the galaxy gas disk. Especially, most of the approaching side of the outflow cone lies behind the galactic disk while vise verse for the receding side. This kind of outflow geometry, especially the relatively perpendicular outflow orientation against the line of sight, is also reflected by the flux distributions and velocity fields of the relatively broader components in [Ne~{\footnotesize II}] and [Ne~{\footnotesize V}] emission lines (i.e., Figure~\ref{NGC5728_Plus_c2}$-$Figure~\ref{NGC3081_Plus_c2}), as well as the emission line profiles of the nuclear spectra (i.e., Figure~\ref{figln}). Consistent with such outflow geometry, these flux distributions and velocity fields, also the emission line profiles exhibit comparable redshifted and blueshifted components, with even stronger redshifted contributions. Moreover, given the relatively low AGN luminosity of the six targets (i.e., $\sim 10^{43.4} - 10^{44.3}\,{\rm erg\,s^{-1}}$) and hence the relatively low outflow strength (\citealt{Fiore.etal.2017}), the outflowing ionized gas in the six targets may not eventually escape the galaxy but may be decelerated by the galaxy gravity to stop and even fall back from certain radius (\citealt{Alonso-Herrero.etal.2023, Davies.etal.2024, Esposito.etal.2024}).

Note that the six targets are {\it selected} with prior outflow rate measurements to study how outflows are launched and driven from AGN with similar luminosities, but significantly different outflow strengths. Therefore, the existence of ionized gas outflows in all the six targets is within expectation, whereas the 100 \% detection rate of outflows is not applicable to a more general sample of AGN (e.g., \citealt{Fischer.etal.2013, Forster-Schreiber.etal.2019, Leung.etal.2019, Ruschel-Dutra.etal.2021}). Some of our targets were also included in previous studies based on slit or IFU spectral observations in optical bands (e.g., \citealt{Fischer.etal.2013, Ruschel-Dutra.etal.2021}), while not all of these targets were claimed to exhibit significant signatures of ionized gas outflows according to the different criteria in these studies. This result could be due to the mismatch between the slit position and the outflow cone (\citealt{Fischer.etal.2013}), or more likely due to the obscuration effect in optical bands as found by \cite{Hermosa-Munoz.etal.2024a}. Accordingly, further study using JWST/MRS observations is promising to reveal relatively weak outflow signatures in different targets that might be missed by optical observations.

Among the four targets with velocity fields showing the evidence of ionized gas outflows, we further note the following points. The $v_{\rm m}$ distribution of [Ne~{\footnotesize II}] emission line in NGC\,5728 exhibits significant discrepancy against the kinematic major axis of the rotating gas disk around the AGN (see Figure~\ref{NGC5728_Plus}a2). This is consistent with the argument in Section~\ref{sec3.1.1} that the AGN ionization (i.e., the corresponding outflow) contributes to these low ionization lines around the AGN as well (see also \citealt{Shimizu.etal.2019} and \citealt{Davies.etal.2024}). The $v_{\rm m}$ distribution of [Ne~{\footnotesize V}] emission line in NGC\,5506 exhibit a complex outflow structure with the relatively weak blueshifted and strong redshifted components, as well as the redshifted velocity blob $\sim$ 600 pc towards the north. Such complex $v_{\rm m}$ distribution in NGC\,5506, especially the redshifted velocity blob towards the north, can be attributed to the nearly perpendicular large-scale ionized gas outflows with the wide open angle as detailed by \cite{Esposito.etal.2024}.\footnote{\cite{Esposito.etal.2024} also found some blueshifted blobs in the south of the velocity field of the broad component in [O~{\footnotesize III}] emission line but these blobs are out of the field of view of the data set studied here.} Specifically, \cite{Esposito.etal.2024} found consistent characteristics as shown here of the ionized gas kinematics in NGC\,5506 based on GTC/MEGARA IFU observations. Further detailed study of NGC\,5506 and also NGC\,3081 in this series will be presented by Delaney et al. (in preparation).

Additionally, the $v_{\rm m}$ distributions of ESO137-G034 all exhibit two significant kinematically distinct regions (KDRs) $\sim 400\,\rm pc$ toward the southeast and northwest, respectively (see Figure~\ref{ESO137-G034_Plus}a2, b2 and Figure~\ref{figc2}). These KDRs severely affect the decomposition of the broad components associated with outflows in [Ne~{\footnotesize II}] and [Ne~{\footnotesize V}] emission lines. Thus, the broad components in [Ne~{\footnotesize II}] and [Ne~{\footnotesize V}] emission lines can not be clearly associated to ionized gas outflows (see Figure~\ref{ESO137-G034_Plus_c2}). Further detailed study of ESO137-G034 in this series will be presented by Haidar et al. (in preparation).

%\begin{figure*}[t]
%\center{\includegraphics[width=0.8\linewidth]{Line_Diagnostics.pdf}}
%\caption{\bf The diagnostic diagram of ionized emission line ratios [Ne~{\footnotesize V}]/[Ne~{\footnotesize II}] versus [Ne~{\footnotesize III}]/[Ne~{\footnotesize II}] for spatially resolved spaxels in each target, with color coded according to the location of each spaxel relative to the nucleus. The greenish, reddish, and purplish contours (from the left to the right) in each panel are the model results for fast radiative shocks, AGN, and SF galaxies, respectively.}\label{LineD}
%\end{figure*}

\begin{figure*}[t]
\center{\includegraphics[width=0.9\linewidth]{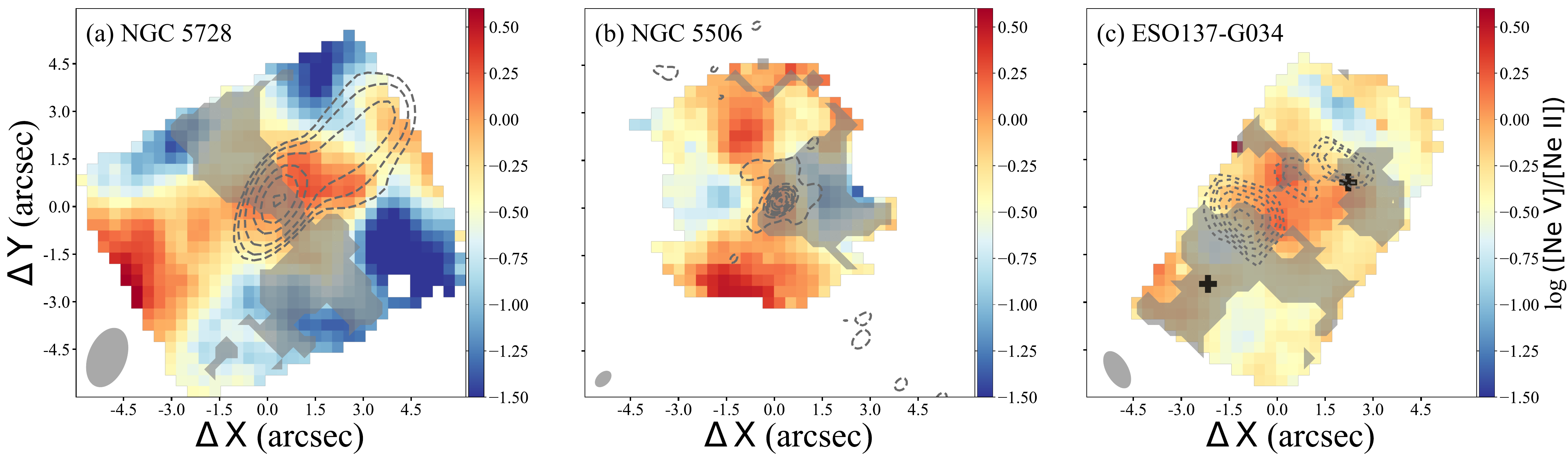}}
\caption{[Ne~{\footnotesize V}]/[Ne~{\footnotesize II}] line ratio distributions as the indicator of AGN excitation strength for central regions of (a) NGC\,5728, (b) NGC\,5506, and (c) ESO137-G034. In each panel, the dashed contours indicate the radio emission as in Figure~\ref{NGC5728_Plus}(b1), \ref{NGC5506_Plus}(b1), and ~\ref{ESO137-G034_Plus}(b1), while the gray-shaded areas delineate the sub-region with $W_{80} > 600\,{\rm km\,s^{-1}}$ measured for the [Ne~{\footnotesize V}] emission line (Section~\ref{sec3.1.3}). For NGC\,5728 in panel (a), VLA image at 4.9 GHz revealed an elongated nuclear radio jet along the AGN ionization cone (\citealt{Durre&Mould2018}). For NGC\,5506 in panel (b), VLA images at 4.9 GHz and 8.5 GHz exhibit an unresolved nuclear core and diffuse wing-like radio emission extending mainly to the northwest and east of the AGN (\citealt{Orienti&Prieto2010}). For ESO137-G034 in panel (c), The Australia Telescope Compact Array (ATCA) image at 8.6 GHz exhibits two off-nuclear radio blobs toward southeast and northwest (\citealt{Morganti.etal.1999}). Additionally, the black plus signs in panel (c) indicate two kinematically distinct regions as in Figure~\ref{ESO137-G034_Plus}(b2).}\label{NeRatioII}
\end{figure*}

\subsubsection{$W80$ Maps}\label{sec3.1.3}
%Broad emission lines (e.g., $W_{80} > 600\,{\rm km\,s^{-1}}$) are proposed to be associated with gas components of non-circular motions, since rotation velocities and velocity dispersions dominated by galaxy potentials {\bf even for most massive galaxies} are not able to produce such broadening 
%{\bf Very broad emission lines are in general associated with gas components of non-circular motions, since rotation velocities and velocity dispersions dominated by galaxy potentials, even for the most massive galaxies, are below certain threshold} (e.g. \citealt{Sun.etal.2017, Kakkad.etal.2020, Ruschel-Dutra.etal.2021, Riffel.etal.2023}). 

Among the six targets, we find NGC\,5728, NGC\,5506, and ESO137-G034 exhibit widely distributed regions of significantly broadened [Ne~{\footnotesize V}] emission line (Figure~\ref{NGC5728_Plus}a3, b3$-$\ref{ESO137-G034_Plus}a3, b3 and see also Figure~\ref{figa3}$-$\ref{figc3}). The other three targets do not exhibit such spatially extended regions of large $W_{80}$, but have some discrete spaxels with large $W_{80}$ values within the field of view (Figure~\ref{NGC7172_Plus}a3, b3$-$\ref{NGC3081_Plus}a3, b3 and see also Figure~\ref{figd3}$-$\ref{figf3}). This result is consistent with the relatively broader emission line profiles of the former three targets as discussed in Section~\ref{sec3.0}. For convenience of discussion here and in Section~\ref{sec5}, we simply denote the regions with $W_{80} > 600\,{\rm km\,s^{-1}}$ as the “highly disturbed regions” (e.g., \citealt{Sun.etal.2017, Kakkad.etal.2020, Ruschel-Dutra.etal.2021}).% This is consistent with the result in Section~\ref{sec3.0} based on individual spectra extracted from a central aperture.

\begin{figure}[t]
\center{\includegraphics[width=0.9\linewidth]{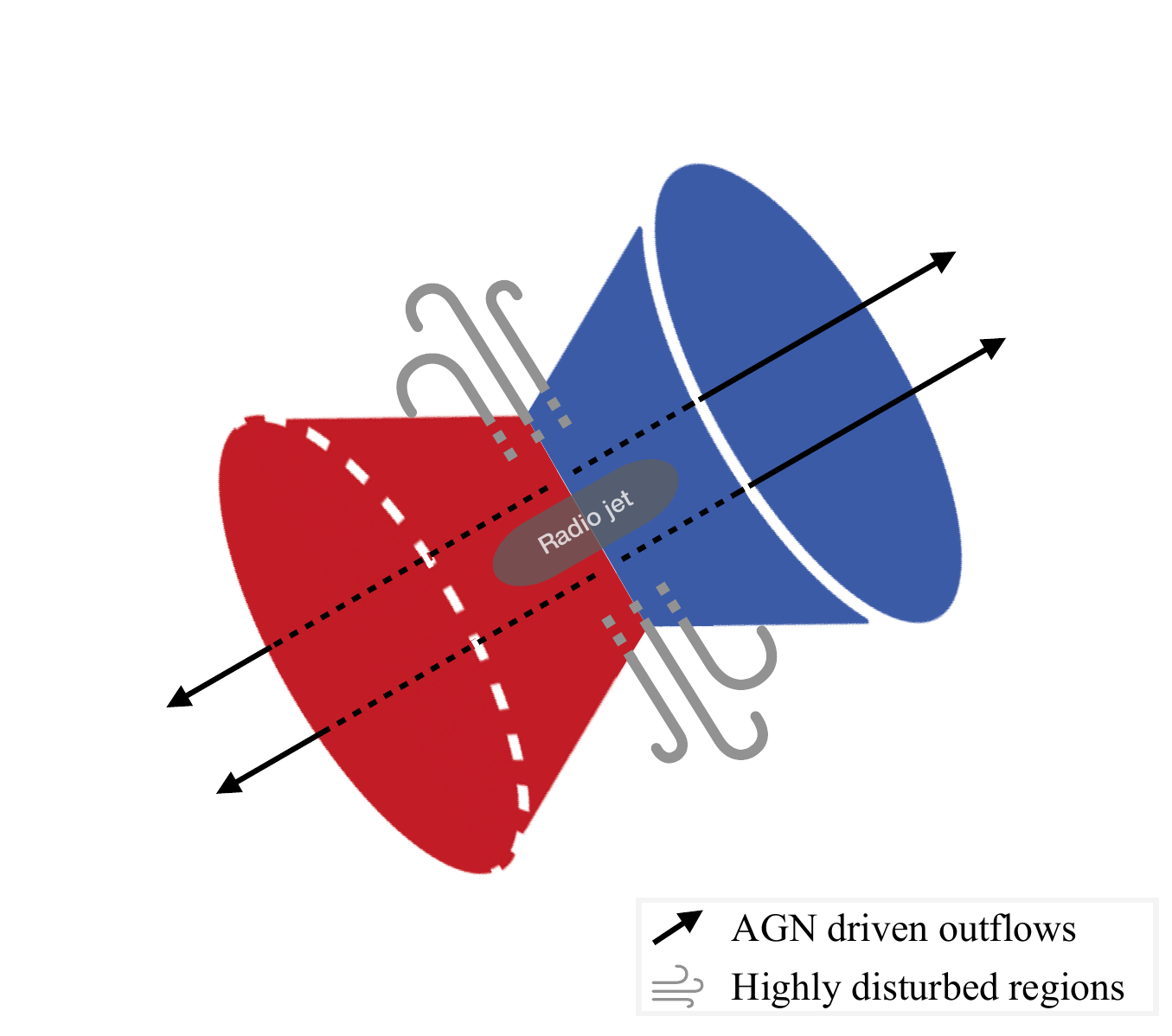}}
\caption{Schematic diagram for the phenomenological model to illustrate AGN driven outflows indicated by black arrows along the AGN ionization cone, and highly disturbed regions indicated by gray curves perpendicular to the AGN ionization cone. The approaching and receding sides of the AGN ionization cone are sketched in blue and red, respectively.}\label{Schematic}
\end{figure}

Besides the large values, the $W_{80}$ distributions of NGC\,5728, NGC\,5506, and ESO137-G034 also exhibit some intriguing features (see also \citealt{Davies.etal.2024, Garcia-Bernete.etal.2024c}). Specifically, their $W_{80}$ distributions of the [Ne~{\footnotesize V}] emission line show that the highly disturbed regions in NGC\,5728 and NGC\,5506 are not aligned with their AGN ionization axes, and hence not aligned with the ionized gas outflows discussed in Section~\ref{sec3.1.2}, but approximately perpendicular to them. For ESO137-G034, the $W_{80}$ distribution of the [Ne~{\footnotesize V}] emission line is even more intriguing. The highly disturbed regions in ESO137-G034 are primarily aligned with the AGN ionization axis, while they also exhibit some minor components roughly perpendicular to the AGN ionization axis. Furthermore, while the highly disturbed regions in NGC\,5728 and NGC\,5506 essentially stretch across their AGN, those of ESO137-G034 are located away from the AGN.

\section{Highly Disturbed Regions and The Triggering Mechanism}\label{sec5}

Section~\ref{sec3.1.3} has shown that NGC\,5728, NGC\,5506, and ESO137-G034 have intriguing highly disturbed regions. Here we further discuss the nature of these disturbed regions. Figure~\ref{NeRatioII} presents AGN excitation strength distributions using the [Ne~{\footnotesize V}]/[Ne~{\footnotesize II}] line ratio as the indicator (e.g., \citealt{Genzel.etal.1998, Dale.etal.2006, Armus.etal.2007}), to disentangle the spatial correlation between these regions and AGN activities. For NGC\,5728 and NGC\,5506, the most highly disturbed regions, i.e., the gray-shaded areas in Figure~\ref{NeRatioII}, are approximately perpendicular to their AGN ionization cones (i.e., reddish areas), while that is not the case for ESO137-G034.% {\bf Note that the highly disturbed regions discussed here are defined based on the $W_{80}$ measured for the [Ne~{\footnotesize V}] emission line, and high $W_{80}$ values are generally associated with but do not ensure the existence of gas outflows.}% In addition, all the three targets show radio emission at different extensions.%NGC\,5728 contains a nuclear radio jet, as depicted by the black-dashed line in Figure~\ref{NeRatioII}(a), propagating along the AGN ionization cone (\citealt{Durre&Mould2018}). NGC\,5506 contains a unresolved radio core, as marked by the black star in Figure~\ref{NeRatioII}(b), as well as diffuse wing-like radio emission extending mainly to the northwest and east of the nucleus as depicted by the black-dashed lines (\citealt{Orienti&Prieto2010}). For ESO137-G034 in Figure~\ref{NeRatioII}(c), two off-nuclear radio blobs are observed, and their edges are depicted by the black-dashed lines (\citealt{Morganti.etal.1999}). In addition, as marked by the black plus signs, the two KDRs in ESO137-G034 are respectively located at the front of the two off-nuclear radio blobs.

\subsection{Highly Disturbed Regions in NGC\,5728}\label{sec5.1}

The highly disturbed regions in NGC\,5728 could be attributed to outflows launched with an angle to the galaxy disk (\citealt{Ruschel-Dutra.etal.2021}). However, in Section~\ref{sec3.1.2} we reveal that the ionized gas outflow around the AGN in NGC\,5728 is mainly aligned with the AGN ionization cone, and for the case of NGC\,5728 is almost within the galaxy disk (\citealt{Shimizu.etal.2019, Davies.etal.2024}). The perturbations induced by central accreting flows might also contribute to these highly disturbed regions since these regions are aligned with the accretion disk plane. Specific kinematic modeling is required to ascertain whether this is the case. 

Very recent studies have reported AGN with low-power ($\lesssim 10^{44}\ \rm erg\,s^{-1}$) radio jets, as is the case here, generally result in intense and extended velocity dispersions perpendicular to the radio jets and hence the ionization cones (e.g., \citealt{Venturi.etal.2021,  Audibert.etal.2023, PeraltadeArriba.etal.2023, Speranza.etal.2024, Venturi.etal.2023, Hermosa-Munoz.etal.2024b, Ulivi.etal.2024}). According to simulations, this kind of perpendicular disturbed regions is due to the more dramatic jet-ISM interaction occurring in AGN with the low-power jets (\citealt{Mukherjee.etal.2018a, Mukherjee.etal.2018b, Meenakshi.etal.2022}). 

Specifically, the strongly interacting propagation of the low-power radio jet through the gas disk will result in significant shocks and dispersions, as well as outflows and the trigger of star formation (\citealt{Nyland.etal.2018}). These processes will strongly disturb the gas disk, especially in the direction  perpendicular to the jet, where has minor resistance. Consistent with this scenario, Figure~\ref{Schematic} provides a phenomenological model for the potential outflow structure in NGC\,5728. This model contains a nuclear radio jet that indirectly drives outflows along the AGN ionization cone, along with shock-driven highly disturbed regions that are perpendicular to the AGN ionization cone. Note that this model is only for a phenomenological explanation, while a conclusive explanation requires specific modeling and/or simulations beyond the scope of this work.

\subsection{Highly Disturbed Regions in NGC\,5506}\label{sec5.2}

The situation is more complicated for NGC\,5506, which contains an unresolved radio core and diffuse wing-like radio emission bisected by the AGN ionization axis. The asymmetry of the highly disturbed regions in NGC\,5506 is likely due to the edge-on view and the asymmetric radio emission. As shown in Figure~\ref{NGC5506_Plus}(b2) (see also Figure~\ref{figb2}), the blueish northeast side, which is more obvious in the [Ne~{\footnotesize II}] $v_{\rm comp. 2}$ field (see Figure~\ref{NGC5506_Plus_c2}a2), represents the approaching side of the AGN ionization cone in NGC\,5506. This geometry explains why the diffuse wing-like radio emission in NGC\,5506 was observed extending mainly to the northwest and east of the nucleus, as these directions are intrinsically toward us.

We can then still use the model in Figure~\ref{Schematic} to illustrate the potential outflow structure in NGC\,5506, but viewing the phenomenological model from the right to the left, and imagining there is a horizontal gas disk with the near side slightly tilted towards the south (\citealt{Fischer.etal.2013}). As a consequence, we can see asymmetrical highly disturbed regions toward the east and the north that are above the tilted gas disk of NGC\,5506 (see Figure~\ref{NGC5506_Plus}b3), while disturbed regions toward the west and the south are blocked by the gas disk. When scrutinizing Figure~\ref{NGC5506_Plus}(b3), we find a few spaxels in the east exhibit lager $W_{80}$ values as well, which are plausibly associated with the blocked disturbed regions.

%In addition, this scenario also explains the redshifted velocity blob $\sim 600\,\rm pc$ towards the north in Figure~\ref{NGC5506_Plus}(b2), {\bf which can be attributed to redshifted materials in the highly disturbed regions above the gas disk (see gray curves in Figure~\ref{Schematic}). Note that the redshifted material in the highly disturbed region with a rather random orientation is different from those in the receding side of the outflow towards the south along the AGN ionization cone.} 

\subsection{Highly Disturbed Regions in ESO137-G034}\label{sec5.3}

Since the highly disturbed regions in ESO137-G034 are located away from the AGN, their triggering mechanism could be associated with some ``delayed'' feedback effects rather than a recent episode of AGN activity. Specifically, we hypothesize the AGN activity in ESO137-G034  triggered the off-nuclear radio blobs first, which then interacted with the ISM gas and increased the gas dispersion as they propagated. Moreover, the propagation of those radio blobs is expected to result in these KDRs on the radio front via enhanced perturbations in the KDR vicinity. In this situation, the phenomenological model in Figure~\ref{Schematic} is still applicable to ESO137-G034, where the minor components of highly disturbed regions perpendicular to the AGN ionization are similar to the highly disturbed regions in NGC\,5728 and NGC\,5506, but already in the fading phase. Again, the analysis here is only for a phenomenological explanation, while more conclusive explanation require specific modelings and/or simulations.

\subsection{Discussion}\label{sec5.4}

Within these highly disturbed regions, some spaxels exhibit the [Ne~{\footnotesize V}] emission line of double-peaked line profiles having comparable blue- and red-shifted components. This is also why we adopt the non-parametric methodology for the analysis. In principle, a mixed motion of the outflowing component and the rotating disk could result in such double-peaked profiles as well. However, this scenario should not be the case here as detailed below.

For NGC\,5728, NGC\,5506, ESO137-G034 and NGC\,3081, the [Ne~{\footnotesize V}] kinematics tend to be dominated by AGN-driven outflowing motions (Section~\ref{sec3.1.2}). As for the [Ne~{\footnotesize V}] broad component, we also checked the velocity fields of the [Ne~{\scriptsize V}] narrow component from the double-Gaussian fitting results for these targets. We find for these targets that the velocity fields of the [Ne~{\scriptsize V}] narrow component exhibit basically the same orientations as the corresponding $v_{m}$ fields of [Ne~{\footnotesize V}] emission line as well, and are different from $v_{m}$ fields of the [Ne~{\footnotesize II}] emission line. This result supports that the [Ne~{\scriptsize V}] kinematics in these targets are dominated by outflows, and the double-peaked profile should be due to mixed outflows in these highly disturbed regions along and against our light of sight (e.g., \citealt{Fischer.etal.2011, Bae&Woo2016}). The case of ESO137-G034 is more complicated given the existence of the KDRs, and will be further discussed by Haidar et al. (in preparation).
%as indicated by the significantly different $v_{m}$ fields of the [Ne~{\footnotesize V}] emission line (Figure~\ref{NGC5728_Plus}b2~\&~\ref{NGC5506_Plus}b2), compared to the $v_{m}$ fields of low ionization lines that are dominated by the disk rotation.

Furthermore, these highly disturbed regions in NGC\,5728 and NGC\,5506 are aligned with the zero-velocity demarcations, i.e., the kinematic minor axis, of their [Ne~{\footnotesize V}] $v_{m}$ fields as well. This invokes another concern that for galaxies with a large central velocity gradient, when averaging out in one resolution element two opposite streams in velocity will also result in an apparent large velocity dispersion. Nevertheless, given the above discussion, such large central velocity gradient, if exists, can only come from outflow motions. Namely, no matter whether they are disturbed or not, these widely distributed regions that are perpendicular to the AGN axis should be associated with outflow relevant processes, albeit they do not mean the outflow itself. % Moreover, such effect can partially contribute to the line broadening but is not able to fully explain these widely distributed regions of significantly broadened emission lines. Not only because these widely distributed regions are far beyond the the kinematic minor axis, but also as most $W_{80}$ values of the broadened emission lines are much larger than the velocity range of corresponding [Ne~{\footnotesize V}] $v_{m}$ filed, i.e., the largest apparent velocity gradient.

\section{Quantification of The Outflow Strength}\label{sec4}

%{\bf upper limits, flatten toward the left?}

%\subsection{The Aperture Integrated Outflow Rate}\label{sec4.3}

In Section~\ref{sec3.1.2}, we showcase the evidence of ionized gas outflows in the six targets. In this section, we provide a rather quantitative comparison of the outflow strength among the six targets.

The most widely used outflow rate is the integrated outflow rate measured from an aperture while assuming a specific outflow geometry. We first estimate for each target a rough upper limit of the integrated ionized gas outflow rate from a $r = 0\farcs9$ aperture (the same as in \citealt{Davies.etal.2020}). These upper limits take the ionized gas mass derived from the total [Ne~{\footnotesize V}] 14.322~$\mum$ flux of all spaxels within the circular aperture as the proxy of ionized gas mass in outflows. The [Ne~{\footnotesize V}] emission line is used for the derivation as the kinematics of this high ionization potential line are dominated by outflows in the six targets. See Appendix~\ref{secB} for the derivation of ionized gas mass, and see also \cite{Davies.etal.2020} for the same strategy of calculating ionized gas mass in outflows but based on the [O~{\footnotesize III}] 5007\AA\ emission. Following previous studies (e.g., \citealt{Kakkad.etal.2022, Riffel.etal.2023}), we take $\dot{M}_{\rm ion} = M_{\rm ion} W_{80} / R$, where $M_{\rm ion}$, $W_{80}$, and $R$ are, respectively, the ionized gas mass derived from the total [Ne~{\footnotesize V}] flux, the flux-weighted $W_{80}$ of [Ne~{\footnotesize V}] emission line, and the flux-weighted distance to the AGN of spaxels within the circular aperture.
 
\begin{figure}[!ht]
\center{\includegraphics[width=0.975\linewidth]{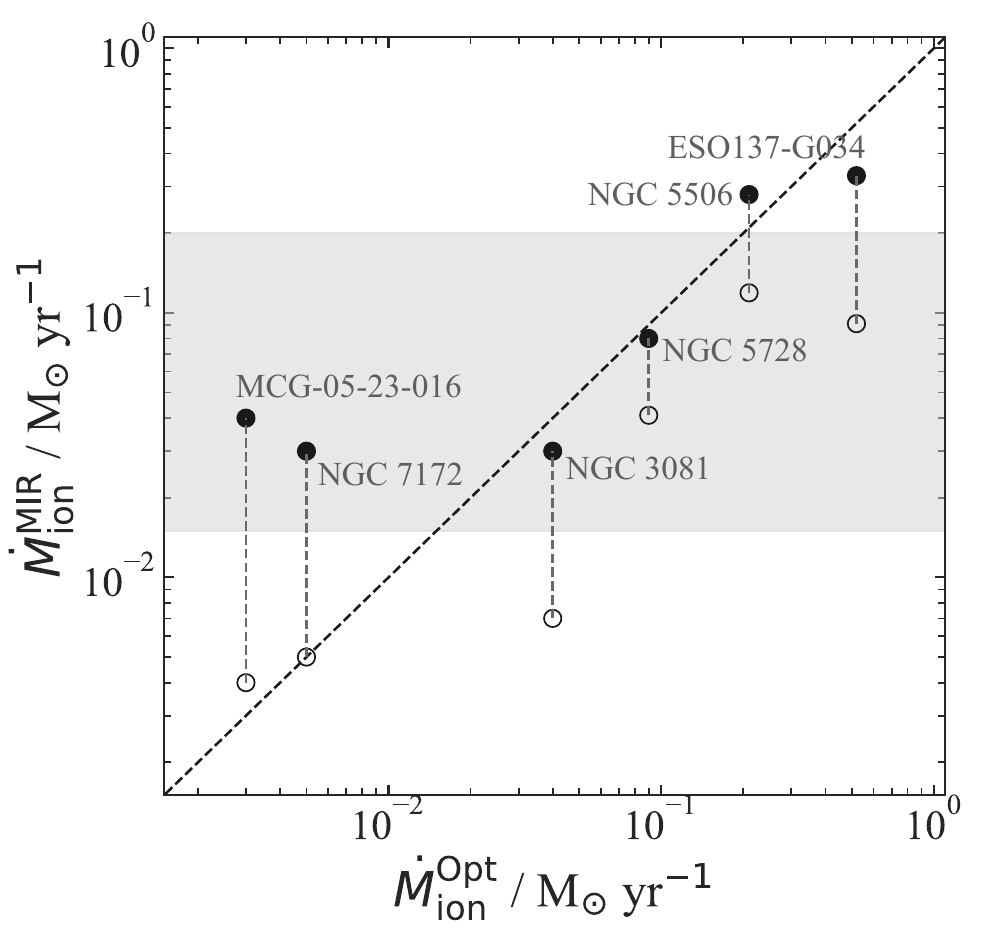}}
\caption{Comparison between ionized gas outflow rates derived from the [Ne~{\footnotesize V}] 14.322~$\mum$ emission line as the y-axis and those from the optical [O~{\footnotesize III}] 5007\AA\ emission line as the x-axis. The filled and blank circles correspond to outflow rates calculated based on the total [Ne~{\footnotesize V}] emission and the broad component in [Ne~{\footnotesize V}] emission, respectively. The gray shadowed region represents the range of ionized gas outflow rates (i.e., $\sim 0.015 - 0.2\,\rm M_{\odot}\,yr^{-1}$) according to the best-fit correlation in Figure~1 of \cite{Fiore.etal.2017}, given the AGN luminosity of the six targets (i.e., $\sim 10^{43.4} - 10^{44.3}\,{\rm erg\,s^{-1}}$).}\label{dotM_aper}
\end{figure}

Meanwhile, we estimate for each target a rough lower limit of the integrated ionized gas outflow rate from the same $r = 0\farcs9$ aperture. These lower limits take the ionized gas mass derived from the flux of the broad component in [Ne~{\footnotesize V}] emission line of all spaxels within the circular aperture as the proxy of ionized gas mass in outflows. In accordance with previous studies (e.g. \citealt{Rupke.etal.2005, Fiore.etal.2017, Fluetsch.etal.2019, Venturi.etal.2023}), we take $\dot{M}_{\rm ion} = M_{\rm ion} v_{\rm out} / R$ with $v_{\rm out} = v_{\rm comp. 2} + 1.18\sigma_{\rm comp. 2}$, where the $\rm comp. 2$ indicates the measurement of the broad component (see Appendix~\ref{secA}). $M_{\rm ion}$, $v_{\rm out}$, and $R$ are, respectively, the ionized gas mass derived from the flux of the broad component in [Ne~{\footnotesize V}] emission line, the flux-weighted $v_{\rm out}$ of the broad component in [Ne~{\footnotesize V}] emission line, and the flux-weighted distance to the AGN of spaxels within the circular aperture. See Table~\ref{taboutrate} for the calculated outflow rates of the six targets.

As shown in Figure~\ref{dotM_aper}, for most of the targets, the outflow rates calculated based on the total [Ne~{\footnotesize V}] flux are consistent with those calculated by \cite{Davies.etal.2020} based on the total [O~{\footnotesize III}] flux, with the discrepancy within their quoted uncertainty (i.e., $0.21\,\rm dex$). However, for NGC\,7172 and MCG-05-23-016, the outflow rates calculated based on the total [Ne~{\footnotesize V}] emission are larger by about an order of magnitude, plausibly due to the factor of dust obscuration (see \citealt{Hermosa-Munoz.etal.2024a} for more detailed discussion). The optical [O~{\footnotesize III}] 5007\AA\  emission line is much more susceptible to dust extinction, whereas the correction of the [Ne~{\footnotesize V}] flux for the six targets is less than $15\%$ with the estimated dust extinction based on the extinction curve measured by \cite{Garcia-Bernete.etal.2024a} combining the measured and theoretical $\rm Pf\alpha$/$\rm Hu\alpha$ ratios.

\afterpage{
\startlongtable
%\begin{longrotatetable}
%\centerwidetable
%\movetableright=-1in
\setlength{\tabcolsep}{8pt}
\begin{deluxetable*}{cccccc}
\tabletypesize{\small}
\tablecolumns{6}
\tablecaption{Integrated Ionized Gas Outflow Rates}
\tablehead{
\colhead{Galaxy} & \colhead{log $\frac{n({\rm Ne^{4+}})}{n ({\rm H})}$} & \colhead{log $M^{a}_{\rm ion}$} & \colhead{$\dot{M}^{a}_{\rm ion}$} & \colhead{log $M^{b}_{\rm ion}$} & \colhead{$\dot{M}^{b}_{\rm ion}$}  \\
\colhead{(-)} & \colhead{[-]} & \colhead{[$\rm M_{\odot}$]} &\colhead{($\rm M_{\odot}\,yr^{-1}$)} &\colhead{[$\rm M_{\odot}$]} & \colhead{($\rm M_{\odot}\,yr^{-1}$)} \\
\colhead{(1)} & \colhead{(2)} & \colhead{(3)} & \colhead{(4)} & \colhead{(5)} & \colhead{(6)} }
\startdata
NGC\,5728 & $-$5.1 & 4.0 & 0.08 & 3.8 & 0.04 \\
NGC\,5506 & $-$5.2 & 4.3 & 0.28 & 4.1 & 0.12 \\
ESO137-G034 & $-$5.3 & 4.7 & 0.33 & 4.4 & 0.09 \\
NGC\,7172 & $-$5.2 & 3.8 & 0.03 & 3.2 & 0.005 \\
MCG-05-23-016 & $-$5.3 & 4.0 & 0.04 & 3.3 & 0.004 \\
NGC\,3081 & $-$4.9 & 4.0 & 0.03 & 3.2 & 0.007 \\
\enddata
\tablecomments{\footnotesize Column (1): Galaxy name. Column (2): $\rm Ne^{4+}$ abundance relative to $\rm H$ ions (see Equation~\ref{equB5}). Column (3)\&(4): Ionized gas mass derived from the total [Ne~{\footnotesize V}] flux and corresponding ionized gas outflow rates from the $r=0\farcs9$ aperture. Column (5)\&(6): Ionized gas mass derived from the flux of the broad component in [Ne~{\footnotesize V}] emission line and corresponding ionized gas outflow rates from the $r=0\farcs9$ aperture.}
\label{taboutrate}
\end{deluxetable*}
}

Figure~\ref{dotM_aper} also shows that the outflow rates reported here are consistent with the best-fit correlation between the ionized gas outflow rate and the AGN luminosity obtained by \cite{Fiore.etal.2017} for AGN with relatively higher luminosity (i.e., $L_{\rm bol} \approx 10^{44.5} - 10^{48.0}\, {\rm erg\,s^{-1}}$). This result verifies the outflow rates reported here are physically reliable. More importantly, averaging the measurements as shown in Figure~\ref{dotM_aper} for reference, the ionized outflow rates of the six targets are converged to a narrower range than previous finding (i.e., \citealt{Davies.etal.2020}). This result indicates the measurement accuracy, especially that of the outflowing ionized gas mass, could be one reason for the observed diversity of ionized outflow rates for AGN with the comparable luminosity. Accordingly, to fully understand the diversity of outflow strength in AGN of the comparable luminosity, more dedicated study of a large sample with accurate outflow rate measurement is indispensable.
% Note that fast radiative shocks occupy the whole central region of NGC\,7172 and MCG-05-23-016, and hence outflow rates of these two targets in Figure~\ref{dotM_aper} are expected to be more close to the upper limits.}

\section{Summary and Conclusions}\label{sec6}

This paper leverages the JWST MIRI/MRS IFU observations to display the diversity of ionized gas distributions and kinematics in central kiloparsec scale regions of six nearby Seyfert galaxies (Section~2). Specifically, we explore the spatially resolved flux distributions and velocity fields of six ionized emission lines covering a large range of ionization potentials (Section~\ref{sec3.1.1}~\&~\ref{sec3.1.2}). 

We find the evidence of ionized gas outflows in the six targets, according to the twisting between velocity fields of six emission lines (for NGC\,5728, NGC\,5506, NGC\,3081, and likely ESO137-G034), and combining the observed ionized line ratios with theoretical calculations (for NGC\,7172, and MCG-05-23-016). Meanwhile, we find NGC\,5728, NGC\,5506, and ESO137-G034 also exhibit some intriguing highly disturbed regions (Section~\ref{sec3.1.3}). For these three targets, their integrated spectra also exhibit broader ionized emission lines than the others (Section~\ref{sec3.0}).

We further discuss case by case the possible triggering mechanisms of such highly disturbed regions in these three targets (Section~\ref{sec5}). We propose that the radio jet associated with AGN activity plausibly plays an important role in triggering such highly disturbed regions. Accordingly, we provide a phenomenological model involving radio jets to illustrate the potential outflow structures of the three targets. To further this work, detailed analysis and modeling of the physical conditions in these regions is required. 

Moreover, we have a rather quantitative comparison of the outflow strength among the six targets (Section~\ref{sec4}). With the outflow rates calculated based on [Ne~{\footnotesize V}] emission, which is relatively immune to dust obscuration, we find the six targets tend to have the ionized outflow rates converged to a narrower range than previous finding. These results have important implication for the diverse outflow properties of AGN with the comparable luminosity, while more convincing conclusions require further dedicated analysis.

%\newpage%\newline

\acknowledgements

{\footnotesize We thank the anonymous referee for detailed comments and suggestions to improve the presentation of our results and corresponding discussions. L.Z., C.P., E.K.S.H, and M.T.L. acknowledge grant support from the Space Telescope Science Institute (ID: JWST-GO-01670.007-A). A.A.H. and L.H.M. acknowledge support from grant PID2021-124665NB-I00 funded by MCIN/AEI/10.13039/501100011033 and by ERDF A way of making Europe. I.G.B. is supported by the Programa Atracci\'on de Talento Investigador ``C\'esar Nombela'' via grant 2023-T1/TEC-29030 funded by the Community of Madrid. I.G.B. and D.R. acknowledge support from STFC through grants ST/S000488/1 and ST/W000903/1. A.J.B acknowledges funding from the ``First Galaxies'' Advanced Grant from the European Research Council (ERC) under the European Union’s Horizon 2020 research and innovation program (Grant agreement No. 789056). O.G.-M. acknowledges support from PAPIIT UNAM IN109123 and to the Ciencia de Frontera project CF-2023-G-100 from CONHACYT. M.P.S. acknowledges support from grant RYC2021-033094-I funded by MICIU/AEI/10.13039/501100011033 and the European Union NextGenerationEU/PRTR. E.L.-R. is supported by the NASA/DLR Stratospheric Observatory for Infrared Astronomy (SOFIA) under the 08\_0012 Program.  SOFIA is jointly operated by the Universities Space Research Association,Inc.(USRA), under NASA contract NNA17BF53C, and the Deutsches SOFIA Institut (DSI) under DLR contract 50OK0901 to the University of Stuttgart. E.L.-R. is also supported by the NASA Astrophysics Decadal Survey Precursor Science (ADSPS) Program (NNH22ZDA001N-ADSPS) with ID 22-ADSPS22-0009 and agreement number 80NSSC23K1585. E.B. acknowledges the Mar{\'i}a Zambrano program of the Spanish Ministerio de Universidades funded by the Next Generation European Union and is also partly supported by grant RTI2018-096188-B-I00 funded by the Spanish Ministry of Science and Innovation/State Agency of Research MCIN/AEI/10.13039/501100011033. C.R.A., A.A. and D.-E.A. acknowledge support by the EU H2020-MSCA-ITN-2019 Project 860744 ``BiD4BESt: Big Data applications for black hole Evolution STudies'' and by project PID2022-141105NB-I00 ``Tracking active galactic nuclei feedback from parsec to kilo-parsec scales'', funded by MICINN-AEI/10.13039/501100011033.}

{\footnotesize C.R. acknowledges support from Fondecyt Regular grant 1230345 and ANID BASAL project FB210003. S.G.B acknowledges support from the Spanish grant PID2022-138560NB-I00, funded by MCIN/AEI/10.13039/501100011033/FEDER, EU. B.G.-L. acknowledges support from the Spanish State Research Agency (AEI-MCINN/10.13039/501100011033) through grants PID2019-107010GB-100 and PID2022-140483NB-C21 PID2022-138560NB-I00, and the Severo Ochoa Program 2020-2023 (CEX2019-000920-S). M.S. acknowledges support by the Ministry of Science, Technological Development and Innovation of the Republic of Serbia (MSTDIRS) through contract no. 451-03-66/2024-03/200002 with the Astronomical Observatory (Belgrade). D.J.R. is supported by the STFC grant ST/X001105/1. This work is based on observations made with the NASA/ESA/CSA James Webb Space Telescope. The data were obtained from the Mikulski Archive for Space Telescopes at the Space Telescope Science Institute, which is operated by the Association of Universities for Research in Astronomy, Inc., under NASA contract NAS 5-03127 for JWST. The specific observations analyzed can be accessed via \dataset[doi: 10.17909/vre3-m991]{https://doi.org/10.17909/vre3-m991}. This paper makes use of the following ALMA data: ADS/JAO.ALMA \#2015.1.00086.S, \#2015.1.00116.S, \#2017.1.00236.S, and \#2019.1.01742.S. ALMA is a partnership of ESO (representing its member states), NSF (USA) and NINS (Japan), together with NRC (Canada), MOST and ASIAA (Taiwan), and KASI (Republic of Korea), in cooperation with the Republic of Chile. The Joint ALMA Observatory is operated by ESO, AUI/NRAO and NAOJ. This paper also makes use of the following VLA images: NGC5728-4.89I1.57-AP0065-1984JAN14-1-27.8U2.72M.fits, NGC5506-4.89I0.50-AW0159-1986MAY26-1-55.7U38.2S.fits, NGC7172-4.89I0.74-AW0093-1983AUG26-1-78.2U36.4S.fits, MCG-05-23-016-4.91I0.63-WILS-1982MAR14-1-78.3U39.3S.fits, NGC3081-4.89I0.50-AW0126-1985FEB05-1-54.5U40.3S. Image credit: \dataset[NRAO/VLA Archive Survey]{http://www.vla.nrao.edu/astro/nvas/}, (c) 2005-2009 AUI/NRAO.}

%\newpage%\newline

\appendix

\section{Characteristics of Outflows with Parametric Measurements}\label{secA}

The parametric method is widely used to study AGN-driven outflows, although caveats are required when used in this work (Section~\ref{sec2.4}). As a supplementary analysis, here we briefly discuss the flux distribution ($f_{\rm comp. 2}$), the velocity field ($v_{\rm comp. 2}$), and the velocity dispersion field ($\sigma_{\rm comp. 2}$), of the relatively broader component, if included (see Section~\ref{sec2.3}), in the best-fit [Ne~{\footnotesize II}] and [Ne~{\footnotesize V}] profiles. For NGC 5728, NGC 5506, ESO137-G034, and NGC\,3081, their $f_{\rm comp. 2}$ distributions of [Ne~{\footnotesize II}] and [Ne~{\footnotesize V}] emission lines are similar to each other, and same for the $v_{\rm comp. 2}$ fields. Moreover, for both  [Ne~{\footnotesize II}] and [Ne~{\footnotesize V}] emission lines, their $f_{\rm comp. 2}$ and $v_{\rm comp. 2}$ distributions are respectively similar to $f_{\rm total}$ and $v_{m}$ distributions of the [Ne~{\footnotesize V}] emission line. This result supports that $f_{\rm total}$ and $v_{m}$ distributions of the [Ne~{\footnotesize V}] emission line mainly trace ionized gas outflow features. The above result is not that evident for the other two targets. We also note that $v_{\rm comp. 2}$ distributions of ESO137-G034 are more disturbed around the two KDRs. This result caution against the use of parametric methodology for systems with complex outflow features. Additionally, for NGC 5728, NGC 5506, and ESO137-G034, their $\sigma_{\rm comp. 2}$ distributions of the [Ne~{\footnotesize V}] emission line also exhibit some highly disturbed regions (i.e., regions with very large $\sigma_{\rm comp. 2}$) that are perpendicular to their AGN ionization cones. This result confirms similar findings as revealed by their $W_{80}$ distributions in Section~\ref{sec3.1.3}.

%\newpage
\section{Derivation of Ionized Gas Mass}\label{secB}

To calculate the ionized gas outflow rate, we need to firstly derive the ionized gas mass from the observed strength of ionized gas emission. This section uses the [Ne~{\footnotesize V}] 14.322~$\mum$ emission line as an example for the derivation, as it is most sensitive to the AGN ionization among the six lines and most likely traces ionized gas outflows.

In theory, the [Ne~{\footnotesize V}] luminosity can be derived by
\begin{align}\label{equB1}
L_{\rm [Ne\,V]} = \int_{V} f n_{e} n({\rm Ne^{4+}}) j_{\rm [Ne\,V]}(n_{e}, T_{e}) dV,
\end{align}
\noindent
where $f$, $n_{e}$, $n({\rm Ne^{4+}})$, and $j_{\rm [Ne\,V]}(n_{e}, T_{e})$ are the filling factor, the electron density, the $\rm Ne^{4+}$ density, and the $\rm [Ne\,V]$ emissivity at given electron density and electron temperature, respectively (\citealt{Draine2011}). Therein, $n({\rm Ne^{4+}})$ can be obtained as $n({\rm Ne^{4+}}) = \frac{n({\rm Ne^{4+}})}{n({\rm H})} \frac{n({\rm H})}{n_{e}} n_{e}$, with $\frac{n({\rm H})}{n_{e}} \approx \frac{n({\rm H})}{n({\rm H}) + 2\times n({\rm He})} = (1.2)^{-1}$ assuming $\frac{n({\rm He})}{n({\rm H})} = 0.1$. Accordingly, the [Ne~{\footnotesize V}] luminosity is derived as
\begin{align}\label{equB2}
L_{\rm [Ne\,V]} = (1.2)^{-1} f \frac{n({\rm Ne^{4+}})}{n({\rm H})} j_{\rm [Ne\,V]}(n_{e}, T_{e}) \langle n_{e}^{2}\rangle V,
\end{align}
\noindent
where $\langle n_{e}^{2}\rangle$ is the volume-averaged squared electron density.

Meanwhile, the ionized gas mass can be derived by
\begin{align}\label{equB3}
M_{\rm ion}  \simeq \int_{V} f n({\rm H}) \overline{m} dV,
\end{align}
\noindent
where $\overline{m}$ is the average molecular-weighted mass. Specifically, we adopt $n({\rm H}) \overline{m} \approx n_{e} \frac{n({\rm H})}{n_{e}} \frac{n({\rm H}) m_{\rm p} + 4\times n({\rm He}) m_{\rm p}}{n({\rm H}) + n({\rm He})} \approx n_{e} (1.2)^{-1} 1.2 m_{\rm p} = n_{e} m_{\rm p}$ (\citealt{Carniani.etal.2015}), where $m_{\rm p}$ is the proton mass, again assuming $\frac{n({\rm He})}{n({\rm H})} = 0.1$. Accordingly, the ionized gas mass is derived as
\begin{align}\label{equB4}
M_{\rm ion}  \simeq f \langle n_{e}\rangle m_{\rm p} V,
\end{align}
\noindent
where $\langle n_{e}\rangle$ is the volume-averaged electron density.

Combing Equation~\ref{equB2} and Equation~\ref{equB4}, the ionized gas mass is finally derived as
\begin{align}\label{equB5}
M_{\rm ion} = \frac{1.2 \frac{\langle n_{e}\rangle^{2}}{\langle n_{e}^{2}\rangle} L_{\rm [Ne\,V]} m_{\rm p}}{\frac{n({\rm Ne^{4+}})}{n({\rm H})} \langle n_{e} \rangle j_{\rm [Ne\,V]}(n_{e}, T_{e})},
\end{align}
\noindent
where we adopt $\frac{\langle n_{e}\rangle^{2}}{\langle n_{e}^{2}\rangle} = 1$ and $\frac{n({\rm Ne^{4+}})}{n({\rm H})} = \frac{I_{\rm [Ne\,V]}}{j_{\rm [Ne\,V]}} / \frac{I_{\rm Pf\alpha}}{j_{\rm Pf\alpha}}$, with $I_{\rm line}$ and $j_{\rm line}$ denote line intensity and line emissivity, respectively. $\rm Pf\alpha$ emission line is used here as it is the strongest hydrogen recombination line covered by our MRS spectral observations and can be replaced by other hydrogen recombination lines if available.

Since the relatively weak $\rm Pf\alpha$ emission is not available for all spatially resolved spaxels, an unresolved $\frac{n({\rm Ne^{4+}})}{n({\rm H})}$ value (see Table~\ref{taboutrate}) is calculated for each target based on the aperture measurements as listed in Table~\ref{tabnucs}. The line emissivity $j_{\rm line}(n_{e}, T_{e})$ is obtained by {\tt PyNeb}, a modern {\tt Python} tool to compute emission line emissivities (\citealt{Luridiana.etal.2015}). Emissivities of these emission lines have very weak dependence on $T_{e}$ for high temperature environments, and hence we assume a standard value of $T_{e} = 10^{4}\,\rm K$ (\citealt{Fernandez-Ontiveros.etal.2021, Perez-Diaz.etal.2022}). $\langle n_{e}\rangle$ is obtained as $n_{e} = 3.2 (\frac{L_{\rm bol}}{10^{45}\, {\rm erg\,s^{-1}}}) (\frac{r}{1\,{\rm kpc}})^{-2} \frac{1}{U}\,{\rm cm^{-3}}$ according to \cite{Baron&Netzer2019}, and see Table~\ref{tabne} for the radial $n_{e}$ profile of each target. Specifically, the ionization parameter $U$ can be derived from measured [S\,{\footnotesize IV}]/[Ne\,{\footnotesize III}] line ratios (i.e., \citealt{Pereira-Santaella.etal.2017}).

According to the emission line ratio diagram (i.e., [Ne~{\footnotesize V}]/[Ne~{\footnotesize II}] versus [Ne~{\footnotesize III}]/[Ne~{\footnotesize II}]) with model results calculated by \cite{Morisset.etal.2015} and \cite{Pereira-Santaella.etal.2024}, the central regions of the six targets are dominated by AGN excitation or fast radiative shocks associated with the AGN. Therefore, we calculate the $U$ based on the correlation between $U$ and  [S\,{\footnotesize IV}]/[Ne\,{\footnotesize III}]  of AGN models in \cite{Pereira-Santaella.etal.2017}, assuming the solar metallicity and $n_{\rm H} = 10^{3}\rm\,cm^{-3}$. The log~$U$ distributions of the six targets is found to have the median values of $\sim -2.9$ to $-2.6$, with the standard deviations of $\sim 0.1$ to $0.3$ dex within their filed of view. These results are consistent with the values derived by \cite{Davies.etal.2020} in an independent way for these targets. This supports the assumptions we adopted here for calculating the ionization parameter. Around the median values of $\sim -2.9$ to $-2.6$, the derived $U$ value varies within $\sim 0.3$ dex for $Z = 0.04 - 2\,Z_{\odot}$ and $n_{\rm H} = 10 - 10^{4}\rm\,cm^{-3}$ (see Figure B2 in \citealt{Pereira-Santaella.etal.2017}). Note that the calculation of $n_{e}$ based on the ionization parameter $U$ is only for the rather quantitative comparison in Section~\ref{sec4}. Specific work in terms of the $n_{e}$ derivation based on emission line pairs (e.g., [Ne~{\footnotesize V}] 14.32 \& 24.32 $\mum$, [Ar~{\footnotesize V}] 7.90 \& 13.10 $\mum$) covered by JWST spectra is in preparation.

\startlongtable
%\begin{longrotatetable}
%\centerwidetable
%\movetableright=-1in
\setlength{\tabcolsep}{3pt}
\begin{deluxetable*}{cccccccccccc}
\tablenum{B}
\tabletypesize{\footnotesize}
\tablecolumns{12}
\tablecaption{The Radial Profile of $n_{e}$ Distribution}
\tablehead{
\colhead{NGC\,5728} & \colhead{} & \colhead{NGC\,5506} & \colhead{} & \colhead{ESO137-G034} & \colhead{} & \colhead{NGC\,7172} & \colhead{} & \colhead{MCG-05-23-016} & \colhead{} & \colhead{NGC\,3081} & \colhead{} \\
\colhead{$r$} & \colhead{log $n_{e}$} & \colhead{$r$} & \colhead{log $n_{e}$} & \colhead{$r$} & \colhead{log $n_{e}$} & \colhead{$r$} & \colhead{log $n_{e}$} & \colhead{$r$} & \colhead{log $n_{e}$} & \colhead{$r$} & \colhead{log $n_{e}$} \\
\colhead{(pc)} & \colhead{[$\rm cm^{-3}$]} & \colhead{(pc)} & \colhead{[$\rm cm^{-3}$]} & \colhead{(pc)} & \colhead{[$\rm cm^{-3}$]} & \colhead{(pc)} & \colhead{[$\rm cm^{-3}$]} & \colhead{(pc)} & \colhead{[$\rm cm^{-3}$]} & \colhead{(pc)} & \colhead{[$\rm cm^{-3}$]} \\
\colhead{(1)} & \colhead{(2)} & \colhead{(3)} & \colhead{(4)} & \colhead{(5)} & \colhead{(6)} & \colhead{(7)} & \colhead{(8)} & \colhead{(9)} & \colhead{(10)} & \colhead{(11)} & \colhead{(12)}}
\startdata
59 & 5.0 & 44 & 5.2 & 57 & 4.2 & 58 & 5.0 & 57 & 5.3 & 56 & 4.8 \\
156 & 4.0 & 105 & 4.4 & 136 & 3.3 & 145 & 4.2 & 148 & 4.4 & 135 & 4.0 \\
280 & 3.5 & 200 & 3.9 & 255 & 2.9 & 269 & 3.6 & 258 & 3.9 & 246 & 3.5 \\
410 & 3.2 & 288 & 3.6 & 370 & 2.7 & 394 & 3.3 & 371 & 3.6 & 358 & 3.2 \\
539 & 3.0 & 375 & 3.3 & 486 & 2.4 & 514 & 3.0 & 484 & 3.3 & 468 & 3.0 \\
667 & 2.9 & 463 & 3.1 & 597 & 2.2 & 632 & 2.8 & 599 & 3.0 & 581 & 2.7 \\
\enddata
\tablecomments{ \footnotesize Column (1): $r$, for the first row is the median radius of the innermost 5 spaxels and for other rows is the median radius for spaxels in a $0\farcs7$ width annulus used to calculate the median electron density. Column (2): log~$n_{e}$, for the first row is the median electron density of the innermost 5 spaxels and for other rows is the median electron density for spaxels in a $0\farcs7$ width annulus.}
\label{tabne}
\end{deluxetable*}
%\end{longrotatetable}

\begin{figure*}[!ht]
\figurenum{A1}
\center{\includegraphics[width=0.8\linewidth]{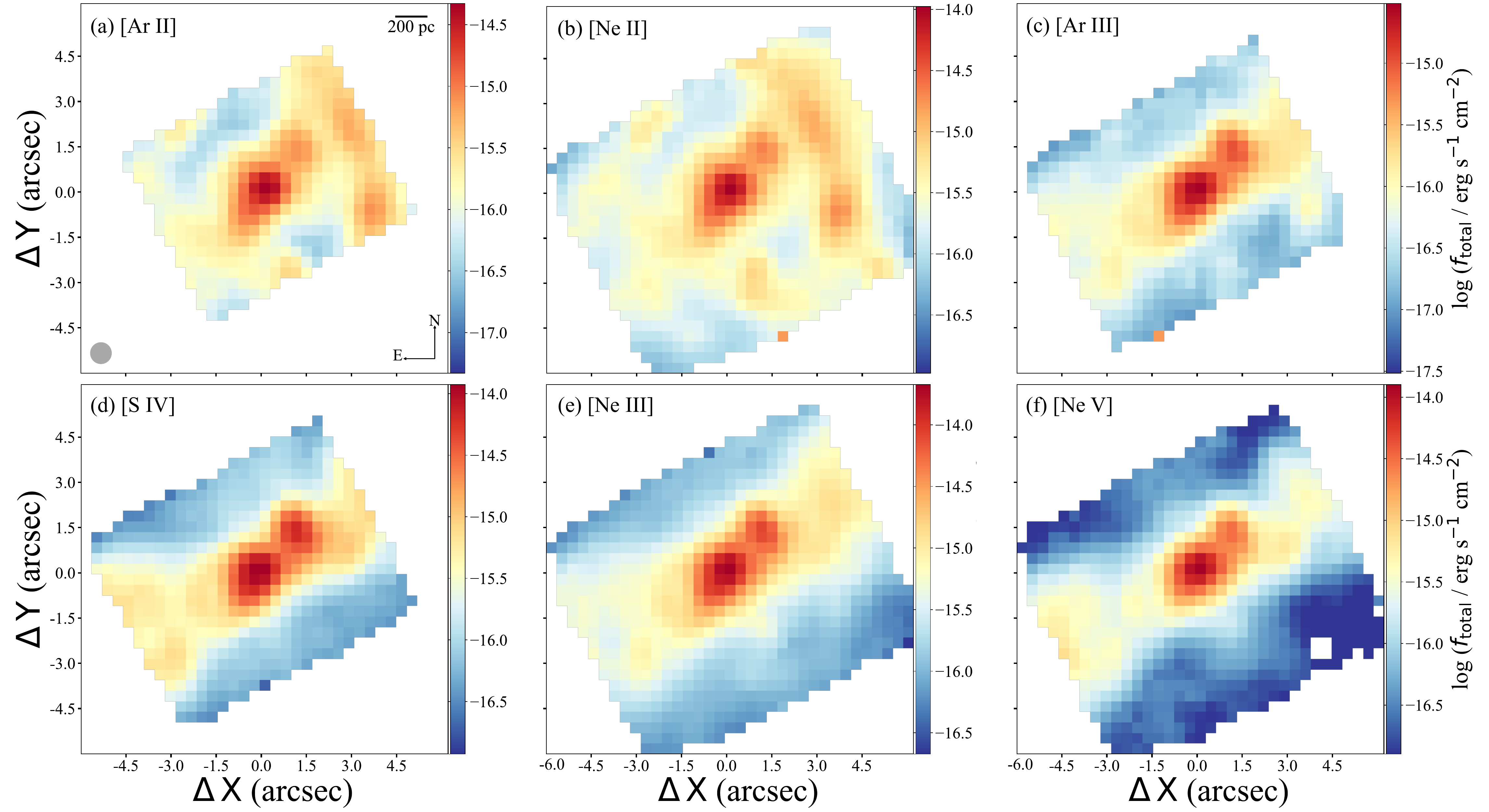}}
\caption{The $f_{\rm total}$ distributions in NGC\,5728 central region for emission lines of (a) [Ar~{\footnotesize II}] 6.985~$\mum$ with the IP of 15.8 eV, (b) [Ne~{\footnotesize II}] 12.814~$\mum$ with the IP of 21.6 eV, (c) [Ar~{\footnotesize III}] 8.991~$\mum$ with the IP of 27.6 eV, (d) [S~{\footnotesize IV}] 10.511~$\mum$ with the IP of 34.8 eV, (e) [Ne~{\footnotesize III}] 15.555~$\mum$ with the IP of 41.0 eV, and (f) [Ne~{\footnotesize V}] 14.322~$\mum$ with the IP of 97.1 eV. Panel (a) features a filled gray circle to indicate the angular resolution of these maps (i.e., 0\farcs7$\times0\farcs7$), a compass with north is up and east is to the left, and a scale bar of 200 pc; they are the same for all panels here, and for all subsequent maps belonging to NGC\,5728.}\label{figa1}
\end{figure*}

\begin{figure*}[!ht]
\figurenum{A2}
\center{\includegraphics[width=0.8\linewidth]{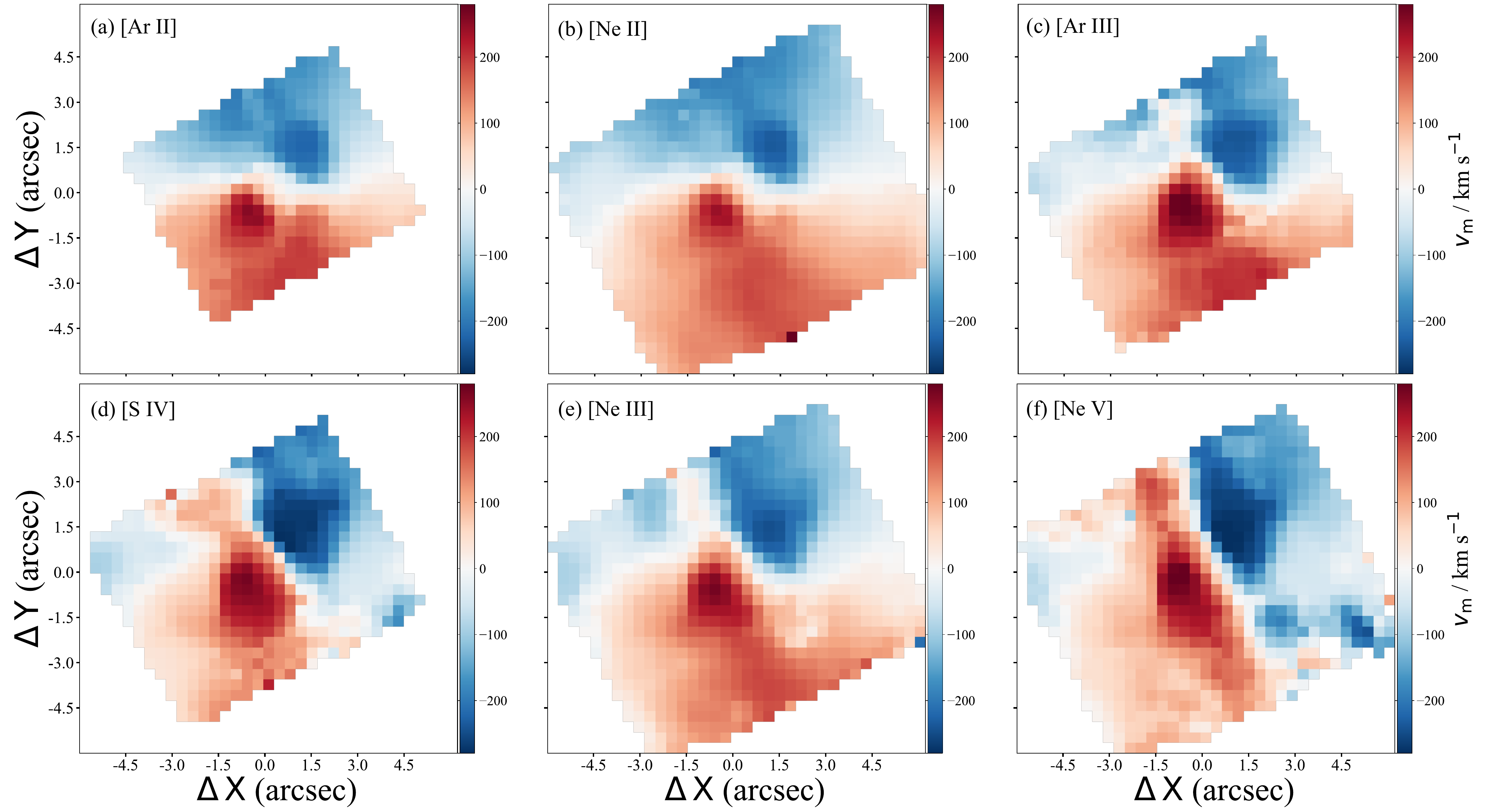}}
\caption{The $v_{\rm m}$ distributions in NGC\,5728 central region for emission lines of (a) [Ar~{\footnotesize II}] 6.985~$\mum$ with the IP of 15.8 eV, (b) [Ne~{\footnotesize II}] 12.814~$\mum$ with the IP of 21.6 eV, (c) [Ar~{\footnotesize III}] 8.991~$\mum$ with the IP of 27.6 eV, (d) [S~{\footnotesize IV}] 10.511~$\mum$ with the IP of 34.8 eV, (e) [Ne~{\footnotesize III}] 15.555~$\mum$ with the IP of 41.0 eV, and (f) [Ne~{\footnotesize V}] 14.322~$\mum$ with the IP of 97.1 eV.}\label{figa2}
\end{figure*}

\begin{figure*}[!ht]
\figurenum{A3}
\center{\includegraphics[width=0.8\linewidth]{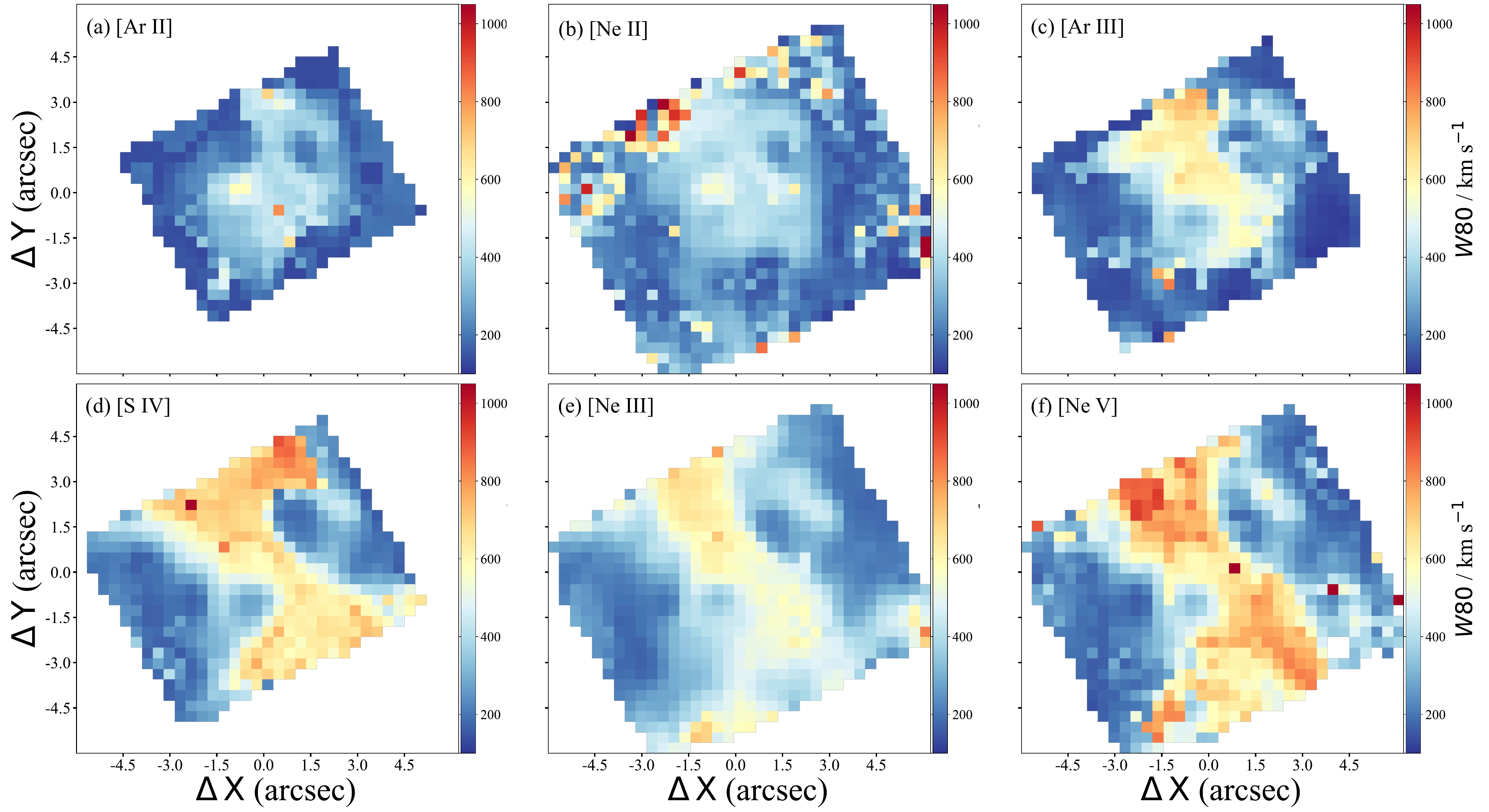}}
\caption{The $W_{80}$ distributions in NGC\,5728 central region for emission lines of (a) [Ar~{\footnotesize II}] 6.985~$\mum$ with the IP of 15.8 eV, (b) [Ne~{\footnotesize II}] 12.814~$\mum$ with the IP of 21.6 eV, (c) [Ar~{\footnotesize III}] 8.991~$\mum$ with the IP of 27.6 eV, (d) [S~{\footnotesize IV}] 10.511~$\mum$ with the IP of 34.8 eV, (e) [Ne~{\footnotesize III}] 15.555~$\mum$ with the IP of 41.0 eV, and (f) [Ne~{\footnotesize V}] 14.322~$\mum$ with the IP of 97.1 eV.}\label{figa3}
\end{figure*}

%\begin{figure*}[!ht]
%\figurenum{A4}
%\center{\includegraphics[width=0.8\linewidth]{NGC5728_Ionmass.pdf}}
%\caption{The $M_{\rm ion}$ distributions in NGC\,5728 central region calculated by Equation~\ref{equ1} based on luminosities and abundances of (a) [Ar~{\footnotesize II}] 6.985~$\mum$ with the IP of 15.8 eV, (b) [Ne~{\footnotesize II}] 12.814~$\mum$ with the IP of 21.6 eV, (c) [Ar~{\footnotesize III}] 8.991~$\mum$ with the IP of 27.6 eV, (d) [S~{\footnotesize IV}] 10.511~$\mum$ with the IP of 34.8 eV, (e) [Ne~{\footnotesize III}] 15.555~$\mum$ with the IP of 41.0 eV, and (f) [Ne~{\footnotesize V}] 14.322~$\mum$ with the IP of 97.1 eV. Note that the ionized gas mass distributions traced by different emission lines are not the same as these lines trace different ionization conditions.}\label{figa4}
%\end{figure*}

\newpage

\begin{figure*}[!ht]
\figurenum{B1}
\center{\includegraphics[width=0.8\linewidth]{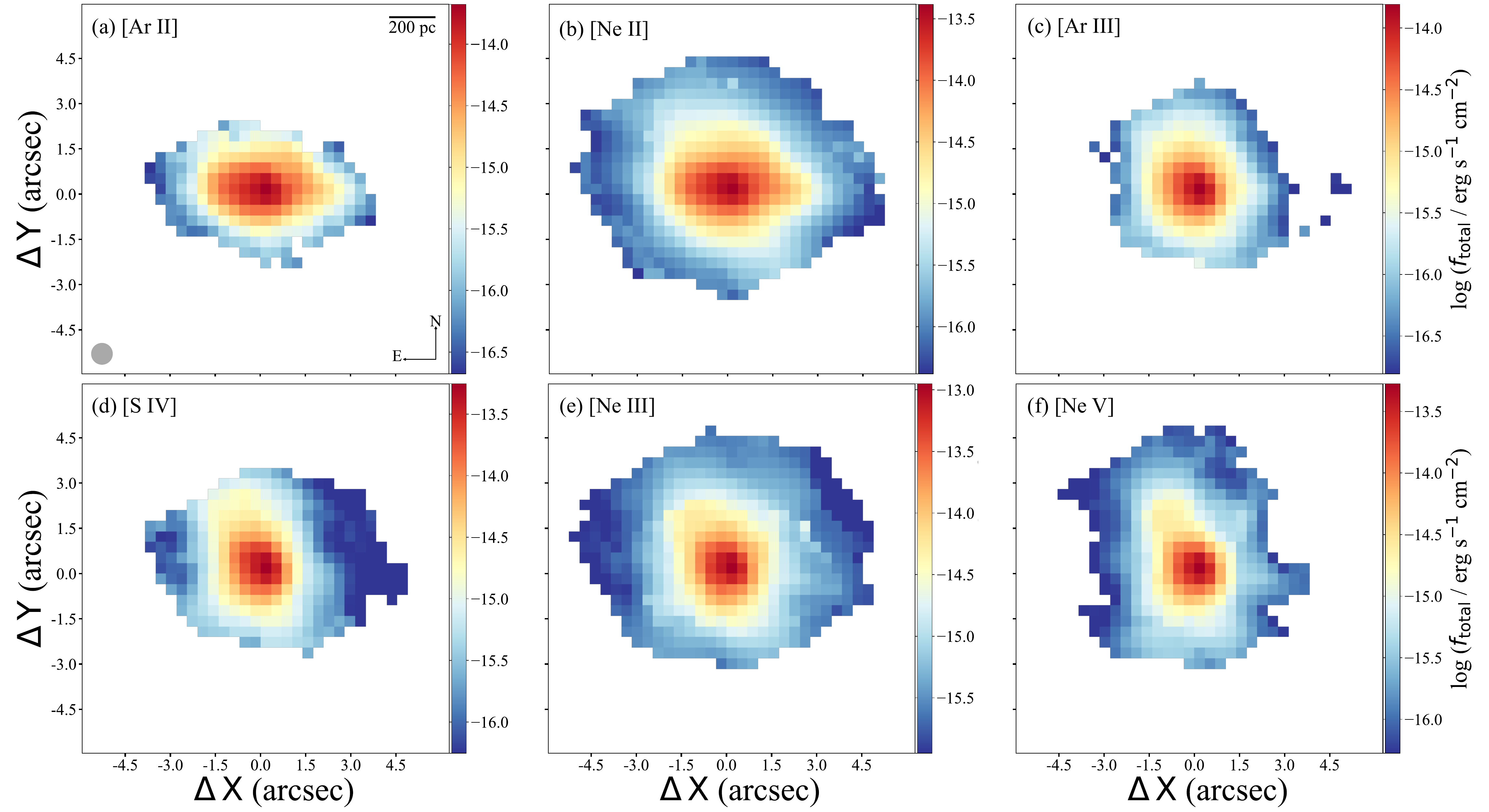}}
\caption{The same as Figure~\ref{figa1} but for NGC\,5506.}\label{figb1}
\end{figure*}

\begin{figure*}[!ht]
\figurenum{B2}
\center{\includegraphics[width=0.8\linewidth]{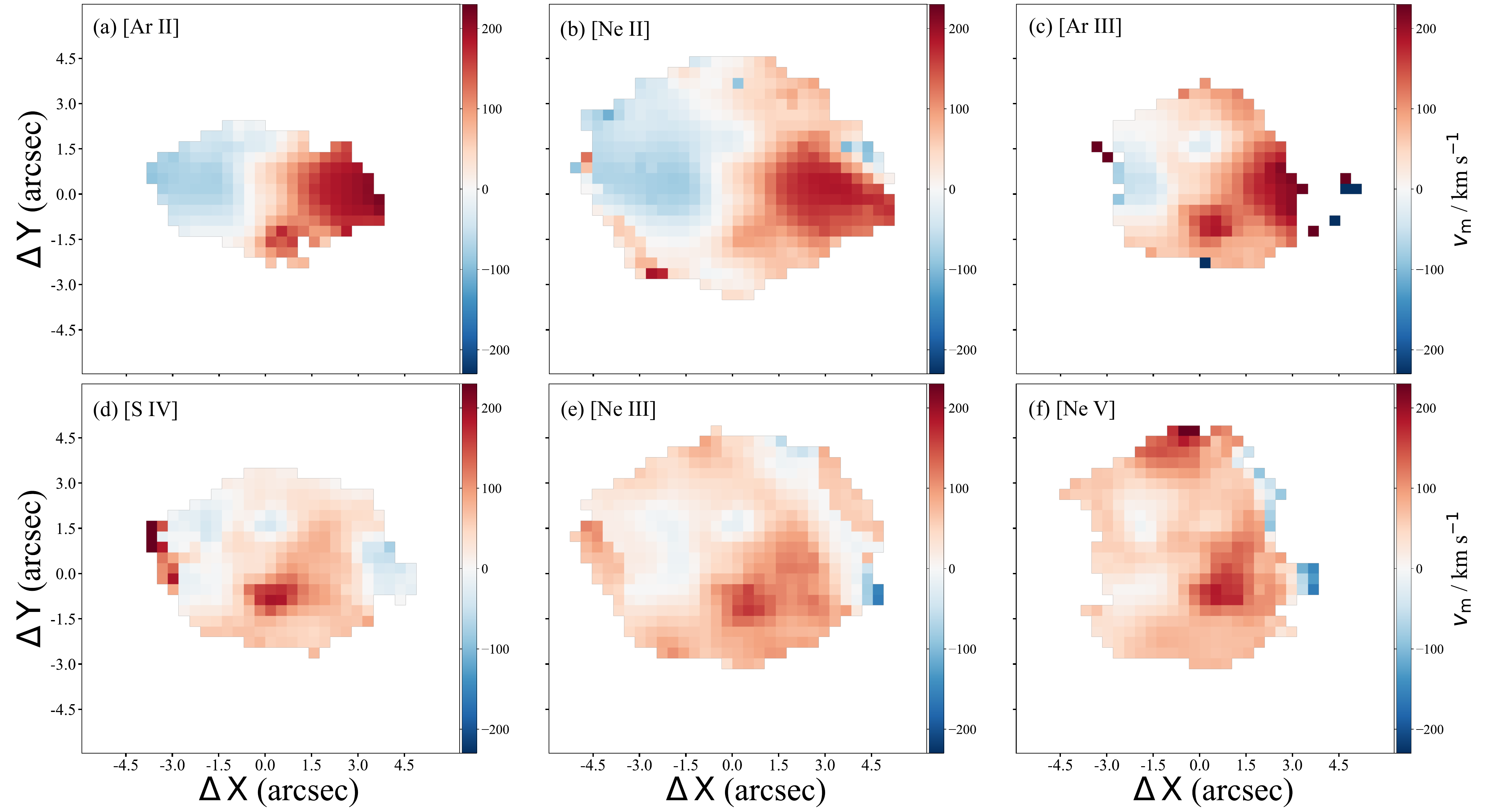}}
\caption{The same as Figure~\ref{figa2} but for NGC\,5506.}\label{figb2}
\end{figure*}

\begin{figure*}[!ht]
\figurenum{B3}
\center{\includegraphics[width=0.8\linewidth]{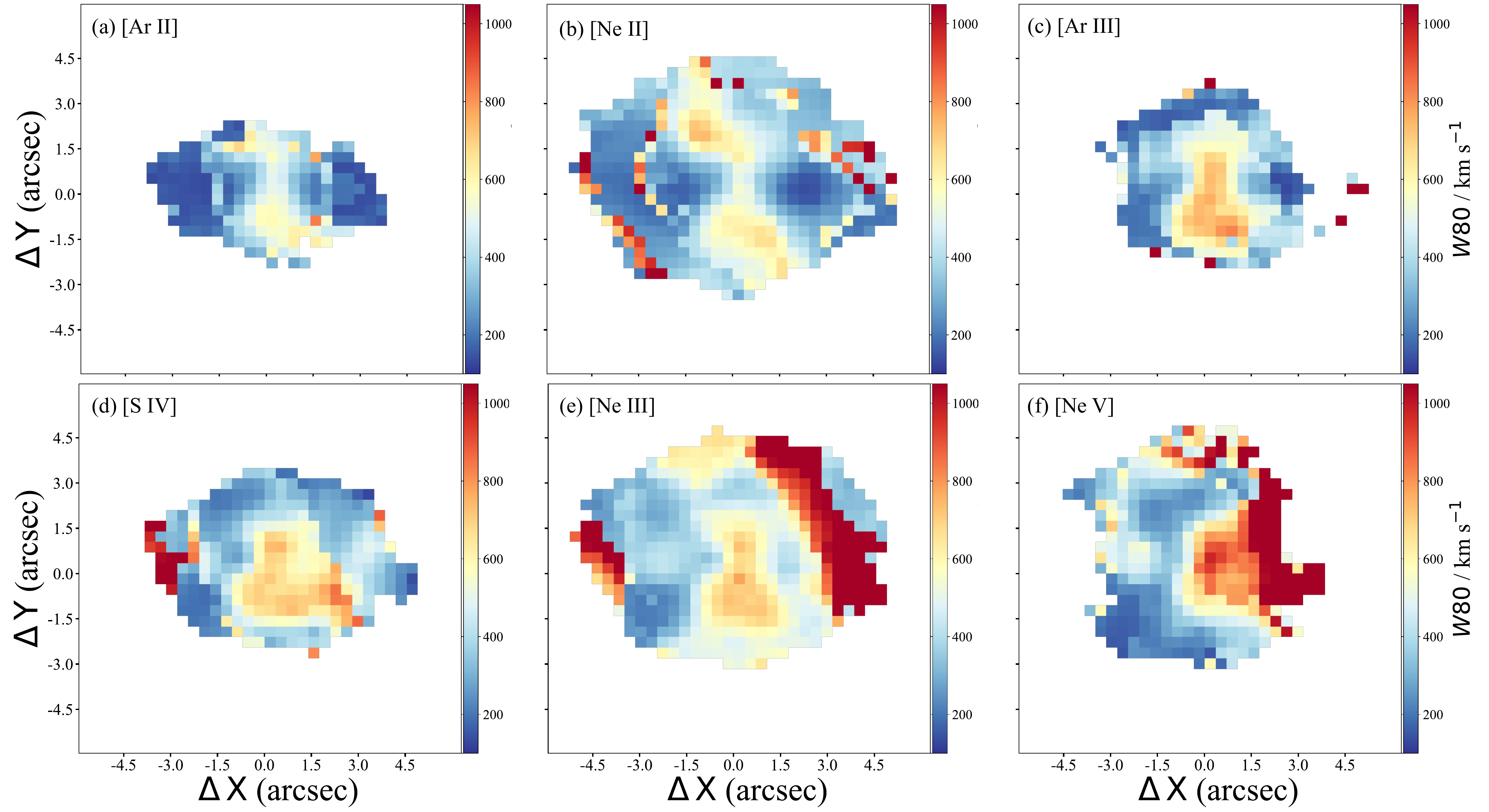}}
\caption{The same as Figure~\ref{figa3} but for NGC\,5506.}\label{figb3}
\end{figure*}

%\begin{figure*}[!ht]
%\figurenum{B4}
%\center{\includegraphics[width=0.8\linewidth]{NGC5506_Ionmass.pdf}}
%\caption{The same as Figure~\ref{figa4} but for NGC\,5506.}\label{figb4}
%\end{figure*}

\begin{figure*}[!ht]
\figurenum{C1}
\center{\includegraphics[width=0.8\linewidth]{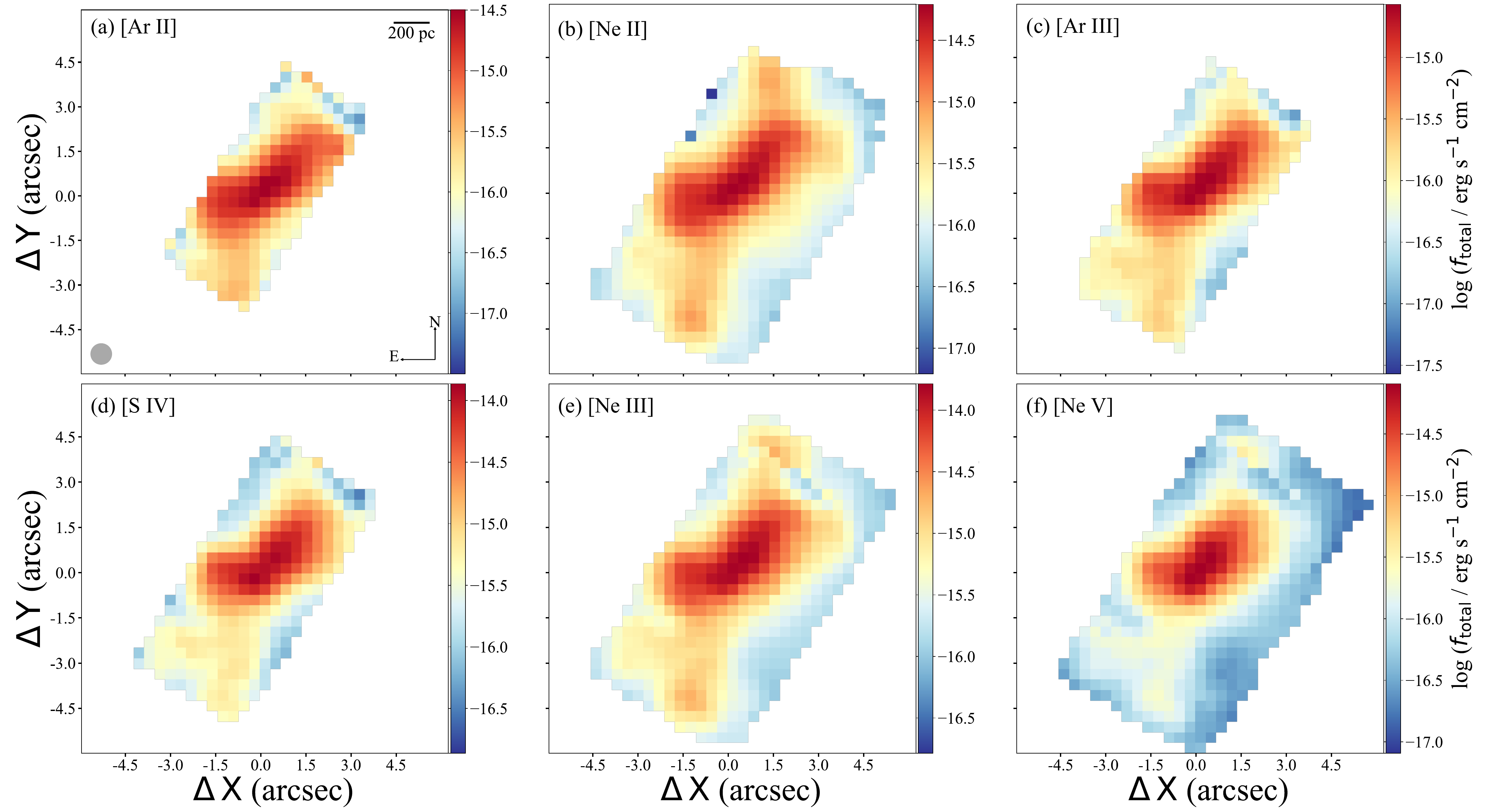}}
\caption{The same as Figure~\ref{figa1} but for ESO137-G034.}\label{figc1}
\end{figure*}

\begin{figure*}[!ht]
\figurenum{C2}
\center{\includegraphics[width=0.8\linewidth]{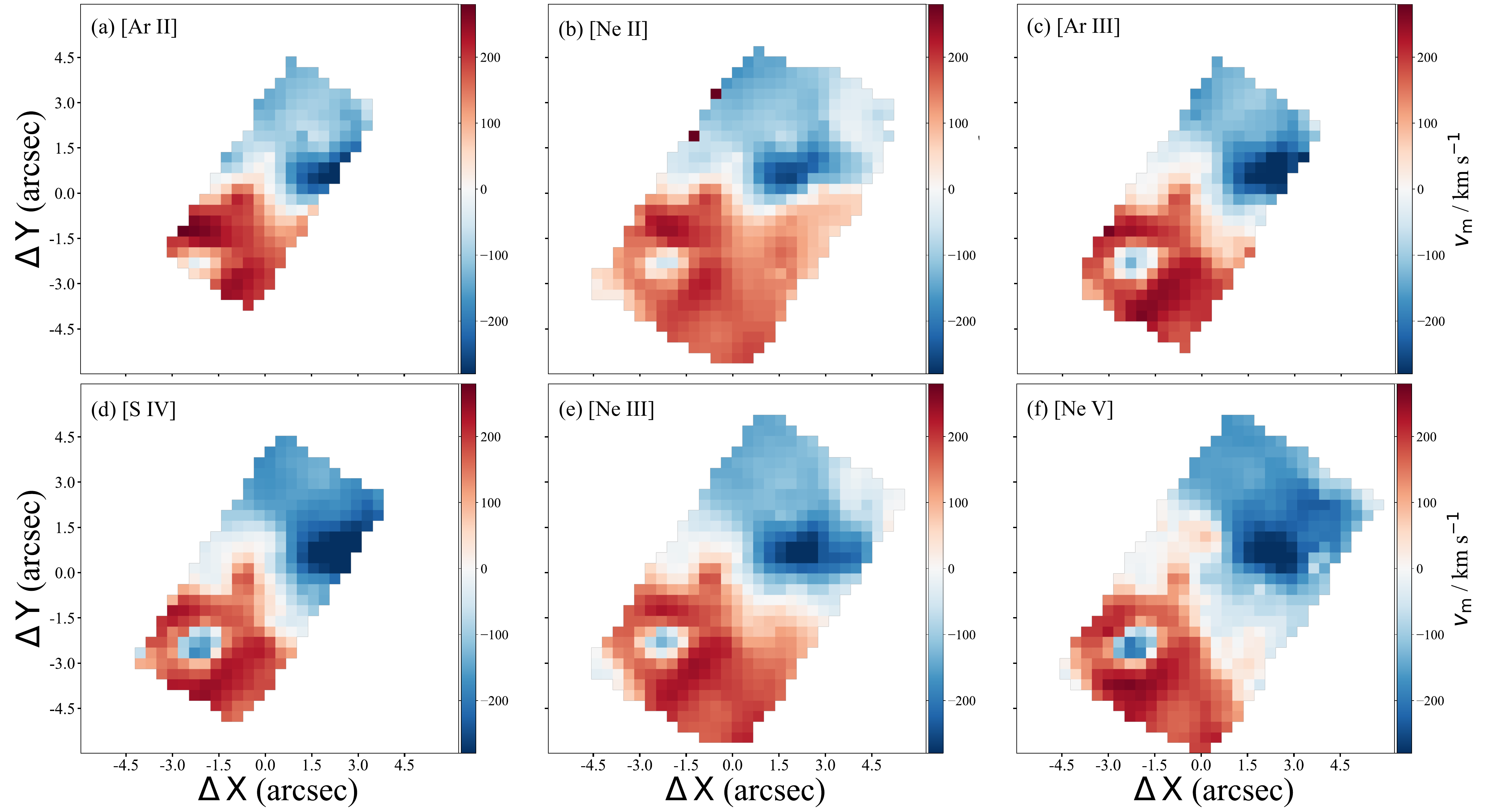}}
\caption{The same as Figure~\ref{figa2} but for ESO137-G034.}\label{figc2}
\end{figure*}

\begin{figure*}[!ht]
\figurenum{C3}
\center{\includegraphics[width=0.8\linewidth]{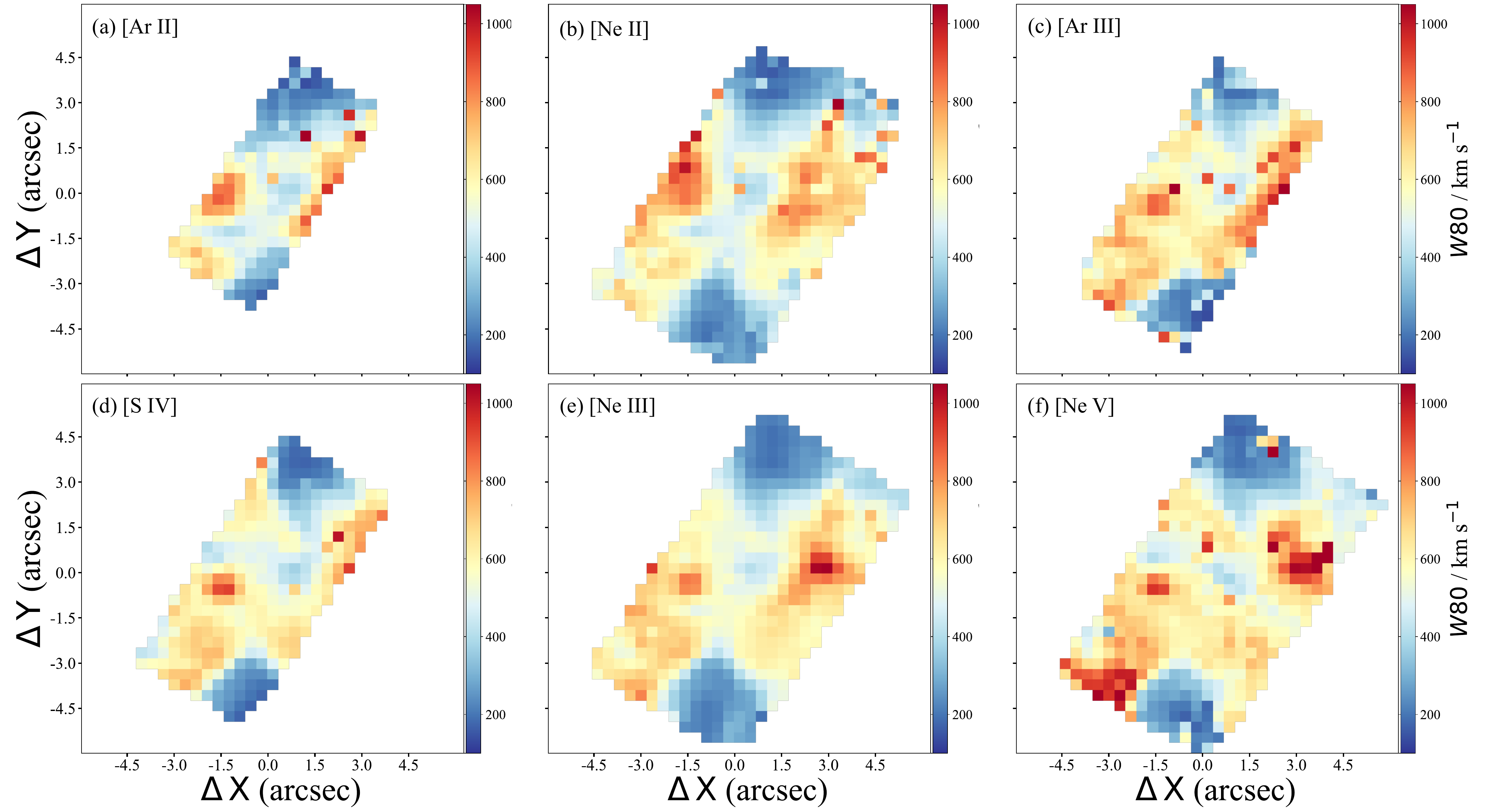}}
\caption{The same as Figure~\ref{figa3} but for ESO137-G034.}\label{figc3}
\end{figure*}

%\begin{figure*}[!ht]
%\figurenum{C4}
%\center{\includegraphics[width=0.8\linewidth]{ESO137-G034_Ionmass.pdf}}
%\caption{The same as Figure~\ref{figa4} but for ESO137-G034.}\label{figc4}
%\end{figure*}

\begin{figure*}[!ht]
\figurenum{D1}
\center{\includegraphics[width=0.8\linewidth]{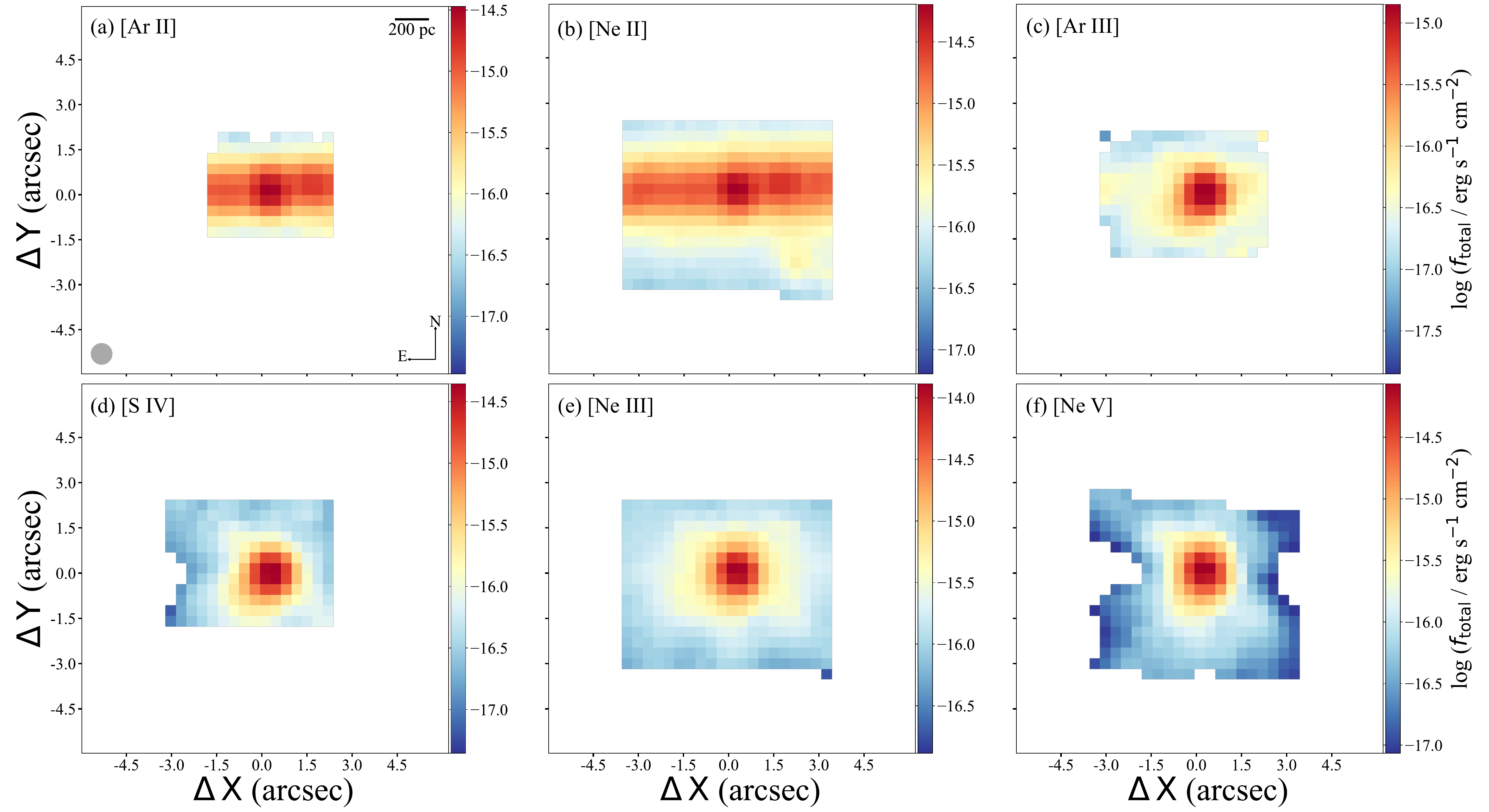}}
\caption{The same as Figure~\ref{figa1} but for NGC\,7172.}\label{figd1}
\end{figure*}

\begin{figure*}[!ht]
\figurenum{D2}
\center{\includegraphics[width=0.8\linewidth]{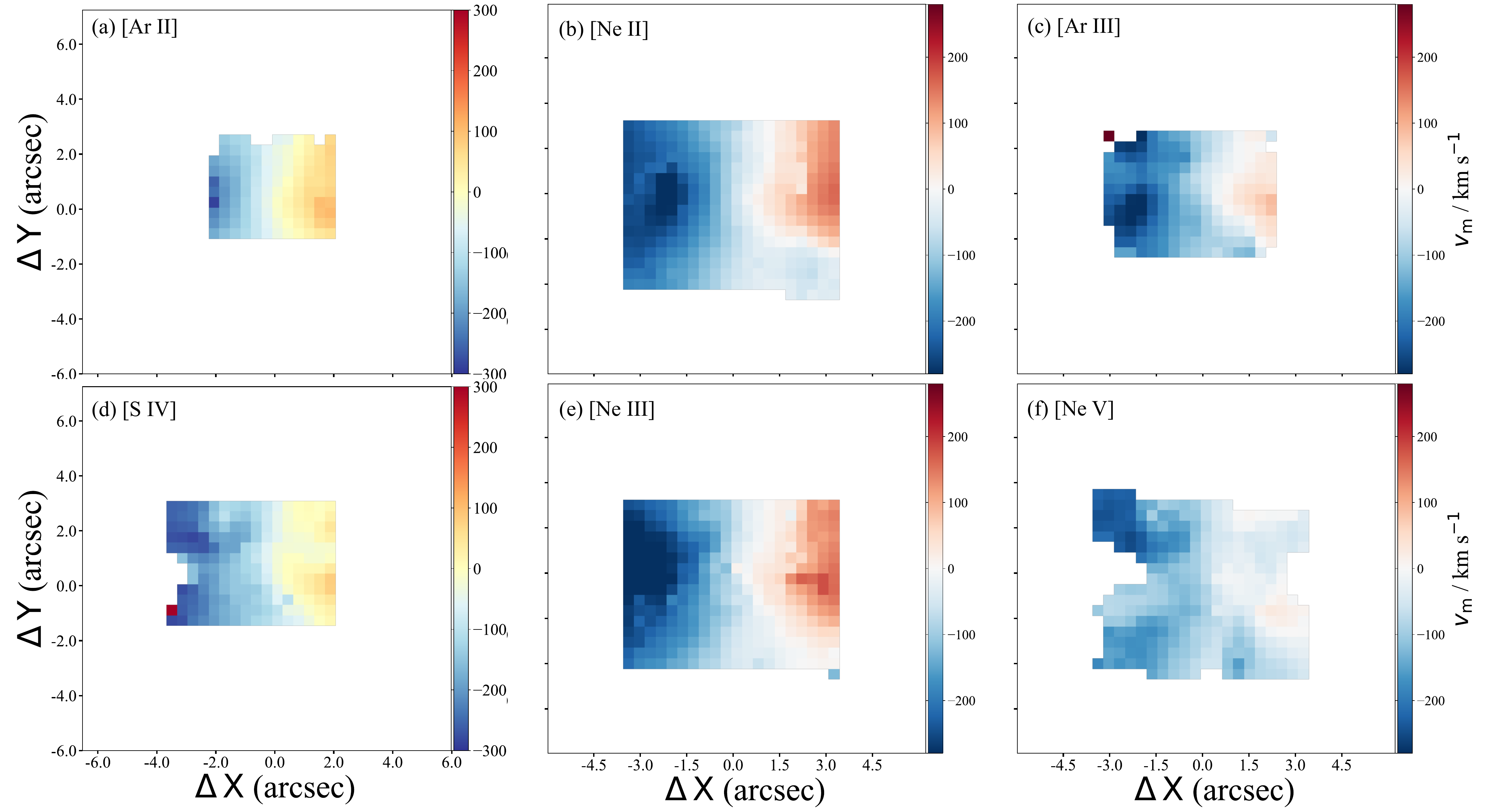}}
\caption{The same as Figure~\ref{figa2} but for NGC\,7172.}\label{figd2}
\end{figure*}

\begin{figure*}[!ht]
\figurenum{D3}
\center{\includegraphics[width=0.8\linewidth]{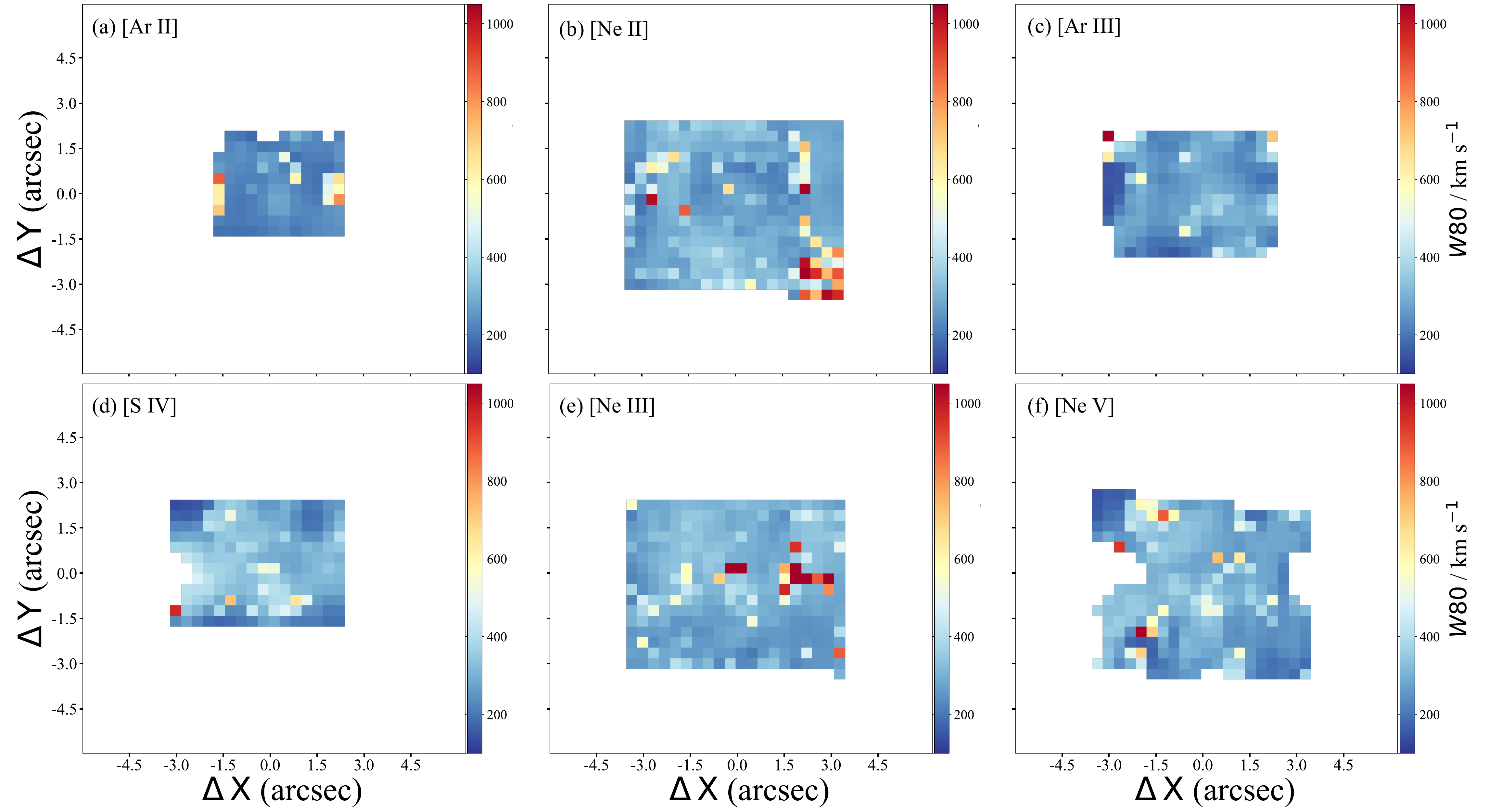}}
\caption{The same as Figure~\ref{figa3} but for NGC\,7172.}\label{figd3}
\end{figure*}

%\begin{figure*}[!ht]
%\figurenum{D4}
%\center{\includegraphics[width=0.8\linewidth]{NGC7172_Ionmass.pdf}}
%\caption{The same as Figure~\ref{figa4} but for NGC\,7172.}\label{figd4}
%\end{figure*}

\begin{figure*}[!ht]
\figurenum{E1}
\center{\includegraphics[width=0.8\linewidth]{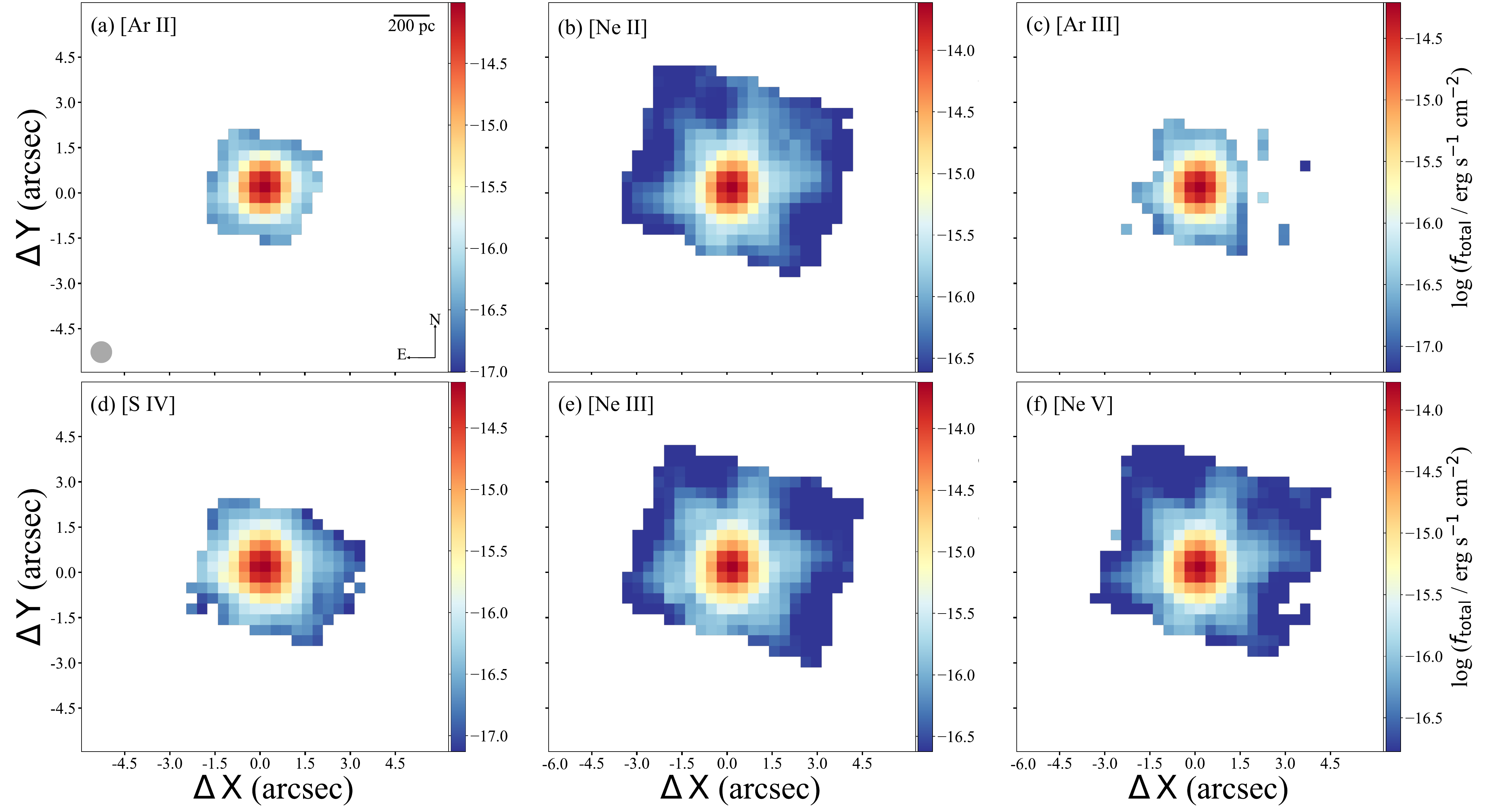}}
\caption{The same as Figure~\ref{figa1} but for MCG-05-23-016.}\label{fige1}
\end{figure*}

\begin{figure*}[!ht]
\figurenum{E2}
\center{\includegraphics[width=0.8\linewidth]{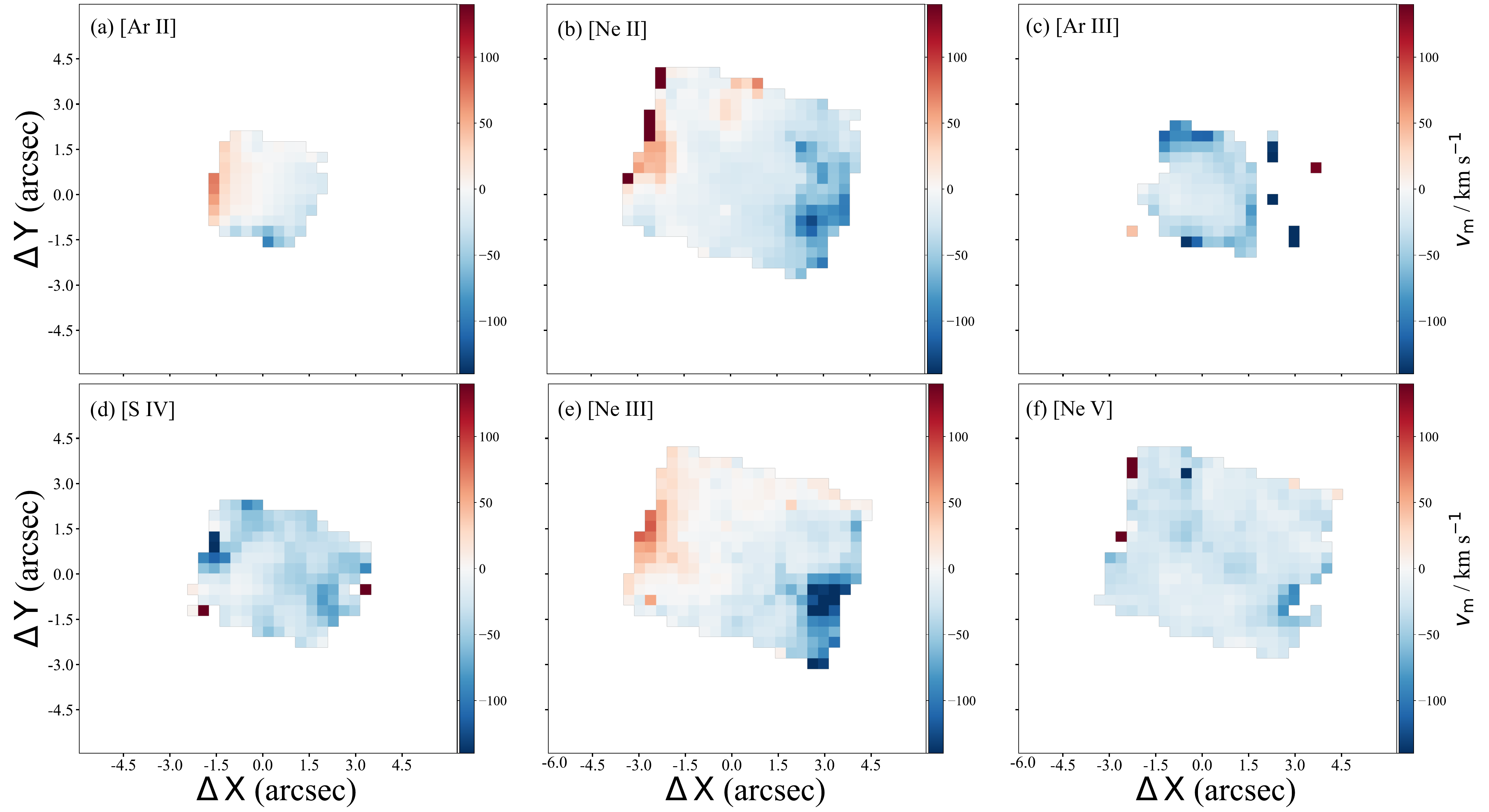}}
\caption{The same as Figure~\ref{figa2} but for MCG-05-23-016.}\label{fige2}
\end{figure*}

\begin{figure*}[!ht]
\figurenum{E3}
\center{\includegraphics[width=0.8\linewidth]{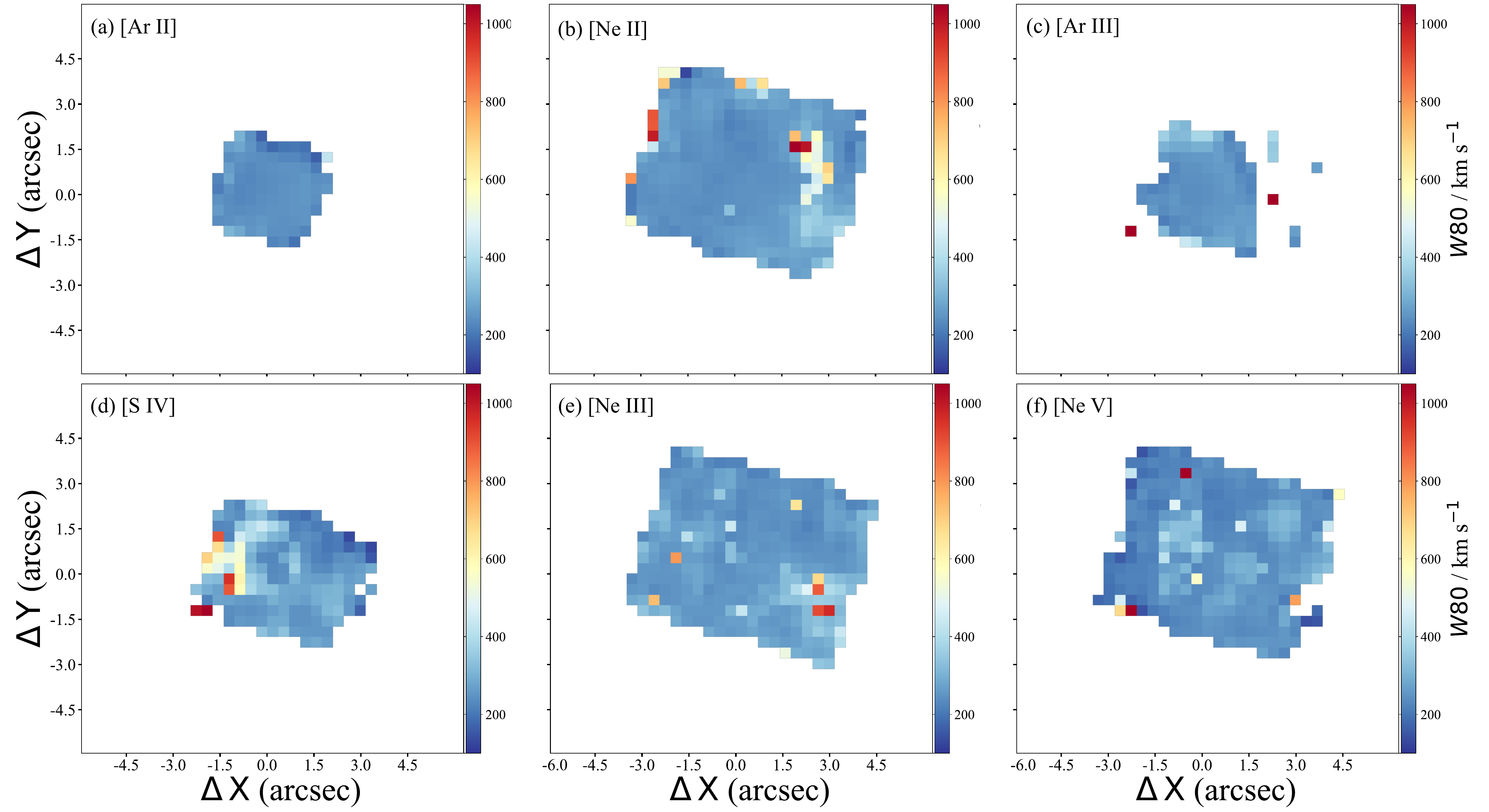}}
\caption{The same as Figure~\ref{figa3} but for MCG-05-23-016.}\label{fige3}
\end{figure*}

%\begin{figure*}[!ht]
%\figurenum{E4}
%\center{\includegraphics[width=0.8\linewidth]{MCG-05-23-016_Ionmass.pdf}}
%\caption{The same as Figure~\ref{figa4} but for MCG-05-23-016.}\label{fige4}
%\end{figure*}

\begin{figure*}[!ht]
\figurenum{F1}
\center{\includegraphics[width=0.8\linewidth]{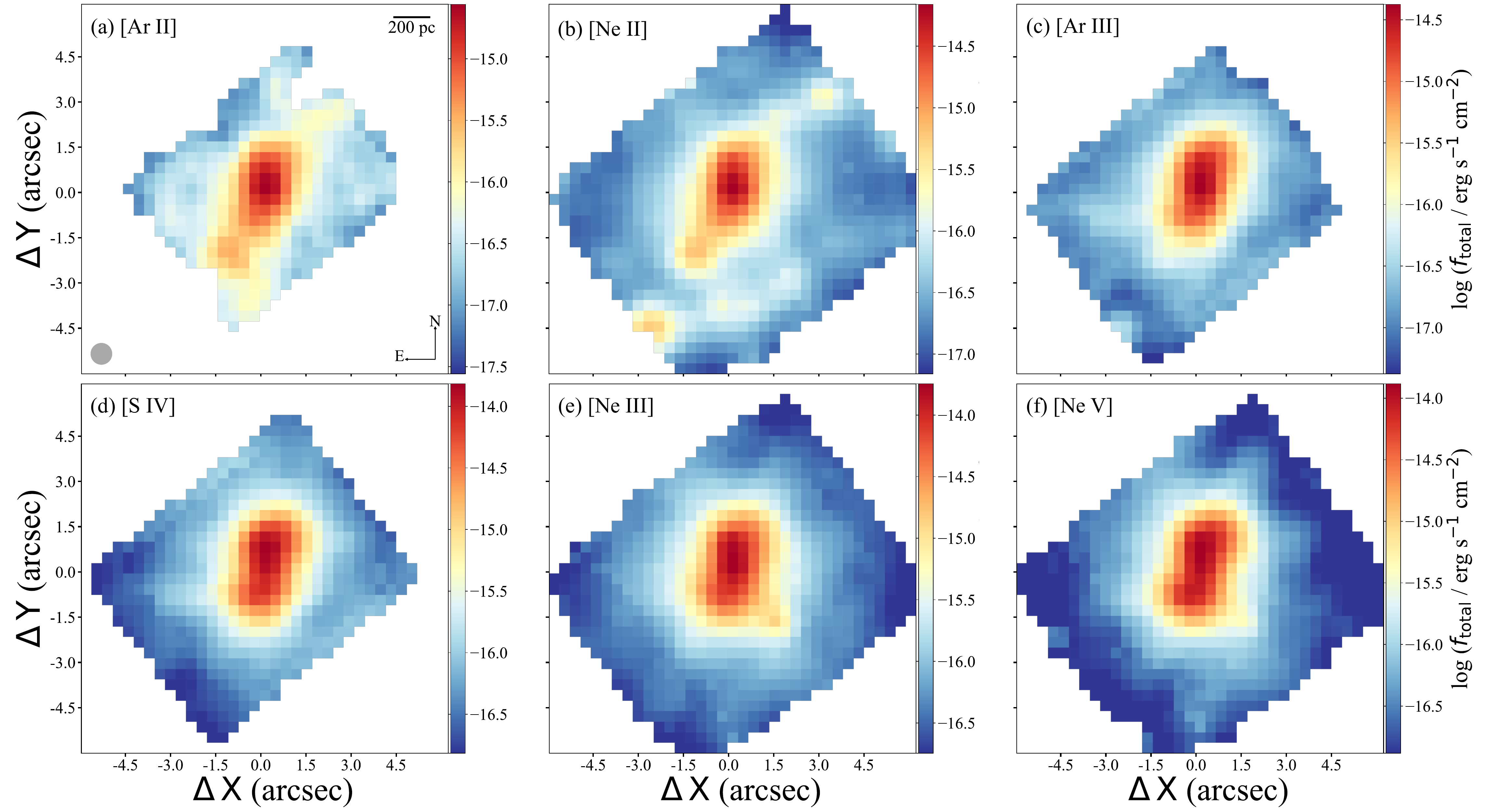}}
\caption{The same as Figure~\ref{figa1} but for NGC\,3081.}\label{figf1}
\end{figure*}

\begin{figure*}[!ht]
\figurenum{F2}
\center{\includegraphics[width=0.8\linewidth]{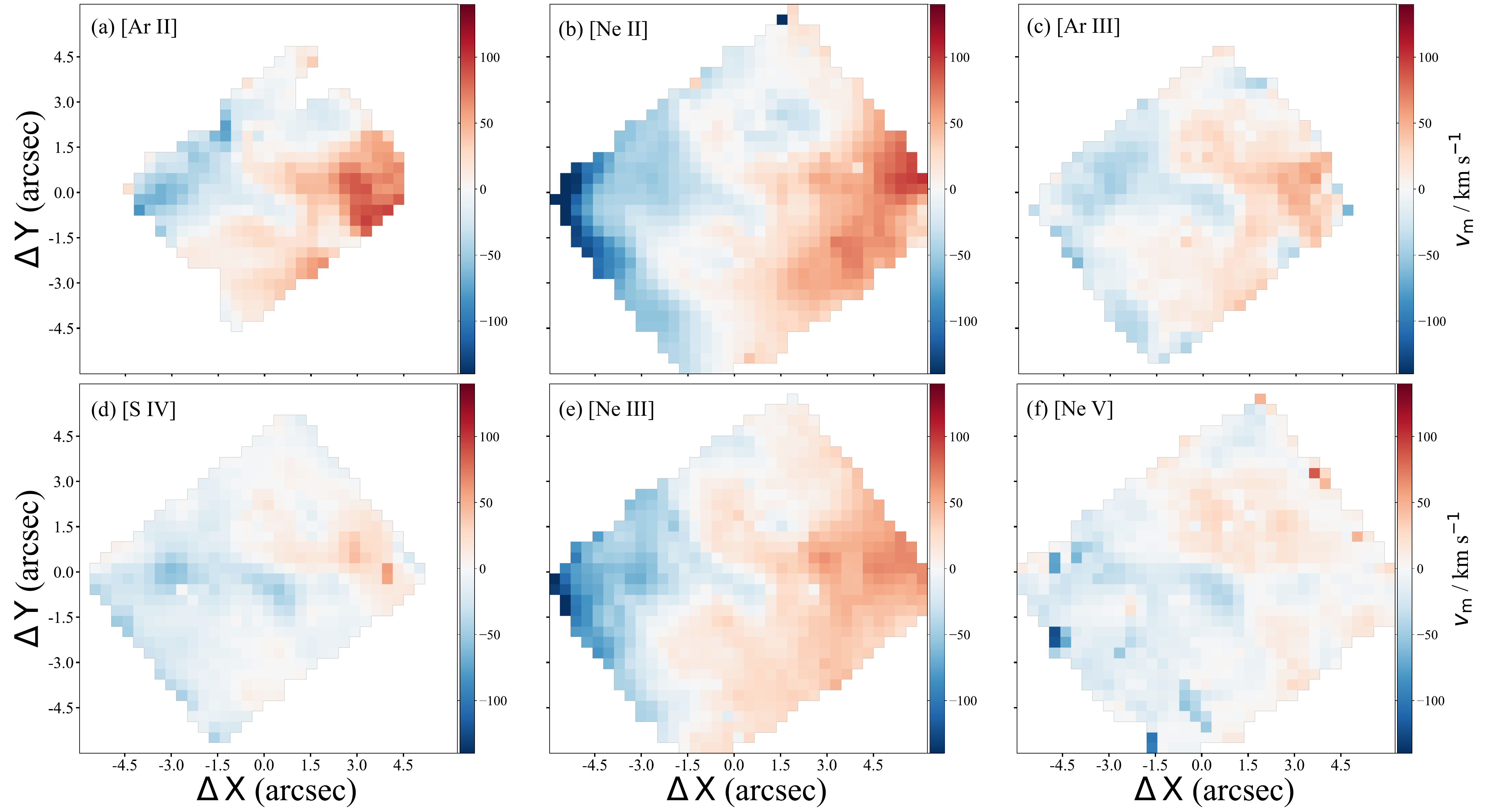}}
\caption{The same as Figure~\ref{figa2} but for NGC\,3081.}\label{figf2}
\end{figure*}

\begin{figure*}[!ht]
\figurenum{F3}
\center{\includegraphics[width=0.8\linewidth]{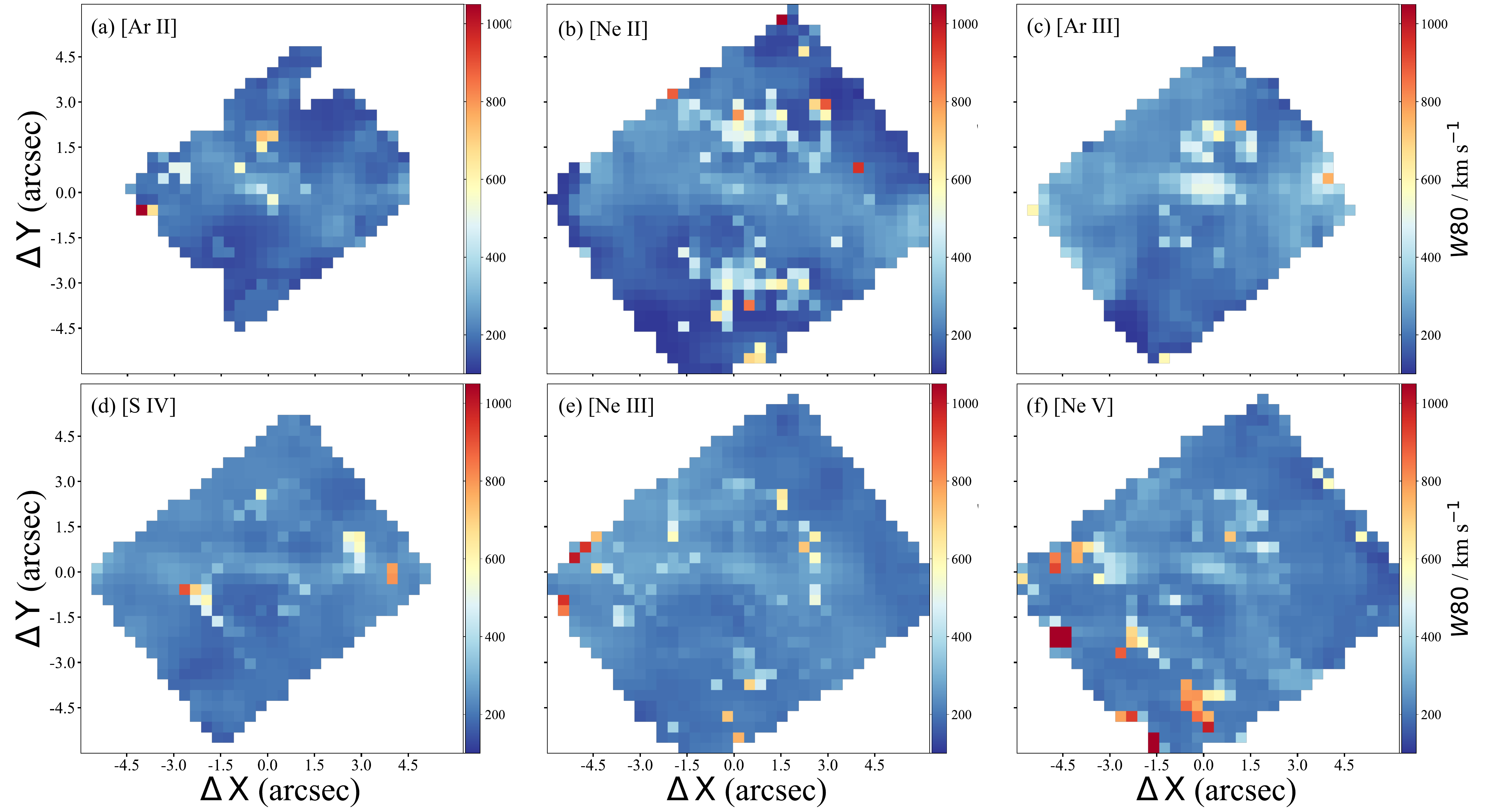}}
\caption{The same as Figure~\ref{figa3} but for NGC\,3081.}\label{figf3}
\end{figure*}

%\begin{figure*}[!ht]
%\figurenum{F4}
%\center{\includegraphics[width=0.8\linewidth]{NGC3081_Ionmass.pdf}}
%\caption{The same as Figure~\ref{figa4} but for NGC\,3081.}\label{figf4}
%\end{figure*}

%\begin{comment}

%\end{comment}


\begin{thebibliography}{}
\expandafter\ifx\csname natexlab\endcsname\relax\def\natexlab#1{#1}\fi

\bibitem[Alarie \& Morisset(2019)]{Alarie&Morisset2019} Alarie, A. \& Morisset, C.\ 2019, \rmxaa, 55, 377

\bibitem[Alonso-Herrero et al.(2021)]{Alonso-Herrero.etal.2021} Alonso-Herrero, A., Garc{\'\i}a-Burillo, S., H{\"o}nig, S.~F., et al.\ 2021, \aap, 652, A99

\bibitem[Alonso-Herrero et al.(2023)]{Alonso-Herrero.etal.2023} Alonso-Herrero, A., Garc{\'\i}a-Burillo, S., Pereira-Santaella, M., et al.\ 2023, \aap, 675, A88

\bibitem[Aniano et al.(2011)]{Aniano.etal.2011} Aniano, G., Draine, B.~T., Gordon, K.~D., et al.\ 2011, \pasp, 123, 1218

\bibitem[Argyriou et al.(2023)]{Argyriou.etal.2023} Argyriou, I., Glasse, A., Law, D.~R., et al.\ 2023, \aap, 675, A111

\bibitem[Argyriou et al.(2020)]{Argyriou.etal.2020} Argyriou, I., Wells, M., Glasse, A., et al.\ 2020, \aap, 641, A150

\bibitem[Armus et al.(2007)]{Armus.etal.2007} Armus, L., Charmandaris, V., Bernard-Salas, J., et al.\ 2007, \apj, 656, 148

\bibitem[Armus et al.(2023)]{Armus.etal.2023} Armus, L., Lai, T., U, V., et al.\ 2023, \apjl, 942, L37

\bibitem[Arribas et al.(2014)]{Arribas.etal.2014} Arribas, S., Colina, L., Bellocchi, E., et al.\ 2014, \aap, 568, A14

\bibitem[Asmus et al.(2014)]{Asmus.etal.2014} Asmus, D., H{\"o}nig, S.~F., Gandhi, P., et al.\ 2014, \mnras, 439, 1648

%\bibitem[Astropy Collaboration et al.(2013)]{AstropyCollaboration2013} Astropy Collaboration, Robitaille, T.~P., Tollerud, E.~J., et al.\ 2013, \aap, 558, A33

\bibitem[Audibert et al.(2023)]{Audibert.etal.2023} Audibert, A., Ramos Almeida, C., Garc{\'\i}a-Burillo, S., et al.\ 2023, \aap, 671, L12

\bibitem[Bae \& Woo(2016)]{Bae&Woo2016} Bae, H.-J. \& Woo, J.-H.\ 2016, \apj, 828, 97

\bibitem[Barbary et al.(2017)]{Barbary.etal.2017} Barbary, K., Boone, K., Craig, M., et al.\ 2017, kbarbary/sep: v1.0.2, Zenodo, doi:10.5281/zenodo.896928

\bibitem[Baumgartner et al.(2013)]{Baumgartner.etal.2013} Baumgartner, W.~H., Tueller, J., Markwardt, C.~B., et al.\ 2013, \apjs, 207, 19

\bibitem[Bellocchi et al.(2013)]{Bellocchi.etal.2013} Bellocchi, E., Arribas, S., Colina, L., et al.\ 2013, \aap, 557, A59

\bibitem[Bellocchi et al.(2019)]{Bellocchi.etal.2019} Bellocchi, E., Villar Mart{\'\i}n, M., Cabrera-Lavers, A., et al.\ 2019, \aap, 626, A89

\bibitem[Baron \& Netzer(2019)]{Baron&Netzer2019} Baron, D. \& Netzer, H.\ 2019, \mnras, 486, 4290

\bibitem[Benson et al.(2003)]{Benson.etal.2003} Benson, A.~J., Bower, R.~G., Frenk, C.~S., et al.\ 2003, \apj, 599, 38

\bibitem[Bertin \& Arnouts(1996)]{Bertin&Arnouts1996} Bertin, E. \& Arnouts, S.\ 1996, \aaps, 117, 393

\bibitem[Bessiere et al.(2024)]{Bessiere.etal.2024} Bessiere, P.~S., Ramos Almeida, C., Holden, L.~R., et al.\ 2024, arXiv:2405.06421

%\bibitem[Bradley et al.(2019)]{Bradley.etal.2019} Bradley, L., Sipocz, B., Robitaille, T., et al.\ 2019, astropy/photutils: v1.9.0 Zenodo, doi:10.5281/zenodo.8248020

\bibitem[Burtscher et al.(2021)]{Burtscher.etal.2021} Burtscher, L., Davies, R.~I., Shimizu, T.~T., et al.\ 2021, \aap, 654, A132

\bibitem[Bushouse et al.(2023)]{Bushouse.etal.2023} Bushouse, H., Eisenhamer, J., Dencheva, N., et al.\ 2023, spacetelescope/jwst: JWST v1.11.4, Zenodo, doi:10.5281/zenodo.8247246

\bibitem[Cano-D{\'\i}az et al.(2012)]{Cano-Diaz.etal.2012} Cano-D{\'\i}az, M., Maiolino, R., Marconi, A., et al.\ 2012, \aap, 537, L8

\bibitem[Caglar et al.(2020)]{Caglar.etal.2020} Caglar, T., Burtscher, L., Brandl, B., et al.\ 2020, \aap, 634, A114

\bibitem[Carniani et al.(2015)]{Carniani.etal.2015} Carniani, S., Marconi, A., Maiolino, R., et al.\ 2015, \aap, 580, A102

%\bibitem[Carrasco et al.(2018)]{Carrasco.etal.2018} Carrasco, E., Gil de Paz, A., Gallego, J., et al.\ 2018, \procspie, 10702, 1070216

\bibitem[Cazzoli et al.(2020)]{Cazzoli.etal.2020} Cazzoli, S., Gil de Paz, A., M{\'a}rquez, I., et al.\ 2020, \mnras, 493, 3656

\bibitem[Cazzoli et al.(2022)]{Cazzoli.etal.2022} Cazzoli, S., Hermosa Mu{\~n}oz, L., M{\'a}rquez, I., et al.\ 2022, \aap, 664, A135

\bibitem[Chatzikos et al.(2023)]{Chatzikos.etal.2023} Chatzikos, M., Bianchi, S., Camilloni, F., et al.\ 2023, \rmxaa, 59, 327

\bibitem[Chisholm et al.(2017)]{Chisholm.etal.2017} Chisholm, J., Tremonti, C.~A., Leitherer, C., et al.\ 2017, \mnras, 469, 4831

\bibitem[Cheung et al.(2016)]{Cheung.etal.2016} Cheung, E., Bundy, K., Cappellari, M., et al.\ 2016, \nat, 533, 504

\bibitem[Cicone et al.(2014)]{Cicone.etal.2014} Cicone, C., Maiolino, R., Sturm, E., et al.\ 2014, \aap, 562, A21

%\bibitem[Comerford et al.(2013)]{Comerford.etal.2013} Comerford, J.~M., Schluns, K., Greene, J.~E., et al.\ 2013, \apj, 777, 64

\bibitem[Concas et al.(2019)]{Concas.etal.2019} Concas, A., Popesso, P., Brusa, M., et al.\ 2019, \aap, 622, A188

\bibitem[D'Agostino et al.(2019)]{DAgostino.etal.2019} D'Agostino, J.~J., Kewley, L.~J., Groves, B.~A., et al.\ 2019, \mnras, 487, 4153

\bibitem[Dale et al.(2006)]{Dale.etal.2006} Dale, D.~A., Smith, J.~D.~T., Armus, L., et al.\ 2006, \apj, 646, 161

\bibitem[Dav{\'e} et al.(2019)]{Dave.etal.2019} Dav{\'e}, R., Angl{\'e}s-Alc{\'a}zar, D., Narayanan, D., et al.\ 2019, \mnras, 486, 2827

%\bibitem[Dav{\'e} et al.(2020)]{Dave.etal.2020} Dav{\'e}, R., Crain, R.~A., Stevens, A.~R.~H., et al.\ 2020, \mnras, 497, 146

\bibitem[Davies et al.(2020)]{Davies.etal.2020} Davies, R., Baron, D., Shimizu, T., et al.\ 2020, \mnras, 498, 4150

\bibitem[Davies et al.(2015)]{Davies.etal.2015} Davies, R.~I., Burtscher, L., Rosario, D., et al.\ 2015, \apj, 806, 127

\bibitem[Davies et al.(2024)]{Davies.etal.2024} Davies, R.~I., Shimizu, T., Pereira-Santaella, M., et al.\ 2024, \aap, accepted; arXiv:2406.17072

\bibitem[Davies et al.(2020)]{DaviesJJ.etal.2020} Davies, J.~J., Crain, R.~A., Oppenheimer, B.~D., et al.\ 2020, \mnras, 491, 4462

\bibitem[Di Matteo et al.(2005)]{DiMatteo.etal.2005} Di Matteo, T., Springel, V., \& Hernquist, L.\ 2005, \nat, 433, 604

\bibitem[Draine(2011)]{Draine2011} Draine, B.~T.\ 2011, Physics of the Interstellar and Intergalactic Medium by Bruce T. Draine. Published by Princeton University Press, 2011. ISBN: 978-0-691-12214-4

\bibitem[Durr{\'e} \& Mould(2018)]{Durre&Mould2018} Durr{\'e}, M. \& Mould, J.\ 2018, \apj, 867, 149

\bibitem[Esparza-Arredondo et al.(2024)]{Esparza-Arredondo.etal.2024} Esparza-Arredondo, D., Ramos Almeida, C., Audibert, A., et al.\ 2024, \aap, submitted

\bibitem[Esposito et al.(2024)]{Esposito.etal.2024} Esposito, F., Alonso-Herrero, A., Garc{\'\i}a-Burillo, S., et al.\ 2024, \aap, 686, A46

\bibitem[Fabian(2012)]{Fabian2012} Fabian, A.~C.\ 2012, \araa, 50, 455

\bibitem[Feltre et al.(2023)]{Feltre.etal.2023} Feltre, A., Gruppioni, C., Marchetti, L., et al.\ 2023, \aap, 675, A74

\bibitem[Ferland et al.(2017)]{Ferland.etal.2017} Ferland, G.~J., Chatzikos, M., Guzm{\'a}n, F., et al.\ 2017, \rmxaa, 53, 385

\bibitem[Fern{\'a}ndez-Ontiveros et al.(2021)]{Fernandez-Ontiveros.etal.2021} Fern{\'a}ndez-Ontiveros, J.~A., P{\'e}rez-Montero, E., V{\'\i}lchez, J.~M., et al.\ 2021, \aap, 652, A23

\bibitem[Fischer et al.(2011)]{Fischer.etal.2011} Fischer, T.~C., Crenshaw, D.~M., Kraemer, S.~B., et al.\ 2011, \apj, 727, 71

\bibitem[Fischer et al.(2013)]{Fischer.etal.2013} Fischer, T.~C., Crenshaw, D.~M., Kraemer, S.~B., et al.\ 2013, \apjs, 209, 1

\bibitem[Fischer et al.(2018)]{Fischer.etal.2018} Fischer, T.~C., Kraemer, S.~B., Schmitt, H.~R., et al.\ 2018, \apj, 856, 102

\bibitem[Fischer et al.(2017)]{Fischer.etal.2017} Fischer, T.~C., Machuca, C., Diniz, M.~R., et al.\ 2017, \apj, 834, 30

\bibitem[Fiore et al.(2017)]{Fiore.etal.2017} Fiore, F., Feruglio, C., Shankar, F., et al.\ 2017, \aap, 601, A143

\bibitem[Fluetsch et al.(2019)]{Fluetsch.etal.2019} Fluetsch, A., Maiolino, R., Carniani, S., et al.\ 2019, \mnras, 483, 4586

\bibitem[F{\"o}rster Schreiber et al.(2019)]{Forster-Schreiber.etal.2019} F{\"o}rster Schreiber, N.~M., {\"U}bler, H., Davies, R.~L., et al.\ 2019, \apj, 875, 21

\bibitem[Freeman et al.(2019)]{Freeman.etal.2019} Freeman, W.~R., Siana, B., Kriek, M., et al.\ 2019, \apj, 873, 102

%\bibitem[Gallagher et al.(2019)]{Gallagher.etal.2019} Gallagher, R., Maiolino, R., Belfiore, F., et al.\ 2019, \mnras, 485, 3409

\bibitem[Garc{\'\i}a-Bernete et al.(2021)]{Garcia-Bernete.etal.2021} Garc{\'\i}a-Bernete, I., Alonso-Herrero, A., Garc{\'\i}a-Burillo, S., et al.\ 2021, \aap, 645, A21

\bibitem[Garc{\'\i}a-Bernete et al.(2024a)]{Garcia-Bernete.etal.2024a} Garc{\'\i}a-Bernete, I., Alonso-Herrero, A., Rigopoulou, D., et al.\ 2024a, \aap, 681, L7

\bibitem[Garc{\'\i}a-Bernete et al.(2024b)]{Garcia-Bernete.etal.2024b} Garc{\'\i}a-Bernete, I., Pereira-Santaella, M., Gonz{\'a}lez-Alfonso, E., et al.\ 2024b, \aap, 682, L5

\bibitem[Garc{\'\i}a-Bernete et al.(2024c)]{Garcia-Bernete.etal.2024c} Garc{\'\i}a-Bernete, I., Rigopoulou, D., Donnan, F. R., et al.\ 2024c, \aap, accepted; arXiv:2409.05686

\bibitem[Garc{\'\i}a-Bernete et al.(2016)]{Garcia-Bernete.etal.2016} Garc{\'\i}a-Bernete, I., Ramos Almeida, C., Acosta-Pulido, J.~A., et al.\ 2016, \mnras, 463, 3531

\bibitem[Garc{\'\i}a-Bernete et al.(2022a)]{Garcia-Bernete.etal.2022a} Garc{\'\i}a-Bernete, I., Rigopoulou, D., Alonso-Herrero, A., et al.\ 2022a, \aap, 666, L5.

\bibitem[Garc{\'\i}a-Bernete et al.(2022b)]{Garcia-Bernete.etal.2022b} Garc{\'\i}a-Bernete, I., Rigopoulou, D., Alonso-Herrero, A., et al.\ 2022b, \mnras, 509, 4256

\bibitem[Garc{\'\i}a-Burillo et al.(2021)]{Garcia-Burillo.etal.2021} Garc{\'\i}a-Burillo, S., Alonso-Herrero, A., Ramos Almeida, C., et al.\ 2021, \aap, 652, A98

\bibitem[Garc{\'\i}a-Burillo et al.(2019)]{Garcia-Burillo.etal.2019} Garc{\'\i}a-Burillo, S., Combes, F., Ramos Almeida, C., et al.\ 2019, \aap, 632, A61

\bibitem[Garc{\'\i}a-Burillo et al.(2014)]{Garcia-Burillo.etal.2014} Garc{\'\i}a-Burillo, S., Combes, F., Usero, A., et al.\ 2014, \aap, 567, A125

\bibitem[Gardner et al.(2023)]{Gardner.etal.2023} Gardner, J.~P., Mather, J.~C., Abbott, R., et al.\ 2023, \pasp, 135, 068001

\bibitem[Gasman et al.(2023)]{Gasman.etal.2023} Gasman, D., Argyriou, I., Sloan, G.~C., et al.\ 2023, \aap, 673, A102

%\bibitem[Gil de Paz et al.(2016)]{GildePaz.etal.2016} Gil de Paz, A., Carrasco, E., Gallego, J., et al.\ 2016, \procspie, 9908, 99081K

\bibitem[Genzel et al.(1998)]{Genzel.etal.1998} Genzel, R., Lutz, D., Sturm, E., et al.\ 1998, \apj, 498, 579

%\bibitem[Glasse et al.(2015)]{Glasse.etal.2015} Glasse, A., Rieke, G.~H., Bauwens, E., et al.\ 2015, \pasp, 127, 686

\bibitem[Hao et al.(2005)]{Hao.etal.2005} Hao, L., Strauss, M.~A., Tremonti, C.~A., et al.\ 2005, \aj, 129, 1783

\bibitem[Harrison(2017)]{Harrison2017} Harrison, C.~M.\ 2017, Nature Astronomy, 1, 0165

\bibitem[Harrison \& Ramos Almeida(2024)]{Harrison&RamosAlmeida2024} Harrison, C.~M. \& Ramos Almeida, C.\ 2024, Galaxies, 12, 17

\bibitem[Harrison et al.(2014)]{Harrison.etal.2014} Harrison, C.~M., Alexander, D.~M., Mullaney, J.~R., et al.\ 2014, \mnras, 441, 3306

\bibitem[Harrison et al.(2016)]{Harrison.etal.2016} Harrison, C.~M., Alexander, D.~M., Mullaney, J.~R., et al.\ 2016, \mnras, 456, 1195

\bibitem[Harrison et al.(2018)]{Harrison.etal.2018} Harrison, C.~M., Costa, T., Tadhunter, C.~N., et al.\ 2018, Nature Astronomy, 2, 198

%\bibitem[Heckman \& Best(2014)]{Heckman&Best2014} Heckman, T.~M. \& Best, P.~N.\ 2014, \araa, 52, 589

\bibitem[Heckman et al.(1990)]{Heckman.etal.1990} Heckman, T.~M., Armus, L., \& Miley, G.~K.\ 1990, \apjs, 74, 833

\bibitem[Hermosa Mu{\~n}oz et al.(2024a)]{Hermosa-Munoz.etal.2024a} Hermosa Mu{\~n}oz, L., Alonso-Herrero, A., Pereira-Santaella, M., \ 2024a, \aap, accepted; arXiv:2407.15807 

\bibitem[Hermosa Mu{\~n}oz et al.(2024b)]{Hermosa-Munoz.etal.2024b} Hermosa Mu{\~n}oz, L., Cazzoli, S., M{\'a}rquez, I., et al.\ 2024b, \aap, 683, A43

\bibitem[Hervella Seoane et al.(2023)]{HervellaSeoane.etal.2023} Hervella Seoane, K., Ramos Almeida, C., Acosta-Pulido, J.~A., et al.\ 2023, \aap, 680, A71

%\bibitem[Ho \& Keto(2007)]{Ho&Keto2007} Ho, L.~C. \& Keto, E.\ 2007, \apj, 658, 314

\bibitem[Hopkins et al.(2008)]{Hopkins.etal.2008} Hopkins, P.~F., Hernquist, L., Cox, T.~J., et al.\ 2008, \apjs, 175, 356

\bibitem[Houck et al.(2004)]{Houck.etal.2004} Houck, J.~R., Roellig, T.~L., van Cleve, J., et al.\ 2004, \apjs, 154, 18

\bibitem[Kakkad et al.(2020)]{Kakkad.etal.2020} Kakkad, D., Mainieri, V., Vietri, G., et al.\ 2020, \aap, 642, A147

\bibitem[Kakkad et al.(2022)]{Kakkad.etal.2022} Kakkad, D., Sani, E., Rojas, A.~F., et al.\ 2022, \mnras, 511, 2105

\bibitem[King \& Pounds(2015)]{King&Pounds2015} King, A. \& Pounds, K.\ 2015, \araa, 53, 115

%\bibitem[Kormendy \& Ho(2013)]{Kormendy&Ho2013} Kormendy, J. \& Ho, L.~C.\ 2013, \araa, 51, 511

\bibitem[Krajnovi{\'c} et al.(2006)]{Krajnovic.etal.2006} Krajnovi{\'c}, D., Cappellari, M., de Zeeuw, P.~T., et al.\ 2006, \mnras, 366, 787

\bibitem[Labiano et al.(2021)]{Labiano.etal.2021} Labiano, A., Argyriou, I., {\'A}lvarez-M{\'a}rquez, J., et al.\ 2021, \aap, 656, A57

\bibitem[Labiano et al.(2016)]{Labiano.etal.2016} Labiano, A., Azzollini, R., Bailey, J., et al.\ 2016, \procspie, 9910, 99102W

\bibitem[Law et al.(2023)]{Law.etal.2023} Law, D.~R., E. Morrison, J., Argyriou, I., et al.\ 2023, \aj, 166, 45

\bibitem[Lamperti et al.(2022)]{Lamperti.etal.2022} Lamperti, I., Pereira-Santaella, M., Perna, M., et al.\ 2022, \aap, 668, A45

\bibitem[Leung et al.(2019)]{Leung.etal.2019} Leung, G.~C.~K., Coil, A.~L., Aird, J., et al.\ 2019, \apj, 886, 11

\bibitem[Liu et al.(2013)]{Liu.etal.2013} Liu, G., Zakamska, N.~L., Greene, J.~E., et al.\ 2013, \mnras, 436, 2576

\bibitem[Lopez-Rodriguez et al.(2024)]{Lopez-Rodriguez.etal.2024} Lopez-Rodriguez, E., et al.\ 2024, \apj, in preparation

\bibitem[Luridiana et al.(2015)]{Luridiana.etal.2015} Luridiana, V., Morisset, C., \& Shaw, R.~A.\ 2015, \aap, 573, A42

%\bibitem[Madau \& Dickinson(2014)]{Madau&Dickinson2014} Madau, P. \& Dickinson, M.\ 2014, \araa, 52, 415

%\bibitem[Maiolino et al.(2017)]{Maiolino.etal.2017} Maiolino, R., Russell, H.~R., Fabian, A.~C., et al.\ 2017, \nat, 544, 202

\bibitem[Maksym et al.(2023)]{Maksym.etal.2023} Maksym, W.~P., Elvis, M., Fabbiano, G., et al.\ 2023, \apj, 951, 146

\bibitem[McCarthy et al.(2011)]{McCarthy.etal.2011} McCarthy, I.~G., Schaye, J., Bower, R.~G., et al.\ 2011, \mnras, 412, 1965

\bibitem[McElroy et al.(2015)]{McElroy.etal.2015} McElroy, R., Croom, S.~M., Pracy, M., et al.\ 2015, \mnras, 446, 2186

\bibitem[McNamara \& Nulsen(2007)]{McNamara&Nulsen2007} McNamara, B.~R. \& Nulsen, P.~E.~J.\ 2007, \araa, 45, 117

\bibitem[Meenakshi et al.(2022)]{Meenakshi.etal.2022} Meenakshi, M., Mukherjee, D., Wagner, A.~Y., et al.\ 2022, \mnras, 516, 766

\bibitem[Melioli et al.(2015)]{Melioli.etal.2015} Melioli, C., Brighenti, F., \& D'Ercole, A.\ 2015, \mnras, 446, 299

\bibitem[Morganti et al.(1999)]{Morganti.etal.1999} Morganti, R., Tsvetanov, Z.~I., Gallimore, J., et al.\ 1999, \aaps, 137, 457

\bibitem[Morisset et al.(2015)]{Morisset.etal.2015} Morisset, C., Delgado-Inglada, G., \& Flores-Fajardo, N.\ 2015, \rmxaa, 51, 103

\bibitem[Mukherjee et al.(2018a)]{Mukherjee.etal.2018a} Mukherjee, D., Bicknell, G.~V., Wagner, A.~Y., et al.\ 2018, \mnras, 479, 5544

\bibitem[Mukherjee et al.(2018b)]{Mukherjee.etal.2018b} Mukherjee, D., Wagner, A.~Y., Bicknell, G.~V., et al.\ 2018, \mnras, 476, 80

\bibitem[M{\"u}ller-S{\'a}nchez et al.(2011)]{Muller-Sanchez.etal.2011} M{\"u}ller-S{\'a}nchez, F., Prieto, M.~A., Hicks, E.~K.~S., et al.\ 2011, \apj, 739, 69

\bibitem[Nagar et al.(1999)]{Nagar.etal.1999} Nagar, N.~M., Wilson, A.~S., Mulchaey, J.~S., et al.\ 1999, \apjs, 120, 209

\bibitem[Nyland et al.(2018)]{Nyland.etal.2018} Nyland, K., Harwood, J.~J., Mukherjee, D., et al.\ 2018, \apj, 859, 23

\bibitem[Orienti \& Prieto(2010)]{Orienti&Prieto2010} Orienti, M. \& Prieto, M.~A.\ 2010, \mnras, 401, 2599

\bibitem[Page et al.(2012)]{Page.etal.2012} Page, M.~J., Symeonidis, M., Vieira, J.~D., et al.\ 2012, \nat, 485, 213

\bibitem[Peralta de Arriba et al.(2023)]{PeraltadeArriba.etal.2023} Peralta de Arriba, L., Alonso-Herrero, A., Garc{\'\i}a-Burillo, S., et al.\ 2023, \aap, 675, A58

\bibitem[Pereira-Santaella et al.(2022)]{Pereira-Santaella.etal.2022} Pereira-Santaella, M., {\'A}lvarez-M{\'a}rquez, J., Garc{\'\i}a-Bernete, I., et al.\ 2022, \aap, 665, L11

\bibitem[Pereira-Santaella et al.(2010)]{Pereira-Santaella.etal.2010} Pereira-Santaella, M., Diamond-Stanic, A.~M., Alonso-Herrero, A., et al.\ 2010, \apj, 725, 2270

\bibitem[Pereira-Santaella et al.(2024)]{Pereira-Santaella.etal.2024} Pereira-Santaella, M., Garc{\'\i}a-Bernete, I., Gonz{\'a}lez-Alfonso, E., et al.\ 2024, \aap, 685, L13

\bibitem[Pereira-Santaella et al.(2017)]{Pereira-Santaella.etal.2017} Pereira-Santaella, M., Rigopoulou, D., Farrah, D., et al.\ 2017, \mnras, 470, 1218

\bibitem[P{\'e}rez-D{\'\i}az et al.(2022)]{Perez-Diaz.etal.2022} P{\'e}rez-D{\'\i}az, B., P{\'e}rez-Montero, E., Fern{\'a}ndez-Ontiveros, J.~A., et al.\ 2022, \aap, 666, A115

%\bibitem[Pillepich et al.(2018)]{Pillepich.etal.2018} Pillepich, A., Springel, V., Nelson, D., et al.\ 2018, \mnras, 473, 4077

\bibitem[Ramos Almeida et al.(2022)]{Ramos Almeida.etal.2022} Ramos Almeida, C., Bischetti, M., Garc{\'\i}a-Burillo, S., et al.\ 2022, \aap, 658, A155

\bibitem[Ramos Almeida et al.(2023)]{Ramos Almeida.etal.2023} Ramos Almeida, C., Esparza-Arredondo, D., Gonz{\'a}lez-Mart{\'\i}n, O., et al.\ 2023, \aap, 669, L5

\bibitem[Ramos Almeida et al.(2011)]{Ramos Almeida.etal.2011} Ramos Almeida, C., S{\'a}nchez-Portal, M., P{\'e}rez Garc{\'\i}a, A.~M., et al.\ 2011, \mnras, 417, L46

\bibitem[Ricci et al.(2017)]{Ricci.etal.2017} Ricci, C., Trakhtenbrot, B., Koss, M.~J., et al.\ 2017, \apjs, 233, 17

\bibitem[Rieke et al.(2015)]{Rieke.etal.2015} Rieke, G.~H., Wright, G.~S., B{\"o}ker, T., et al.\ 2015, \pasp, 127, 584

\bibitem[Riffel et al.(2023)]{Riffel.etal.2023} Riffel, R.~A., Storchi-Bergmann, T., Riffel, R., et al.\ 2023, \mnras, 521, 1832

\bibitem[Rigby et al.(2023)]{Rigby.etal.2023} Rigby, J., Perrin, M., McElwain, M., et al.\ 2023, \pasp, 135, 048001

\bibitem[Roberts-Borsani et al.(2020)]{Roberts-Borsani.etal.2020} Roberts-Borsani, G.~W., Saintonge, A., Masters, K.~L., et al.\ 2020, \mnras, 493, 3081

\bibitem[Rubin et al.(2014)]{Rubin.etal.2014} Rubin, K.~H.~R., Prochaska, J.~X., Koo, D.~C., et al.\ 2014, \apj, 794, 156

\bibitem[Rupke \& Veilleux(2011)]{Rupke&Veilleux2011} Rupke, D.~S.~N. \& Veilleux, S.\ 2011, \apjl, 729, L27

\bibitem[Rupke et al.(2005)]{Rupke.etal.2005} Rupke, D.~S., Veilleux, S., \& Sanders, D.~B.\ 2005, \apj, 632, 751

\bibitem[Ruschel-Dutra et al.(2021)]{Ruschel-Dutra.etal.2021} Ruschel-Dutra, D., Storchi-Bergmann, T., Schnorr-M{\"u}ller, A., et al.\ 2021, \mnras, 507, 74

\bibitem[Sajina et al.(2022)]{Sajina.etal.2022} Sajina, A., Lacy, M., \& Pope, A.\ 2022, Universe, 8, 356

%\bibitem[Schaye et al.(2015)]{Schaye.etal.2015} Schaye, J., Crain, R.~A., Bower, R.~G., et al.\ 2015, \mnras, 446, 521

\bibitem[Schawinski et al.(2007)]{Schawinski.etal.2007} Schawinski, K., Thomas, D., Sarzi, M., et al.\ 2007, \mnras, 382, 1415

%\bibitem[Schnorr-M{\"u}ller et al.(2016)]{Schnorr-Muller.etal.2016} Schnorr-M{\"u}ller, A., Storchi-Bergmann, T., Robinson, A., et al.\ 2016, \mnras, 457, 972

\bibitem[Shapley et al.(2003)]{Shapley.etal.2003} Shapley, A.~E., Steidel, C.~C., Pettini, M., et al.\ 2003, \apj, 588, 65

\bibitem[Shimizu et al.(2019)]{Shimizu.etal.2019} Shimizu, T.~T., Davies, R.~I., Lutz, D., et al.\ 2019, \mnras, 490, 5860

\bibitem[Smaji{\'c} et al.(2012)]{Smajic.etal.2012} Smaji{\'c}, S., Fischer, S., Zuther, J., et al.\ 2012, \aap, 544, A105

\bibitem[Somerville \& Dav{\'e}(2015)]{Somerville&Dave2015} Somerville, R.~S. \& Dav{\'e}, R.\ 2015, \araa, 53, 51

\bibitem[Speranza et al.(2024)]{Speranza.etal.2024} Speranza, G., Ramos Almeida, C., Acosta-Pulido, J.~A., et al.\ 2024, \aap, 681, A63

%\bibitem[Storchi-Bergmann \& Schnorr-M{\"u}ller(2019)]{Storchi-Bergmann&Schnorr-Muller2019} Storchi-Bergmann, T. \& Schnorr-M{\"u}ller, A.\ 2019, Nature Astronomy, 3, 48

\bibitem[Sun et al.(2017)]{Sun.etal.2017} Sun, A.-L., Greene, J.~E., \& Zakamska, N.~L.\ 2017, \apj, 835, 222

\bibitem[Sun et al.(2018)]{Sun.etal.2018} Sun, S., Guainazzi, M., Ni, Q., et al.\ 2018, \mnras, 478, 1900

\bibitem[Sutherland \& Dopita(2017)]{Sutherland&Dopita2017} Sutherland, R.~S. \& Dopita, M.~A.\ 2017, \apjs, 229, 34

\bibitem[Thean et al.(2000)]{Thean.etal.2000} Thean, A., Pedlar, A., Kukula, M.~J., et al.\ 2000, \mnras, 314, 573

\bibitem[Theureau et al.(2007)]{Theureau.etal.2007} Theureau, G., Hanski, M.~O., Coudreau, N., et al.\ 2007, \aap, 465, 71

\bibitem[Thomas et al.(2017)]{Thomas.etal.2017} Thomas, A.~D., Dopita, M.~A., Shastri, P., et al.\ 2017, \apjs, 232, 11

\bibitem[U et al.(2022)]{U.etal.2022} U, V., Lai, T., Bianchin, M., et al.\ 2022, \apjl, 940, L5

\bibitem[Ulivi et al.(2024)]{Ulivi.etal.2024} Ulivi, L., Venturi, G., Cresci, G., et al.\ 2024, \aap, 685, A122

\bibitem[Veilleux et al.(2005)]{Veilleux.etal.2005} Veilleux, S., Cecil, G., \& Bland-Hawthorn, J.\ 2005, \araa, 43, 769

\bibitem[Veilleux et al.(2020)]{Veilleux.etal.2020} Veilleux, S., Maiolino, R., Bolatto, A.~D., et al.\ 2020, \aapr, 28, 2

\bibitem[Veilleux et al.(2013)]{Veilleux.etal.2013} Veilleux, S., Mel{\'e}ndez, M., Sturm, E., et al.\ 2013, \apj, 776, 27

\bibitem[Venturi et al.(2021)]{Venturi.etal.2021} Venturi, G., Cresci, G., Marconi, A., et al.\ 2021, \aap, 648, A17

\bibitem[Venturi et al.(2023)]{Venturi.etal.2023} Venturi, G., Treister, E., Finlez, C., et al.\ 2023, \aap, 678, A127

\bibitem[Weinberger et al.(2017)]{Weinberger.etal.2017} Weinberger, R., Springel, V., Hernquist, L., et al.\ 2017, \mnras, 465, 3291

\bibitem[Wells et al.(2015)]{Wells.etal.2015} Wells, M., Pel, J.-W., Glasse, A., et al.\ 2015, \pasp, 127, 646

\bibitem[Weiner et al.(2009)]{Weiner.etal.2009} Weiner, B.~J., Coil, A.~L., Prochaska, J.~X., et al.\ 2009, \apj, 692, 187

\bibitem[Werner et al.(2004)]{Werner.etal.2004} Werner, M.~W., Roellig, T.~L., Low, F.~J., et al.\ 2004, \apjs, 154, 1

\bibitem[Winkel et al.(2023)]{Winkel.etal.2023} Winkel, N., Husemann, B., Singha, M., et al.\ 2023, \aap, 670, A3

\bibitem[Wright et al.(2023)]{Wright.etal.2023} Wright, G.~S., Rieke, G.~H., Glasse, A., et al.\ 2023, \pasp, 135, 048003

\bibitem[Wright et al.(2015)]{Wright.etal.2015} Wright, G.~S., Wright, D., Goodson, G.~B., et al.\ 2015, \pasp, 127, 595

\bibitem[Zhang \& Ho(2023)]{Zhang&Ho2023} Zhang, L. \& Ho, L.~C.\ 2023, \apjl, 953, L9

\bibitem[Zhang et al.(2024)]{Zhang.etal.2024} Zhang, L., Garc{\'\i}a-Bernete, I., Packham, C., et al.\ 2024, ApJL, submitted

\bibitem[Zhang et al.(2022)]{Zhang.etal.2022} Zhang, L., Ho, L.~C., \& Li, A.\ 2022, \apj, 939, 22

\bibitem[Zhang et al.(2021)]{Zhang.etal.2021} Zhang, L., Ho, L.~C., \& Xie, Y.\ 2021, \aj, 161, 29

\bibitem[Zoghbi et al.(2017)]{Zoghbi.etal.2017} Zoghbi, A., Matt, G., Miller, J.~M., et al.\ 2017, \apj, 836, 2

%\bibitem[Zhuang et al.(2019)]{Zhuang.etal.2019} Zhuang, M.-Y., Ho, L.~C., \& Shangguan, J.\ 2019, \apj, 873, 103

%\bibitem[Zhuang et al.(2021)]{Zhuang.etal.2021} Zhuang, M.-Y., Ho, L.~C., \& Shangguan, J.\ 2021, \apj, 906, 38

\bibitem[Zubovas \& Nardini(2020)]{Zubovas&Nardini2020} Zubovas, K. \& Nardini, E.\ 2020, \mnras, 498, 3633

%\bibitem[Zubovas et al.(2013)]{Zubovas.etal.2013} Zubovas, K., Nayakshin, S., King, A., et al.\ 2013, \mnras, 433, 3079


\end{thebibliography}
\end{document}